\documentclass{article} 
\usepackage{iclr2027_conference,times}

\usepackage{microtype}
\usepackage{graphicx}
\usepackage{subcaption}
\usepackage{booktabs}
\usepackage{longtable} 
\usepackage{multirow}
\usepackage{amsmath,amssymb,mathtools,amsthm}
\usepackage{xcolor}
\usepackage{float} 
\usepackage{hyperref}
\usepackage{url}
\usepackage{algorithm}
\usepackage[noend]{algpseudocode}
\algrenewcommand\algorithmicrequire{\textbf{Input:}}
\algrenewcommand\algorithmicensure{\textbf{Output:}}
\algrenewcommand{\algorithmiccomment}[1]{\hfill{\color{black!55}\(\triangleright\)~#1}}
\algnewcommand{\AlgStage}[1]{\Statex{\color{black!60}\textit{#1}}}
\definecolor{citeblue}{rgb}{0.0,0.2,0.6}
\hypersetup{colorlinks=true,linkcolor=magenta,citecolor=citeblue,urlcolor=citeblue,linktoc=section}
\usepackage[capitalize,noabbrev]{cleveref}
\usepackage{enumitem}
\usepackage{tikz}
\usetikzlibrary{positioning,arrows.meta,fit,backgrounds,calc}
\usepackage{pgfplots}
\pgfplotsset{compat=1.18}
\usepgfplotslibrary{groupplots}
\usepackage{wrapfig}
\usepackage{needspace}
\usepackage{placeins}
\definecolor{cNemo}{HTML}{CC3311}
\definecolor{cGemma}{HTML}{0077BB}
\definecolor{cGpt}{HTML}{009988}
\definecolor{cMistral}{HTML}{EE7733}
\definecolor{cQwen}{HTML}{AA4499}
\definecolor{cCond}{HTML}{0077BB}
\definecolor{cAgn}{HTML}{888888}
\definecolor{cShuf}{HTML}{EE7733}

\newcommand{\NumDatasets}{six}
\newcommand{\NumReceivers}{five}
\newcommand{\ProbeRepeats}{6}
\newcommand{\RewriteCandidates}{4}

\newcommand{\NumMessages}{6{,}629}
\newcommand{\NumTrials}{33{,}145}
\newcommand{\NumProbeCalls}{198{,}870}
\newcommand{\NaturalPIRRate}{4.4\%}

\newcommand{\SemanticVerifierPassRate}{92.8\%}

\newcommand{\RewriteModel}{DeepSeek V4 Flash}
\newcommand{\EmbeddingModel}{Qwen3-Embedding-8B}

\newcommand{\CorrectInterpretFailureRate}{56.4\%}
\newcommand{\MisinterpretSuccessRate}{1.4\%}

\newcommand{\MisreadPassShare}{32\%}
\newcommand{\ReceiverYIMin}{1.7\%}
\newcommand{\ReceiverYIMax}{10.7\%}
\newcommand{\RepeatStabilityMaxDiff}{0.07}

\newcommand{\BaseRateAUPRC}{0.045}
\newcommand{\PIRAUROC}{0.732}
\newcommand{\PIRAUPRC}{0.135}

\newcommand{\PIRECE}{0.010}
\newcommand{\AgnosticAUROC}{0.729}

\newcommand{\AgnosticECE}{0.032}

\newcommand{\ExecutionPredAUROC}{0.511}

\newcommand{\MessageOnlyAUROC}{0.552}
\newcommand{\EquivalenceAUROC}{0.500}

\newcommand{\ReceiverTypeNLL}{1.584}
\newcommand{\ReceiverTypeBrier}{0.790}

\newcommand{\HistoryZeroAcc}{20.0\%}
\newcommand{\HistoryShortAcc}{22.6\%}
\newcommand{\HistoryMediumAcc}{24.9\%}
\newcommand{\HistoryLongAcc}{29.0\%}
\newcommand{\UniformPriorNLL}{1.609}
\newcommand{\ShuffledHistoryAccRange}{18.4--21.3\%}

\newcommand{\ConfirmationQuestions}{2,400}
\newcommand{\ConfirmationEpisodes}{36,000}

\newcommand{\QueryQuestions}{6,629}
\newcommand{\QueryEpisodes}{99,435}
\newcommand{\QueryBankSize}{192}
\newcommand{\QueryCost}{0.001}
\newcommand{\QueryBootstrapResamples}{2,000}
\newcommand{\QueryReceiverCalls}{3,159,599}          
\newcommand{\NeverQueryPIRRate}{3.84\%}              
\newcommand{\OracleQueryPIRRate}{3.77\%}             
\newcommand{\AlwaysQueryPIRRate}{3.80\%}             
\newcommand{\DongVoIPIRRate}{4.42\%}                 
\newcommand{\VoIIBudgetPIR}{3.79\%}                  
\newcommand{\IGBudgetPIR}{3.83\%}                    
\newcommand{\RandomBudgetPIR}{3.84\%}                
\newcommand{\VoIIQueryRate}{18.2\%}                  
\newcommand{\VoIIFinalPIR}{3.79\%}                   
\newcommand{\IGMatchedPIR}{3.83\%}                   
\newcommand{\RandomMatchedPIR}{3.84\%}               
\newcommand{\VoIIUtilityGain}{\ensuremath{+0.0004}}  
\newcommand{\VoIIStrictDeltaNever}{\ensuremath{-0.046}} 
\newcommand{\OracleGainRecovered}{72\%}              
\newcommand{\VoIIPositiveShare}{32.9\%}              

\theoremstyle{plain}
\newtheorem{theorem}{Theorem}[section]

\theoremstyle{definition}
\newtheorem{definition}[theorem]{Definition}

\theoremstyle{remark}

\newcommand{\E}{\mathbb{E}}
\newcommand{\Pp}{\mathbb{P}}
\newcommand{\1}{\mathbf{1}}
\newcommand{\Rset}{\mathcal{R}}

\newcommand{\Aset}{\mathcal{A}}
\newcommand{\Qset}{\mathcal{Q}}
\newcommand{\Zset}{\mathcal{Z}}
\newcommand{\qI}{q_\phi^I}
\newcommand{\qC}{q_\phi^C}
\newcommand{\yI}{Y^I}
\newcommand{\yA}{Y^A}
\newcommand{\pir}{\mathcal{R}_I}

\title{Prospective Interpretation Risk:  Principled
Communication Control Between LLMs}

\author{
\begin{tabular}{ccccc}
\textbf{Wanrong Yang}\textsuperscript{1} &
\textbf{Rehan Deen}\textsuperscript{2} &
\textbf{Julian Ma}\textsuperscript{4} &
\textbf{Yuheng Fan}\textsuperscript{1} &
\textbf{Yaoyu Jin}\textsuperscript{3}
\\[0.35em]
\textbf{Taher Jafferjee}\textsuperscript{2} &
\textbf{Ziquan Liu}\textsuperscript{3} &
\textbf{Dominik Wojtczak}\textsuperscript{1} &
\textbf{Yalin Zheng}\textsuperscript{1} &
\textbf{David Henry Mguni}\textsuperscript{1}\thanks{Corresponding author: \textlangle d.mguni@qmul.ac.uk\textrangle}
\end{tabular}
\\[1.3em]
\begin{tabular}{c}
\textsuperscript{1}University of Liverpool
\quad
\textsuperscript{2}Independent Researcher
\quad
\textsuperscript{3}Queen Mary University of London
\\[.5em]
\textsuperscript{4}University College London
\end{tabular}
}

\iclrfinalcopy

\begin{document}

\maketitle

\addtocontents{toc}{\protect\setcounter{tocdepth}{-1}}

\begin{abstract}
Large language model (LLM) based agentic systems increasingly rely on models communicating with one another, yet existing uncertainty and multi-agent methods rarely estimate how a particular receiver will interpret a message before it is sent. This matters in heterogeneous systems, where capable receivers can reconstruct different tasks from the same message. We model this as a sender--receiver problem with a latent receiver type and define \emph{prospective interpretation risk} (PIR): the probability that a receiver reconstructs a task other than intended. Rather than model an LLM's full input--output behaviour, we use black-box probes relating messages, intended tasks, and receiver-specific reconstructions, yielding scalable supervision while separating interpretation from downstream capability failure. Offline, heterogeneous frozen receivers provide supervision for receiver-conditioned risk and the effects of predefined mutable message features. At deployment, history induces a posterior over receiver types, guiding message revision and selection. We further introduce \emph{value of interpretation information} (VoII), querying for receiver information only when its expected communication benefit exceeds its cost. Our theory characterises when receiver information has decision value and bounds such queries. Empirically, interpretation-failure rates vary by 4--13$\times$ across receivers. Receiver information reduces PIR calibration error by 68\% relative to a receiver-agnostic predictor, largely by correcting receiver-specific risk levels. PIR-guided revision reduces interpretation failure by 44\% relative to the original message and 40\% relative to a generic rewrite, mostly through a repair that helps every receiver. VoII outperforms information-gain and random querying at matched cost on the interpretation objective it optimises, lowering interpretation failure from 3.84\% to 3.79\% while querying 18.2\% of episodes.\looseness=-1
\end{abstract}

\section{Introduction}
\label{sec:intro}

Large language model (LLM) applications are increasingly built as systems rather than single models. Planners delegate to executors, specialist agents exchange intermediate results, and tool-using components pass natural-language instructions to one another. Frameworks such as CAMEL \citep{li2023camel}, AutoGen \citep{wu2024autogen}, MetaGPT \citep{hong2024metagpt}, and ChatDev \citep{qian2024chatdev} make this pattern explicit. In these systems, communication is part of the execution path: a downstream model acts on what an upstream model meant to convey.

\begin{figure}[t]
\centering
\includegraphics[width=\textwidth]{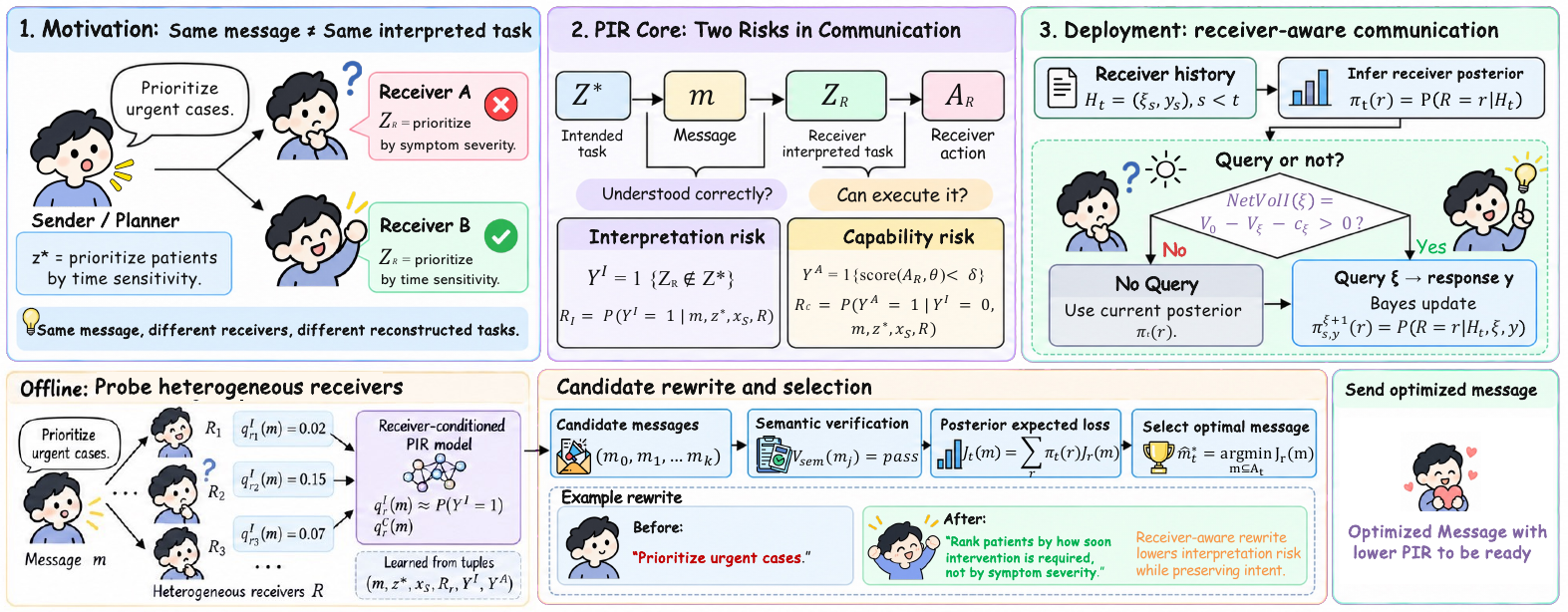}
\caption{\textbf{Framework overview.} Top: receivers read one message as different tasks (1), PIR separates interpretation from capability risk (2), and the sender infers the receiver and queries only if worthwhile (3). Bottom: offline probes train the PIR model, which picks the verified rewrite to send.}
\label{fig:overview}\vspace{-1em}
\end{figure}

This creates a reliability problem that is easy to miss when models are evaluated in isolation. LLM systems are heterogeneous by nature: receivers can differ in architecture, pre-training, fine-tuning and post-training \citep{ouyang2022training}, system prompts, tools, memory, and update history \citep{chen2024chatgpt}. A natural-language handoff is therefore less stable than a typed interface. Different receivers can read the same message as different tasks \citep{sclar2024quantifying,mizrahi2024state}, and the mismatch can spread even when each component is capable on its own \citep{cemri2025fail}. In a healthcare triage pipeline, for example, one model may read ``prioritise urgent cases'' as symptom severity and another as time sensitivity (\cref{fig:overview}, panel~1). Experiment~1 confirms that this happens in practice: on the same test messages, interpretation-failure rates vary widely across real receiver LLMs, and their ordering changes across datasets (\cref{fig:receiver_dependence}).

Current approaches cover parts of this problem but not the quantity needed to control it. Uncertainty methods mainly estimate confidence in a model's own answer \citep{kadavath2022mostly,farquhar2024semantic}, ambiguity \citep{cole2023selectively,kim2024ambiguity} and clarification \citep{zhang2025clarify,agentask2026,dong2026voi,suri2026structured} methods ask whether a message has multiple readings or needs more information, and multi-agent diagnostics study broader communication failures and how they spread \citep{cemri2025fail,mguni2026learning,lin2026before}. Listener-aware methods show that generation can adapt to the listener \citep{fried2018speaker,zhu2021tom,wang2021listeners,choi2025a2atom}. What is largely missing is a calibrated estimate, available \emph{before} sending, of whether \emph{this receiver} will recover the intended task from \emph{this message}, and a way to act on it.

An LLM's full text input--output space is too large to model in advance. We instead model a smaller, task-relevant object: the mapping from a message to the semantic task a receiver reconstructs. Each benchmark instance gives an intended task $z^\star$ and nearby alternative tasks. Independent black-box probes \citep{ribeiro2020beyond} ask which task the receiver infers, and a separate rollout measures execution. This gives a scalable behavioural label without modelling free-form outputs or treating chain-of-thought \citep{wei2022chain} as ground truth \citep{turpin2023language}. It also lets us separate \emph{interpretation failure}, which message revision can address, from downstream \emph{capability failure}, which wording alone may not fix (\cref{fig:overview}, panel~2). Experiment~1 confirms this separation. Experiment~2 shows that interpretation risk is predictable in advance, with strongly receiver-dependent calibration.

We call this quantity \emph{prospective interpretation risk} (PIR). Offline, sending the same semantic handoffs to heterogeneous frozen LLM receivers gives privileged supervision \citep{vapnik2009lupi} on how interpretation and capability vary by receiver type (\cref{fig:overview}, bottom left). We distil this response surface into receiver-conditioned heads and receiver-specific risk effects of predefined mutable message features \citep{hinton2015distilling}. At deployment, this population-wide supervision disappears: behavioural history gives a posterior over a closed set of receiver types \citep{sun2025idiosyncrasies,pasquini2025llmmap}, and risk is averaged over that posterior instead of conditioned on a known model identity. Experiment~3 shows that ordinary behavioural histories sharpen this posterior.

Beyond prediction, PIR changes how the sender communicates. The sender can rewrite a message \citep{zhou2023large} to reduce posterior expected interpretation loss (\cref{fig:overview}, bottom centre), or first gather receiver information and revise under the updated posterior. Querying, however, is itself costly: in larger agent networks, clarifying by default can cause \emph{query flooding}, which has long limited the scalability of broadcast-style search in peer-to-peer systems \citep{lv2002search}. We therefore introduce \emph{value of interpretation information} (VoII, \cref{fig:overview}, panel~3) and query only when the expected improvement in the next message decision exceeds its cost \citep{howard1966information,rao2018learning}.

This setting can be viewed as a Bayesian game \citep{harsanyi1967games}. The sender chooses a message under uncertainty about a latent receiver type, much as cheap-talk \citep{crawford1982strategic} and Bayesian persuasion \citep{kamenica2011bayesian} models study communication under private information. Here, however, the main friction is receiver-dependent interpretation, not conflicting preferences. When receiver information is costly, the problem links to information theory and rational inattention \citep{sims2003rational,matejka2015rational}: the sender should gather only the receiver information useful for its next message. Our two theoretical results make this precise. The receiver-conditioning theorem shows when heterogeneity can change the optimal message, and the information-to-communication frontier bounds the decision value of any query at a given receiver-information budget.

The framework therefore connects measurement, privileged receiver modelling, posterior inference, message revision, and selective querying. The experiments follow this logic. Experiment~1 shows that interpretation varies by receiver and differs from execution. Experiment~2 tests prospective prediction and the value of receiver conditioning. Experiment~3 tests how much ordinary history reveals about the latent receiver type. Experiment~4 tests whether posterior-guided revision reduces \emph{measured} misinterpretation, and Experiment~5 tests whether VoII improves query allocation for the interpretation objective. Because interpretation failures are rare, any enrichment stays in training or analysis data and is tracked by provenance \citep{king2001logistic,saerens2002adjusting}. Calibration \citep{guo2017calibration} and headline evaluation use untouched natural-prevalence validation and test sets.

\Cref{app:related} surveys uncertainty, ambiguity, clarification, partner modelling, multi-agent communication, strategic communication, costly information, and privileged information.

\section{Problem Setup and Measurement}
\label{sec:setup}

\Needspace*{3\baselineskip}
\textbf{Sender Intent, Interpretation, and Action}

Let $\theta\sim p(\theta)$ denote an underlying task instance and $x_S$ the task/environment information visible to the sender. The sender's intended semantic task is
\(
z^\star=g(\theta,x_S)\in\Zset,
\)
where $\Zset$ is the semantic task space and $g$ maps the underlying task and sender-visible context to the semantic task the sender intends the receiver to perform. A sender $S$ expresses this intent in a natural-language message $m$. Receiver $R$ first reconstructs a task $Z_R \in \Zset$ and then acts on it, producing action $A_R$:
\begin{equation}
 z^\star \longrightarrow m \longrightarrow Z_R \longrightarrow A_R.
 \label{eq:chain}
\end{equation}
We model the intermediate task variable $Z_R$, not the receiver's full distribution over text outputs. This reduces a huge black-box input--output space to the semantic object that matters for communication reliability: which task the receiver believes it has been asked to do. Receiver type $R$ is likewise defined by behaviour: it stands for a given frozen model and configuration and needs no access to the model's internals.
Because several surface descriptions can express the same task, let $\Zset^\star$ denote the semantic equivalence class of $z^\star$. We separate failure at the two stages of \cref{eq:chain}:
\begin{align}
 \yI = \1\{Z_R\notin\Zset^\star\}, \quad
 \yA = \1\{\operatorname{score}(A_R,\theta)<\delta\}.
 \label{eq:labels}
\end{align}
Thus $\yI=0,\yA=1$ means the receiver understood the task but its action scored below the pass threshold $\delta$, whereas $\yI=1,\yA=0$ can occur when a wrong interpretation happens to pass this check. The distinction matters because only the former is a true capability failure.

\begin{definition}[Prospective interpretation risk]
For a prospective receiver $R$, sender intent $z^\star$, context $x_S$, and candidate message $m$,
\begin{equation}
 \pir(m,z^\star,x_S,R)
 :=
 \Pp(\yI=1\mid m,z^\star,x_S,R).
 \label{eq:pir}
\end{equation}
\end{definition}
To keep execution difficulty separate, we also define conditional capability risk
\begin{equation}
 \mathcal R_C
 :=
 \Pp(\yA=1\mid \yI=0,m,z^\star,x_S,R).
 \label{eq:caprisk}
\end{equation}

\Needspace*{3\baselineskip}
\textbf{Independent Interpretation Probe}
\label{sec:probe}

Because $Z_R$ is latent, we measure it through behaviour, using benchmark-defined task identities rather than self-reported reasoning. Each probe presents an independently worded description of $z^\star$, a small set of nearby but truly different tasks $z_1^-,\ldots,z_K^-$, and a \textsc{None of these}/unclear option. The receiver picks the task that best matches the original message. This keeps supervision finite and black-box: it needs neither access to model internals nor a list of possible text outputs, and probe calls grow linearly with message--receiver pairs and can run in parallel across receivers. Probe wording is kept separate from the sender message, and option order is randomised \citep{zheng2024large}.
Interpretation and execution are measured in independent rollouts from the same $(m,x_S,R)$: a task-identification probe gives $\widehat Z_R$, and task execution gives $A_R$.
The execution branch never sees the probe alternatives. A specific wrong task and \textsc{None of these} are kept as separate raw outcomes, but both count as $\yI=1$ for binary PIR.
We repeat the probe $K_p=\ProbeRepeats$ times per message--receiver pair, in separate calls with independently generated wording and option order. For message--receiver pair $i$, let $\mathcal P_i$ contain the
indices of its parsed probe responses. For $|\mathcal P_i|>0$,
the soft training target is
\begin{equation}
 \widetilde y_i^I
 =
 \frac{1}{|\mathcal P_i|}
 \sum_{k\in\mathcal P_i}
 \1\{\widehat Z_{R,i}^{(k)}\notin\Zset_i^\star\}.
 \label{eq:softlabel}
\end{equation}
These repeats reduce measurement noise and are not treated as independent examples. An unparsed probe response is dropped from the average, and \textsc{None of these} counts as a wrong task. Training uses the resulting soft target directly. \Cref{app:taskbank} describes the task bank and splits.

\section{Method}
\label{sec:method}

The measurement protocol turns open-ended LLM behaviour into receiver-specific interpretation and execution outcomes. We use these compact labels in two stages: offline, to learn the decision-relevant response surface across receiver types, and at deployment, to infer the current receiver and choose a communication action without seeing its identity or predicting all of its future text.

\textbf{Privileged Receiver Supervision.}

Offline, each candidate message can be sent to several frozen receivers. Let $r$ index receiver types and $R_r$ denote the corresponding frozen receiver. For $R_r$ we collect
\(
(m,z^\star,x_S,R_r,\widetilde y_r^I,y_r^A)
\),
where $y_r^A\in\{0,1\}$ is the observed action-failure label for that message--receiver pair. Hereon, subscripts $r$ and posterior terms $\pi_t(r)$ refer to the receiver-type index. Sending one semantic handoff to several frozen LLMs gives privileged supervision on heterogeneity arising jointly from architecture, training, post-training, and configuration. Receiver identity indexes the behavioural type, so these sources need not be separated. The responses are distilled into two receiver-conditioned heads with parameters $\phi$, $q_r^I(m)=\qI(m,z^\star,x_S,e_r)$ and $q_r^C(m)=\qC(m,z^\star,x_S,e_r)$, where $e_r$ is a learned embedding of type $r$. These separate networks are trained with binary cross-entropy (BCE) losses $\mathcal L^I=\mathrm{BCE}(q^I,\widetilde y^I)$ and $\mathcal L^C=(1-\widetilde y^I)\,\mathrm{BCE}(q^C,y^A)$, whose weight, the share of probes read correctly, softens the condition $\yI=0$ of \cref{eq:caprisk}. Both are calibrated on natural-prevalence validation data. PIR is the estimand, not one fixed predictor: Experiment~2 tests the heads on natural messages, and Experiments~4--5 fit candidate-level estimators to measured rewrites whose distribution differs.\looseness=-1

\Needspace*{3\baselineskip}
\textbf{Closed-Set Receiver Inference.}
\label{sec:receiverinfer}

Privileged supervision is available only offline. At deployment the true receiver is hidden but belongs to the same finite type set,
\(
 \Rset=\{R_1,\ldots,R_K\},
 \qquad \Rset_{\rm train}=\Rset_{\rm test}
\).
The sender instead observes a history of calibration queries and responses,
\(H_t=\{(\xi_s,y_s)\}_{s<t},
\)
from which a small history encoder gives the calibrated posterior $\pi_t(r)=P(R=r\mid H_t)$.
For one-shot evaluation, $H_0$ is built from genuine stored responses of the hidden receiver to separate calibration-bank items, and every receiver type uses the same sampling rule.
The type model need only separate receiver profiles with different communication risks, not predict the receiver's future text.
Rather than a maximum a posteriori (MAP) guess, we carry receiver uncertainty into the decision. For any message, $\bar q_t^I(m)=\sum_r\pi_t(r)q_r^I(m)$ and $\bar q_t^C(m)=\sum_r\pi_t(r)q_r^C(m)$, and define per-receiver loss
\begin{equation}
 J_r(m)=c_M(m)+L_Iq_r^I(m)
 +L_C(1-q_r^I(m))q_r^C(m),
 \label{eq:loss}
\end{equation}
where $L_I,L_C\ge0$ weight interpretation and capability failure, with $L_C=0$ when we isolate interpretation risk, and $c_M(m)$ the cost of sending message $m$ (the subscript $M$ denotes message cost, e.g. token or latency cost). The controller acts on the posterior expected loss
\begin{equation}
 \bar J_t(m)=\sum_r\pi_t(r)J_r(m).
 \label{eq:posteriorloss}
\end{equation}

\Needspace*{3\baselineskip}
\textbf{Posterior-Guided Message Revision.}
\label{sec:rewrite}

Receiver awareness matters only if it changes what the sender does. The interpretation/capability split tells the sender what to change: better wording can often reduce interpretation risk, but paraphrasing cannot fix a receiver's execution capability. Our first intervention therefore targets \emph{mutable} message features linked to interpretation failure: the feature set is predefined, and what is learned is each receiver's risk effect $s_{rk}$ of feature $k$. Let $m_0$ be the initial message. Task-inherent features may predict risk but are not valid rewrite targets. Under the current posterior we aggregate $\bar s_{tk}=\sum_r\pi_t(r)s_{rk}$ and give $m_0$, $z^\star$, $x_S$, and the highest-risk mutable features to the rewrite model, which produces a fixed set of $K_m=\RewriteCandidates$ candidates.
Reducing predicted PIR is useful only if the rewrite preserves the task, so every candidate $m'_j$ first passes a semantic-equivalence filter $V_{\rm sem}$. This filter is a separate, stateless, batched call to the rewriter's model version. The verifier sees the intended task/context, original message, and candidates, but not the receiver posterior, PIR scores, feature weights, or candidate preference. A candidate passes only if it preserves the operation, target/referent, scope, constraints, conditions, and required output. The valid action set is $\Aset_t=\{m_0\}\cup\{m'_j:V_{\rm sem}(m'_j)=\textsc{Pass}\}$. We then choose $\widehat m_t^\star=\arg\min_{m\in\Aset_t}\bar J_t(m)$.
The original message is always kept, and the fixed generation budget rules out a ``rewrite until the score falls'' loop.

\Needspace*{3\baselineskip}
\textbf{Receiver Queries and VoII.}
\label{sec:voii_method}

Revision adapts the message to the current receiver belief. The second intervention asks whether that belief is worth refining. Before sending, the sender may therefore ask one costly query $\xi$ from a fixed calibration bank $\Qset$, kept separate from the test tasks. Repeated offline responses give a smoothed empirical likelihood
$\widehat P(Y_\xi=y\mid R=r,\xi)$,
where $Y_\xi$ denotes the observable response signature elicited by query $\xi$ (binary execution success in Experiment~5). Under $\pi_t$, the predictive response distribution is $\widehat P(y\mid H_t,\xi)=\sum_r\pi_t(r)\widehat P(y\mid R=r,\xi)$, and observing a reply $y$ gives the update
\begin{equation}
 \pi_{t+1}^{\xi,y}(r)
 =
 \frac{\widehat P(y\mid R=r,\xi)\pi_t(r)}
 {\sum_{r'}\widehat P(y\mid R=r',\xi)\pi_t(r')}.
 \label{eq:bayesupdate}
\end{equation}
Expected entropy reduction, or information gain,
\(
 \mathrm{IG}(\xi)=H(\pi_t)-\E_y H(\pi_{t+1}^{\xi,y})
\)
measures how informative a query is and can shortlist the bank. Being informative, however, does not imply decision value: identifying the receiver better does not help if the preferred message stays the same. We therefore compare the best available action before and after a hypothetical query. Holding $\Aset_t$ fixed, the no-query value is $V_0=\min_{m\in\Aset_t}\sum_r\pi_t(r)J_r(m)$, and the expected post-query value is
\begin{equation}
 V_\xi=
 \sum_y\widehat P(y\mid H_t,\xi)
 \left[
 \min_{m\in\Aset_t}
 \sum_r\pi_{t+1}^{\xi,y}(r)J_r(m)
 \right].
 \label{eq:queryvalue}
\end{equation}
The corresponding gross VoII is
$\operatorname{VoII}_t(\xi)=V_0-V_\xi$.
Let $c_\xi\ge0$ denote the decision cost of issuing query $\xi$
(e.g. token, latency, or compute cost). The net value is
\begin{equation}
\operatorname{NetVoII}(\xi)
=
\operatorname{VoII}_t(\xi)-c_\xi
=
V_0-V_\xi-c_\xi.
\label{eq:netvoii}
\end{equation}
The sender asks $\xi^\star=\arg\max_\xi\operatorname{NetVoII}(\xi)$ only when this value is positive. After seeing the response, it updates $\pi_t$, re-scores the same valid action set, and sends the message with the lowest expected loss. Keeping the candidate set fixed in the hypothetical branches separates the value of receiver information from random differences in generation.

\section{Decision-Theoretic Results}
\label{sec:theory}

The controller above is a finite Bayesian sender--receiver decision problem: the receiver has a latent type, the sender holds a posterior over that type, and the sender chooses a message after possibly observing another signal. It raises two questions: when can receiver information change the best message, and how much can a query be worth? The results below answer these questions and connect the controller to an information-theoretic view of LLM communication. Proofs are in \cref{app:proofs}.

\Needspace*{3\baselineskip}
\textbf{Value of Receiver Conditioning.}

Fix context $C=(z^\star,x_S)$. For candidate $m_j$ with cost $c_j$ and failure indicator $F_j$, let $p_j(e)=\Pp(F_j=1\mid E=e,C)$ and $J_j(e)=c_j+L_Ip_j(e)$, where $E$ is the sender's receiver information. A receiver-agnostic sender chooses before observing $E$, a receiver-aware one after. Their difference is
\begin{equation}
 \Delta_E
 :=
 \min_j\E[J_j(E)]
 -\E\!\left[\min_jJ_j(E)\right].
 \label{eq:deltaE}
\end{equation}

\begin{theorem}[Receiver-conditioning value]
\label{thm:conditioning}
For $J$ candidates and bounded PIR losses:
\begin{align}
 \Delta_E
 =\E[J_{j_0}(E)-J_{j^\star(E)}(E)]\ge0,\quad
 \Delta_E=0
 \quad\Longleftrightarrow\quad
 J_{j_0}(E)=\min_jJ_j(E)\ \text{a.s.}
\end{align}
for some receiver-agnostic minimiser $j_0 := \arg \min_j \mathbb{E} [J_j (E)]$ and receiver-aware minimiser $j^{\star}(E) := \arg \min_j J_j (E)$. If $\Pp\!\left(J_{j_0}(E)-\min_jJ_j(E)\ge\gamma\right)\ge\alpha$, then $\Delta_E\ge\alpha\gamma$. Moreover,
\begin{align}
 \Delta_E
 \le L_I\sqrt{\sum_{j=1}^J\operatorname{Var}[p_j(E)]},\quad
 \Delta_E
 \le L_I\sqrt{\sum_{j=1}^J I(F_j;E\mid C)}.
 \label{eq:mi-bound}
\end{align}
\end{theorem}

The result separates heterogeneity from usefulness. PIR can vary widely across receivers while the same message stays optimal for all of them, and then receiver conditioning has no decision value. Positive value appears only when receiver information moves probability mass across message decision regions. The variance and mutual-information bounds show how large that value can be.

\textbf{Information-to-Communication Frontier}
The first theorem assumes receiver information is already available. The second asks how much can be gained by acquiring it. Let the latent payoff-relevant receiver state be the discrete receiver type $R$ and condition on $I_t=(H_t,C)$. Define the best loss attainable over message policies $\rho$ when at most $B$ nats of receiver information may influence the final message,
\begin{equation}
 D_t(B)
 =
 \min_{{\rho(m\mid r, I_t)}:\,I(R;M\mid I_t)\le B}
 \E[L(M,R)\mid I_t],
 \label{eq:frontier}
\end{equation}
where $M$ takes values in $\Aset_t$, with $L(m,r)=J_r(m)$, and let $G_t(B)=D_t(0)-D_t(B)$.

\begin{theorem}[Information-to-communication frontier]
\label{thm:frontier}
For finite receiver and message spaces and bounded loss, $D_t(B)$ is non-increasing and convex in $B$, and $G_t(B)$ is non-decreasing and concave. For any query $\xi$ with response $Y_\xi$, followed by any message-selection rule based on that response,
\begin{equation}
\operatorname{VoII}_t(\xi)
\le
G_t\!\left(I(R;Y_\xi\mid I_t)\right).
 \label{eq:voiibound}
\end{equation}
\end{theorem}
Conditioned on $I_t$, the Markov chain
$R\rightarrow Y_\xi\rightarrow M$
gives the bound by data processing \citep{cover2006elements}. This explains why IG and VoII differ: a query can identify the receiver well and still have no value if the optimal communication action does not change. The concavity of $G_t$ also captures the diminishing returns of more receiver information.

\section{Experimental Design}
\label{sec:experiments}

\textbf{Datasets.}
We use \NumDatasets\ knowledge-base question-answering (KBQA) collections: FreebaseQA \citep{jiang2019freebaseqa}, WebQSP \citep{yih2016webqsp}, GrailQA \citep{gu2021grailqa}, SimpleQuestions-Wikidata (SQ-WD) \citep{bordes2015large,diefenbach2017sqwd}, PopQA \citep{mallen2023popqa}, and Mintaka \citep{sen2022mintaka}. The message $m$ is the question rendered by the fixed PromptSource template \texttt{qa\_template\_basic} \citep{bach2022promptsource}. The sender is fixed on purpose: every receiver sees the same message, which isolates receiver-dependent interpretation in Experiments~1--3. Experiments~4--5 then test LLM revision and query control. The intended task $z^\star$ is the benchmark task, and each message carries one contrast task from a fixed set (entity versus category for entity-valued questions, value versus attribute for class-valued questions). The test split has \NumMessages\ messages and \NumTrials\ message--receiver pairs (per-dataset sizes in \cref{tab:decomp}, \cref{app:res_e1}, and splits in \cref{app:taskbank}).

\textbf{Receivers.}
The receiver pool has \NumReceivers\ open-weight models from different families: Nemotron-3 Nano 30B \citep{nvidia2025nemotron3nano}, Gemma 4 26B \citep{gemmateam2026gemma4}, gpt-oss-20b \citep{openai2025gptoss}, Mistral Small 3.2, and Qwen3.6 35B. All run frozen through vLLM \citep{kwon2023efficient} at temperature 0, so each label is a deterministic function of message and receiver.

\textbf{Baselines.}
Following \cref{tab:baseline_defs} (\cref{app:ablations}), A0 uses global and per-receiver base rates, A1 is a message-only ambiguity judge, A2 is a semantic-equivalence judge that sees $(m,z^\star,x_S)$ but not the receiver, A3 is our model without $e_R$, A4 is our model trained on $\yA$ and scored against $\yI$, and A5 is a theory-of-mind (ToM) listener that predicts the receiver's probe choice and uses $1-P(z^\star)$ as risk \citep{zhu2021tom}. A shuffled-receiver control permutes receiver identities across training pairs.

\textbf{Protocols.}
Each probe offers the intended task, the contrast task, and \textsc{None of these}. Execution is a separate answer call scored by gold-alias containment. Prediction is scored by areas under the receiver operating characteristic and precision--recall curves (AUROC, AUPRC), negative log-likelihood (NLL), Brier score, and expected calibration error (ECE). Labels, the predictor, and full protocols for Experiments~3--5 are in \cref{app:protocols}, with metrics, statistics, and scope in \cref{app:metrics,app:scope}.

\section{Results}
\label{sec:results}

\Needspace*{16.4\baselineskip}
\textbf{Interpretation and Execution Failures Separate Empirically}
\label{sec:res_e1}

\begin{wrapfigure}[14]{r}{0.62\textwidth}
\vspace{-10pt}
\centering
\begin{tikzpicture}
\pgfplotsset{
  e1rec/.style={only marks, mark options={draw=white, line width=0.3pt}},
}
\begin{groupplot}[
  group style={group size=2 by 1, horizontal sep=8pt, yticklabels at=edge left},
  scale only axis,
  height=76pt,
  ymin=0.5, ymax=6.5, y dir=reverse,
  ytick={1,2,3,4,5,6},
  yticklabels={FreebaseQA,WebQSP,GrailQA,SQ-WD,PopQA,Mintaka},
  tick label style={font=\scriptsize},
  every tick label/.append style={inner sep=1.5pt},
  axis x line*=bottom,
  axis y line*=left,
  y axis line style={draw=none},
  ytick style={draw=none},
  axis line style={black!40, line width=0.4pt},
  xtick style={black!40, line width=0.4pt},
  major tick length=2pt,
  tick align=outside,
  title style={font=\scriptsize, anchor=south west, at={(0,1)}, inner xsep=0pt, inner ysep=2pt, yshift=-5pt},
  clip=false,
]
\nextgroupplot[
  width=64pt,
  xmin=0, xmax=100, xtick={0,50,100},
  title={\textbf{(a)} $\yA$ (\%) by reading},
]
\fill[black!7, rounded corners=2.2pt] (axis cs:20.9,0.58) rectangle (axis cs:31.3,1.42);
\fill[black!7, rounded corners=2.2pt] (axis cs:50.3,1.58) rectangle (axis cs:70.6,2.42);
\fill[black!7, rounded corners=2.2pt] (axis cs:86.5,2.58) rectangle (axis cs:93.2,3.42);
\fill[black!7, rounded corners=2.2pt] (axis cs:55.5,3.58) rectangle (axis cs:65.0,4.42);
\fill[black!7, rounded corners=2.2pt] (axis cs:72.7,4.58) rectangle (axis cs:78.9,5.42);
\fill[black!7, rounded corners=2.2pt] (axis cs:37.2,5.58) rectangle (axis cs:48.2,6.42);
\addplot[only marks, mark=*, mark size=1.5pt, black!75] coordinates {(28.3,1) (67.6,2) (90.2,3) (58.5,4) (75.9,5) (45.2,6)};
\addplot[only marks, mark=*, mark size=1.5pt, mark options={fill=white, draw=black!75, line width=0.6pt}] coordinates {(23.9,1) (53.3,2) (89.5,3) (62.0,4) (75.7,5) (40.2,6)};
\node[anchor=east, font=\tiny, text=black!65, inner sep=1pt] at (axis cs:49.8,2) {correct read};
\node[anchor=west, font=\tiny, text=black!65, inner sep=1pt] at (axis cs:71.1,2) {misread};
\nextgroupplot[
  width=110pt,
  xmode=log, xmin=0.7, xmax=24,
  xtick={1,2,5,10,20}, xticklabels={1,2,5,10,20},
  title={\textbf{(b)} $\yI$ (\%) by receiver},
  legend to name=e1legend,
  legend columns=5,
  legend style={draw=none, fill=none, font=\scriptsize, inner sep=0pt,
    /tikz/every even column/.append style={column sep=5pt}},
  legend image code/.code={\draw[mark repeat=2, mark phase=2] plot coordinates {(0cm,0cm) (0.12cm,0cm) (0.24cm,0cm)};},
]
\fill[black!7, rounded corners=2.2pt] (axis cs:0.82,0.55) rectangle (axis cs:9.02,1.45);
\fill[black!7, rounded corners=2.2pt] (axis cs:2.44,1.55) rectangle (axis cs:13.20,2.45);
\fill[black!7, rounded corners=2.2pt] (axis cs:3.63,2.55) rectangle (axis cs:20.38,3.45);
\fill[black!7, rounded corners=2.2pt] (axis cs:0.81,3.55) rectangle (axis cs:13.07,4.45);
\fill[black!7, rounded corners=2.2pt] (axis cs:0.79,4.55) rectangle (axis cs:8.26,5.45);
\fill[black!7, rounded corners=2.2pt] (axis cs:1.31,5.55) rectangle (axis cs:7.04,6.45);
\addplot[e1rec, cNemo, mark=*, mark size=2.0pt] coordinates {(8.21,1) (12.02,2) (18.56,3) (11.90,4) (7.52,5) (6.41,6)};
\addplot[e1rec, cGemma, mark=square*, mark size=1.75pt] coordinates {(1.07,1) (3.65,2) (4.82,3) (1.99,4) (1.14,5) (1.94,6)};
\addplot[e1rec, cGpt, mark=triangle*, mark size=2.6pt] coordinates {(0.98,1.33) (3.20,2.33) (4.19,3.33) (3.06,4.33) (1.60,5.33) (1.52,6.33)};
\addplot[e1rec, cMistral, mark=diamond*, mark size=2.5pt] coordinates {(1.41,1) (6.01,2) (7.88,3) (8.27,4) (5.35,5) (1.46,6)};
\addplot[e1rec, cQwen, mark=triangle*, mark size=2.6pt, every mark/.append style={rotate=180}] coordinates {(0.95,0.67) (2.82,1.67) (4.28,2.67) (0.94,3.67) (0.92,4.67) (1.60,5.67)};
\legend{Nemotron, Gemma 4, gpt-oss, Mistral 3.2, Qwen3.6}
\node[anchor=west, font=\tiny, text=black!60, inner sep=1pt] at ({rel axis cs:1,0} |- {axis cs:1,1}) {$9\times$};
\node[anchor=west, font=\tiny, text=black!60, inner sep=1pt] at ({rel axis cs:1,0} |- {axis cs:1,2}) {$4\times$};
\node[anchor=west, font=\tiny, text=black!60, inner sep=1pt] at ({rel axis cs:1,0} |- {axis cs:1,3}) {$4\times$};
\node[anchor=west, font=\tiny, text=black!60, inner sep=1pt] at ({rel axis cs:1,0} |- {axis cs:1,4}) {$13\times$};
\node[anchor=west, font=\tiny, text=black!60, inner sep=1pt] at ({rel axis cs:1,0} |- {axis cs:1,5}) {$8\times$};
\node[anchor=west, font=\tiny, text=black!60, inner sep=1pt] at ({rel axis cs:1,0} |- {axis cs:1,6}) {$4\times$};
\node[anchor=south, font=\tiny, text=black!60, inner xsep=1pt, inner ysep=2pt, yshift=2.2pt] at ([xshift=6pt]{rel axis cs:1,1}) {max/min};
\end{groupplot}
\node[anchor=south, inner sep=0pt] at ([yshift=11pt]$(group c1r1.north west)!0.5!(group c2r1.north east)$) {\pgfplotslegendfromname{e1legend}};
\end{tikzpicture}
\vspace{-8pt}
\caption{Interpretation failure is separate from execution failure and depends on the receiver (test split). (a)~$\yA$ after a correct read ($\circ$) and a misread ($\bullet$) of the probe. (b)~$\yI$ per receiver. Bands span the lowest to the highest receiver.}
\label{fig:receiver_dependence}
\vspace{-6pt}
\end{wrapfigure}
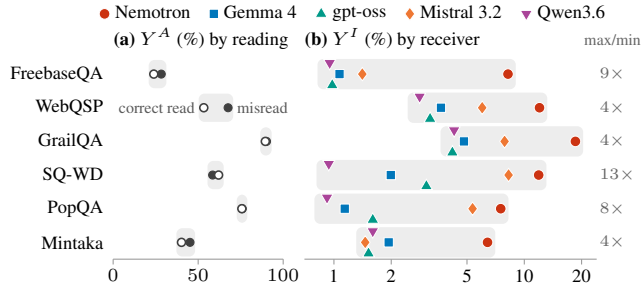

Three facts support treating $\yI$ as its own target. First, the two failures do not coincide. \Cref{tab:decomp} in \cref{app:res_e1} gives the $\yI\times\yA$ decomposition. Both off-diagonal cells are non-zero on every dataset: \CorrectInterpretFailureRate\ of pairs read the task correctly but still fail, and \MisinterpretSuccessRate\ read it wrongly yet pass the answer check. Pooled over datasets, \MisreadPassShare\ of all misread pairs still pass, so an answer-based score would call them successes. Second, the execution outcome carries almost no information about interpretation (\cref{fig:receiver_dependence}a): pooled over receivers, the execution-failure rates after a misread and after a correct read differ by at most five points on five of the six datasets (14 points on WebQSP, the smallest set), while the rate itself spans 24--90\% across datasets.

Third, the receiver matters (\cref{fig:receiver_dependence}b). Pooled $\yI$ ranges from \ReceiverYIMin\ (Qwen3.6) to \ReceiverYIMax\ (Nemotron) on the same messages, and within each dataset the best and worst receiver differ by 4--13$\times$. Nemotron is highest on every dataset, but the lowest receiver changes (Qwen3.6 on four datasets, gpt-oss on GrailQA, Mistral on Mintaka), so receiver risk is not a single number per receiver. Heterogeneity survives a control for option position: with the intended task listed first or second, the highest and lowest receivers still differ by 3.8$\times$ (95\% confidence interval [CI] 3.4--4.3), and by 1.7--14$\times$ within datasets (\cref{app:position}, which traces most of Nemotron's excess to the last listed option). Five temperature-0 repeats reproduce every dataset's $\yI$ within \RepeatStabilityMaxDiff\ points.

\textbf{Receiver Conditioning Improves Prospective PIR Prediction}
\label{sec:res_e2}

\begin{wraptable}[15]{r}{0.6\textwidth}
\vspace{-8pt}
\caption{Prospective PIR prediction (test split, six-dataset means, seed 0). Ours is best on all seven metrics (bold, second best underlined, A0 excluded).}
\label{tab:prediction}
\vspace{-8pt}
\centering
\scriptsize
\setlength{\tabcolsep}{3.088pt}
\begin{tabular}{@{}lccccccc@{}}
\toprule
& \multicolumn{2}{c}{Per receiver$\uparrow$} & \multicolumn{2}{c}{Across receivers$\uparrow$} & \multicolumn{3}{c}{Calibration$\downarrow$}\\
\cmidrule(lr){2-3}\cmidrule(lr){4-5}\cmidrule(lr){6-8}
Method & AUROC & AUPRC & AUROC & AUPRC & Brier & NLL & ECE\\
\midrule
A0 Global prior & .500 & .045 & .500 & .045 & .0428 & .180 & .029\\
A0 Receiver prior & .500 & .045 & .697 & .086 & .0415 & .168 & .002\\
A1 Message judge & .552 & .057 & .540 & .055 & .0428 & .180 & .031\\
A2 Equiv.\ judge & .500 & .045 & .500 & .045 & .0428 & .180 & .029\\
A4 $\yA$ predictor & .511 & .048 & .530 & .050 & .4229 & 1.329 & .540\\
A5 Listener/ToM & .716 & .104 & .782 & .129 & .0412 & .160 & .013\\
\midrule
A3 No receiver & .729 & .128 & .701 & .115 & .0414 & .168 & .032\\
A3 + recv.\ calib. & .728 & .128 & \underline{.796} & \underline{.158} & \underline{.0399} & \underline{.155} & \underline{.012}\\
Shuffled receiver & \underline{.730} & \underline{.130} & .700 & .115 & .0414 & .168 & .032\\
\midrule
\textbf{Ours} & \textbf{.732} & \textbf{.135} & \textbf{.800} & \textbf{.164} & \textbf{.0398} & \textbf{.154} & \textbf{.010}\\
\bottomrule
\end{tabular}
\vspace{-8pt}
\end{wraptable}
\Cref{tab:prediction} compares the receiver-conditioned predictor with the baselines of \cref{tab:baseline_defs}. Five points follow. (i) Interpretation risk is predictable before sending: with pairs ranked within each receiver and then averaged (per receiver), AUROC is \PIRAUROC\ and AUPRC \PIRAUPRC\ against a positive rate of \BaseRateAUPRC. (ii) The signal is specific to interpretation. The action-failure predictor (A4), trained on $\yA$ with the same inputs, scores \ExecutionPredAUROC\ on $\yI$ and is below chance on three datasets. The message-only judge (A1, \MessageOnlyAUROC) and the equivalence judge (A2, \EquivalenceAUROC) stay near chance. Neither general difficulty nor message ambiguity explains the target.
\newline
(iii) Receiver information mainly improves calibration. Removing $e_R$ (A3) leaves within-receiver AUROC almost unchanged (\AgnosticAUROC\ vs.\ \PIRAUROC) but triples ECE (\AgnosticECE\ vs.\ \PIRECE) and worsens NLL and Brier on all six datasets, with paired intervals that exclude zero (\cref{tab:receiver_ablation,fig:calibration} in \cref{app:res_e2}). When all pairs are ranked together (across receivers), the same information lifts AUROC from 0.701 to 0.800 and AUPRC from 0.115 to 0.164, because the model must then rank one message across receivers, which is exactly the sender's decision. (iv) Most of the gain comes from learning each receiver's risk level, including Nemotron's last-option effect (\cref{app:position}): recalibrating A3 with the full model's per-receiver map (A3 + recv.\ calib.) recovers 93\% of the ECE reduction and 96\% of the across-receiver AUROC gain. The full model still ranks better within receivers (AUROC $+0.004$ and AUPRC $+0.007$, with paired intervals excluding zero), so message--receiver interaction adds its own ranking gain (\cref{app:res_e2}). (v) The gain needs the correct receiver: the shuffled-receiver control matches A3 within 0.002 on every averaged metric. The listener model (A5), the strongest non-ablation baseline, is worse than ours on all seven averaged metrics and on Brier in every dataset (\cref{tab:e2_per_dataset}).

\Needspace*{5\baselineskip}
\textbf{Behavioural History Sharpens Receiver Beliefs}
\label{sec:res_e3}

The receiver is never named at test time. The sender only sees stored responses to calibration tasks. \Cref{fig:history} in \cref{app:res_e3} shows that these responses carry receiver information that grows with history length. Accuracy rises from the uniform prior (\HistoryZeroAcc, five receivers) to \HistoryShortAcc\ after one response, \HistoryMediumAcc\ after five, and \HistoryLongAcc\ after twenty, while NLL falls from \UniformPriorNLL\ to \ReceiverTypeNLL\ and Brier from 0.800 to \ReceiverTypeBrier. The shuffled-history control stays at chance (\ShuffledHistoryAccRange) at every length, so the gain comes from the receiver's own behaviour, not from the tasks. The confusion matrix (\cref{tab:confusion} in \cref{app:res_e3}) shows where the evidence is: predictions concentrate on Nemotron and Mistral, the two receivers with the highest interpretation risk (Nemotron histories are recognised 71\% of the time), whereas Gemma is almost never predicted. This is the pattern the framework needs, since the posterior only has to separate receivers whose risk differs. Because the posterior stays broad even at twenty responses, the controller keeps the full posterior $\pi_t$ instead of committing to a MAP receiver. Experiment~4 compares it with a uniform prior and the true identity.

\Needspace*{16.4\baselineskip}
\textbf{Risk-Guided Revision Reduces Measured Misinterpretation}
\label{sec:res_e4}

\begin{wraptable}[14]{r}{0.57\textwidth}
\vspace{-9pt}
\caption{Message revision on 1,579 held-out questions: ours beats every pre-registered reference on $\yI$ ($^\ddagger$\,Holm $p<0.001$). Gen.: learned risk guides generation ($\pi_t$: posterior only). Sel.: it selects the sent message.}
\label{tab:rewrite}
\vspace{-8pt}
\centering
\scriptsize
\setlength{\tabcolsep}{2.6pt}
\begin{tabular}{@{}lccccc@{}}
\toprule
& & & $\yI\downarrow$ & $\Delta\yI$ (ours$-$row) & $\yA\downarrow$\\
Condition & Gen. & Sel. & (\%) & points [95\% CI] & (\%)\\
\midrule
Original message & -- & -- & 4.16 & $-1.85$ [$-2.09$, $-1.62$]$^\ddagger$ & 59.33\\
Generic rewrite & -- & -- & 3.85 & $-1.54$ [$-1.78$, $-1.30$]$^\ddagger$ & 58.15\\
Selection only & -- & \checkmark & 3.47 & $-1.15$ [$-1.41$, $-0.93$]$^\ddagger$ & 58.20\\
Posterior only & $\pi_t$ & \checkmark & 3.43 & $-1.11$ [$-1.33$, $-0.90$]$^\ddagger$ & 58.59\\
\midrule
Guidance only & \checkmark & -- & 2.29 & $+0.03$ [$+0.01$, $+0.05$]\phantom{$^\ddagger$} & 58.33\\
\textbf{Ours} & \checkmark & \checkmark & \textbf{2.31} & -- & 58.33\\
True identity & \checkmark & \checkmark & 2.30 & $+0.02$ [$-0.08$, $+0.12$]\phantom{$^\ddagger$} & 58.58\\
One feature per step & \checkmark & \checkmark & 2.12 & $+0.20$ [$+0.03$, $+0.37$]\phantom{$^\ddagger$} & 58.75\\
\bottomrule
\end{tabular}
\vspace{-8pt}
\end{wraptable}

Experiment~4 runs the full method of \cref{sec:rewrite} on 1,579 held-out questions, with $s_{rk}$ learned on 600 separate training questions and the Experiment~3 posterior after five stored responses (\cref{app:protocol_e4}). All rewrite conditions share the rewrite model, intended task, four candidates, and verifier. They differ only in where learned risk is used. Revision lowers $\yI$ from 4.16\% to 2.31\%, a 44\% drop, and beats all four pre-registered references by 1.11 to 1.85 points ($p<0.001$ each after \citet{holm1979simple} correction, \cref{tab:rewrite}). It does not hurt execution: $\yA$ is 1.00 point lower than for the original and within a 1-point non-inferiority margin for every reference. The drop holds for all five receivers and six datasets and is largest where the original risk is highest: 4.96 points for Nemotron, 2.62 for Mistral, and 0.40--0.69 for the other three (\cref{app:res_e4}).

The gain comes from guidance, not selection: guided candidates sent without selection are 1.56 points below the generic rewrite, while selection among generic candidates gains 0.38. Of the guided rewrites, 97\% keep the question and add a line that states the output (e.g., ``Give the name of the Italian fashion designer.''), while 95\% of generic rewrites leave it unstated. Most of the gain thus comes from one repair that helps every receiver, and the framework shows when receiver-specific adaptation is needed rather than assuming it always is. Here all five receivers share the top risk feature, an unstated output, so a uniform belief changes only 2 of 7,895 decisions and the true identity gives almost the same $\yI$ (2.30\% vs.\ 2.31\%): the case $\Delta_E\approx0$ of \cref{thm:conditioning}, where receiver information barely changes the best message. One feature per step lowers $\yI$ further, to 2.12\%, at the cost of four dependent rewrite rounds and about twice the tokens.

\Needspace*{5\baselineskip}
\textbf{VoII Allocates Receiver Queries by Decision Value}
\label{sec:res_e5}

A query is worth asking only if its answer can change the message. Experiment~5 tests this rule. \Cref{tab:e5_policies} reports the primary cohort (\QueryQuestions\ test questions, five receivers, three fresh repeats, \QueryEpisodes\ episodes) and a replication on a second cohort of new questions. Each episode starts from five stored responses and the Experiment~3 posterior, and every policy asks at most one query from the same bank and sends the message chosen by a risk head (\cref{app:protocol_e5}). Without a query, measured interpretation failure is \NeverQueryPIRRate. With the true receiver identity it would be \OracleQueryPIRRate, a gap of 0.06 points. At a 20\% budget, NetVoII reaches \VoIIBudgetPIR\ and closes 78\% of this gap, while IG and random querying reach \IGBudgetPIR\ and \RandomBudgetPIR\ and close 5\% and 3\%. The paired differences are consistent: NetVoII lowers $\yI$ by 0.03 to 0.06 points against both controls at every budget from 10\% to 50\% (\cref{fig:e5_query}b in \cref{app:res_e5}), and all eight pre-registered contrasts in net utility $1-\yI-c\,r$ (query rate $r$, query cost $c$) survive Holm correction ($p\le0.012$). The strict rule queries only where $\operatorname{NetVoII}>0$ at $c=\QueryCost$. It queries \VoIIQueryRate\ of episodes and ends at \VoIIFinalPIR, \VoIIStrictDeltaNever\ points below no query (95\% CI $[-0.068,-0.025]$), which is \OracleGainRecovered\ of the way to the true-identity reference. At the same query count, IG and random querying stay at \IGMatchedPIR\ and \RandomMatchedPIR, so the strict rule improves net utility by \VoIIUtilityGain\ over the stronger of the two. The adapted value-of-information (VoI) rule \citep{dong2026voi}, a secondary comparator, replaces the risk heads by LLM utility estimates. These are so coarse that one candidate is best, or tied for best, for every receiver in every question, so no query can change the message. Ties, broken toward the original as for every policy, keep the original in 81\% of episodes (18\% with the heads), and $\yI$ stays at \DongVoIPIRRate\ (\cref{app:res_e5}).

\Needspace*{16.8\baselineskip}
\begin{wraptable}[16]{r}{0.60\textwidth}
\vspace{-8pt}
\caption{Query policies (interpretation objective, $c=\QueryCost$). Gap: share of the no-query to true-identity $\yI$ gap that is closed. Ident.: posterior mode is correct. IG: information gain. $^\dagger$\,Secondary comparator. Bold: best deployable value.}
\label{tab:e5_policies}
\vspace{-8pt}
\centering
\scriptsize
\setlength{\tabcolsep}{7.4419pt}
\begin{tabular}{@{}lcccccc@{}}
\toprule
& \multicolumn{4}{c}{Primary cohort} & \multicolumn{2}{c@{}}{Replication}\\
\cmidrule(lr){2-5}\cmidrule(lr){6-7}
& $\yI$ & Gap & Net & Ident. & $\yI$ & Gap\\
Policy (query rate) & (\%) & (\%) & utility & (\%) & (\%) & (\%)\\
\midrule
No query (0\%) & 3.84 & 0 & .9616 & 24.6 & 3.93 & 0\\
\multicolumn{7}{@{}l}{\emph{Exact 20\% quota}}\\
\quad Random (3 seeds) & 3.84 & 3 & .9614 & 25.7 & 3.93 & 4\\
\quad IG & 3.83 & 5 & .9615 & 28.5 & 3.94 & ${<}0$\\
\quad \textbf{NetVoII (ours)} & \textbf{3.79} & \textbf{78} & \textbf{.9619} & 28.3 & \textbf{3.91} & \textbf{76}\\
\multicolumn{7}{@{}l}{\emph{No quota}}\\
\quad \textbf{NetVoII strict (18\%)} & \textbf{3.79} & 72 & \textbf{.9619} & 28.0 & 3.92 & 47\\
\quad Always, IG (100\%) & 3.80 & 55 & .9610 & \textbf{44.0} & 3.96 & ${<}0$\\
\quad Adapted VoI$^\dagger$ (0\%) & 4.42 & ${<}0$ & .9558 & 24.6 & 4.27 & ${<}0$\\
\midrule
True-identity reference & 3.77 & 100 & .9623 & 100.0 & 3.91 & 100\\
\bottomrule
\end{tabular}%
\vspace{-8pt}
\end{wraptable}

\Cref{fig:e5_query} shows why decision value, not information, is the right criterion. IG identifies the receiver at least as well as NetVoII: at 20\% the posterior mode is the true receiver in 28.5\% of episodes for IG and 28.3\% for NetVoII, and in 44.0\% when every episode is queried (panel c). Yet querying every episode lowers $\yI$ by only 0.04 points, to \AlwaysQueryPIRRate, and on the replication cohort it raises $\yI$ (3.96\% vs.\ 3.93\%). Entropy falls, but the preferred message rarely changes. NetVoII spends its queries where a changed posterior would change the message. It queries about half of the PopQA and Mintaka episodes and none of WebQSP or GrailQA, and nearly all of its gain comes from those two datasets (\cref{tab:e5_per_dataset} in \cref{app:res_e5}).

The size of this gain is set by how much decision-relevant receiver information is left after five stored responses (\cref{app:res_e5}, which also reports task success). On a base of \NeverQueryPIRRate, the true identity would lower $\yI$ by 0.06 points, and NetVoII lowers it by about 0.05. On the replication cohort, NetVoII at 20\% again closes 76\% of the gap (\cref{tab:e5_policies}), a directional replication.

\section{Conclusion}
\label{sec:conclusion}

We introduced PIR, the probability, before sending, that a receiver reconstructs a task other than intended. Interpretation failure is receiver-specific, distinct from execution failure, and predictable before sending. Our theory shows that receiver information pays off exactly when it changes the preferred message. Empirically, receiver information reduces PIR calibration error by 68\% relative to a receiver-agnostic predictor. PIR-guided revision lowers measured interpretation failure by 44\% relative to the original message and 40\% relative to a generic rewrite. For interpretation, VoII allocates queries better than IG and random querying at matched cost, closing 78\% of the 0.06-point gap to the true-identity reference at a 20\% budget. More broadly, PIR suggests that multi-agent LLM systems should check messages for likely misreading before sending, rather than after failure. Beyond these results, our framework could become part of the communication infrastructure for future multi-agent systems, supporting reliable massive agentic interaction by making interpretation failure something agents can anticipate and mitigate before sending.

\subsection*{AI use statement}
We used generative artificial intelligence (AI) tools for language editing, drafting support, literature search, and to support some implementation methods. The authors reviewed all AI-assisted text, technical claims, derivations, experimental specifications, and code, and take full responsibility for the final content.


\subsection*{Reproducibility statement}
The paper specifies the PIR labels, receiver-inference model, query-response model, VoII controller, rewrite procedure, baselines, and ablations. Full proofs are in \cref{app:proofs}, and probe construction, receiver inference, message revision, experimental protocols, baselines, pseudocode, and full experimental details are in \cref{app:probe,app:receiver,app:rewrite,app:protocols,app:ablations,app:pseudocode,app:implementation}.

\bibliography{iclr2026_conference}
\bibliographystyle{iclr2027_conference}

\clearpage
\appendix
\addtocontents{toc}{\protect\setcounter{tocdepth}{3}}
\renewcommand{\contentsname}{Appendix}
{\setlength{\parskip}{0pt}\tableofcontents}
\clearpage
\section{Related Work}
\label{app:related}

\textbf{Uncertainty, ambiguity, and clarification.}
Most language-model uncertainty methods estimate confidence in a model's own output through verbalised confidence, sampling consistency, or semantic uncertainty \citep{kadavath2022mostly,farquhar2024semantic}. Ambiguity work instead studies inputs with multiple plausible readings or trains models to request missing information \citep{min2020ambigqa,saparina2025disambiguate,cole2023selectively,kim2024ambiguity,hu2026concept,yang2026underspecification,zhang2025clarify}. These are natural points of comparison, but neither asks whether a given receiver will reconstruct the sender's intended task. PIR can therefore be low for a linguistically ambiguous message if a receiver resolves it consistently, or high for a message that looks clear but that this receiver tends to misread.

\textbf{Partner modelling and multi-agent communication.}
Pragmatic speakers and listener-aware generation model how a listener may respond \citep{fried2018speaker,zhu2021tom,wang2021listeners}, and recent agent-to-agent work studies interlocutor awareness between LLMs \citep{choi2025a2atom}. Other work classifies multi-agent communication failures \citep{cemri2025fail} or predicts how hallucinations spread before interaction \citep{lin2026before}. Our target is more specific: a calibrated, receiver-specific probability of task misinterpretation measured independently from execution and used directly to choose a communication action.

\textbf{Strategic communication and costly information.}
Our sender--receiver formulation is related to Bayesian communication and cheap-talk games \citep{crawford1982strategic}, but we do not require conflicting preferences: uncertainty enters through the receiver's latent behavioural type and type-dependent interpretation. The query decision is closer to value-of-information \citep{dong2026voi} and rational-inattention \citep{sims2003rational,matejka2015rational} problems, where information is acquired only when its decision value justifies its cost. This distinction matters at scale. Always asking for clarification is not a neutral baseline in a networked system: query flooding is a classic cause of traffic and scalability problems in distributed peer-to-peer search \citep{lv2002search}. VoII makes the same trade-off explicit for LLM communication.

\textbf{Privileged information.}
The training setup is closest in spirit to learning using privileged information \citep{vapnik2009lupi} and distillation \citep{hinton2015distilling}. Offline, we observe a receiver-indexed response surface: how the same handoff is interpreted and executed across heterogeneous real LLMs. Deployment sees none of these population-wide outcomes, only ordinary behavioural history for the current receiver. The goal is therefore not to model each LLM's full text behaviour, but to distil the lower-dimensional, decision-relevant variation that matters for communication control before sending.

\section{Notation}
\label{app:notation}

\Cref{tab:notation} lists every symbol defined in the paper. Subscripts $r$ and $t$ index a receiver type and a decision step, superscripts $I$ and $C$ mark interpretation and capability (execution), and a bar marks an average over the receiver, as in $\bar J_t(m)=\sum_r\pi_t(r)J_r(m)$. A few letters also have a second, local meaning, among them $K$, $g$, $G_t$, $R$, $r$, $B$, and $\alpha$. The table lists each use.

\newcommand{\symrow}[3]{\raggedright\relpenalty=10000 \binoppenalty=10000 #1 & \raggedright #2 & \raggedright #3\tabularnewline}
\newcommand{\symgroup}[1]{\multicolumn{3}{@{}l}{\emph{#1}}\tabularnewline}
{\small
\setlength{\LTcapwidth}{\textwidth}
\renewcommand{\arraystretch}{1.1}
\begin{longtable}{@{}p{0.2\textwidth}p{\dimexpr0.68\textwidth-4\tabcolsep\relax}p{0.12\textwidth}@{}}
\caption{Notation, grouped by the part of the framework that introduces each symbol. The last column gives where the symbol is defined, and -- marks standard notation.}
\label{tab:notation}\\
\toprule
Symbol & Meaning & Defined in\\
\midrule
\endfirsthead
\multicolumn{3}{@{}l}{\emph{\tablename~\thetable\ (continued)}}\\
\toprule
Symbol & Meaning & Defined in\\
\midrule
\endhead
\bottomrule
\endfoot
\bottomrule
\endlastfoot
\symgroup{General}
\symrow{$\E$, $\Pp$ (or $P$), $\operatorname{Var}$}{Expectation, probability, and variance}{--}
\symrow{$\1\{\cdot\}$}{Indicator: 1 if the condition holds and 0 otherwise}{--}
\symrow{$H(\cdot)$, $I(\cdot\,;\cdot\mid\cdot)$}{Shannon entropy and conditional mutual information, in nats}{--}
\symrow{$D_{\mathrm{KL}}(\cdot\Vert\cdot)$}{Kullback--Leibler divergence}{--}
\symrow{$\mathrm{Ber}(p)$}{Bernoulli distribution with mean $p$}{--}
\symrow{$\mathrm{BCE}(q,y)$}{Binary cross-entropy, $-y\log q-(1-y)\log(1-q)$}{--}
\addlinespace[4pt]
\symgroup{Tasks, messages, and outcomes}
\symrow{$\theta\sim p(\theta)$}{Task and its distribution}{Sec.~\ref{sec:setup}}
\symrow{$S$, $x_S$}{Sender and the task/environment information visible to it (in the experiments, the original question, used as a proxy)}{Sec.~\ref{sec:setup}, App.~\ref{app:labels}}
\symrow{$z^\star$, $\Zset$}{Intended semantic task of the sender and the task space, with $z^\star\in\Zset$}{Sec.~\ref{sec:setup}}
\symrow{$g$}{Intent map from the task and the sender's information to the intended task, $z^\star=g(\theta,x_S)$}{Sec.~\ref{sec:setup}}
\symrow{$\Zset^\star$}{Semantic equivalence class of $z^\star$: all tasks with the same meaning}{Sec.~\ref{sec:setup}}
\symrow{$m$}{Natural-language message from the sender}{Sec.~\ref{sec:setup}}
\symrow{$R$}{Receiver type: a frozen model and configuration, defined by its behaviour and latent at deployment}{Sec.~\ref{sec:setup}}
\symrow{$r$, $R_r$}{A value of $R$, which indexes receivers in sums and subscripts, and the receiver of type $r$}{Sec.~\ref{sec:method}}
\symrow{$Z_R$}{Task that the receiver reconstructs from $m$ (latent)}{Eq.~\eqref{eq:chain}}
\symrow{$A_R$ ($A_r$)}{Action or output of the receiver}{Eq.~\eqref{eq:chain}}
\symrow{$\operatorname{score}(A_R,\theta)$, $\delta$}{Task score of the output and its success threshold}{Eq.~\eqref{eq:labels}}
\symrow{$\yI$}{Interpretation failure, $\1\{Z_R\notin\Zset^\star\}$}{Eq.~\eqref{eq:labels}}
\symrow{$\yA$}{Execution failure, $\1\{\operatorname{score}(A_R,\theta)<\delta\}$, so $1-\yA$ is task success}{Eq.~\eqref{eq:labels}}
\symrow{$\pir(m,z^\star,x_S,R)$}{Prospective interpretation risk (PIR), $\Pp(\yI=1\mid m,z^\star,x_S,R)$}{Eq.~\eqref{eq:pir}}
\symrow{$\mathcal R_C$}{Conditional capability risk, $\Pp(\yA=1\mid\yI=0,m,z^\star,x_S,R)$}{Eq.~\eqref{eq:caprisk}}
\addlinespace[4pt]
\symgroup{Interpretation probe}
\symrow{$z_1^-,\ldots,z_K^-$}{Nearby but different contrast tasks offered as probe options, where $K$ counts them (one in the experiments)}{Sec.~\ref{sec:setup}}
\symrow{$\widehat Z_R$}{Task that the receiver picks in the probe, a measurement of $Z_R$}{Sec.~\ref{sec:setup}}
\symrow{$K_p$}{Probe repeats per message--receiver pair ($K_p=\ProbeRepeats$)}{Sec.~\ref{sec:setup}}
\symrow{$i$, $\widehat Z_{R,i}^{(k)}$, $\Zset_i^\star$}{Index of a message--receiver pair, the task picked in its $k$-th probe, and the equivalence class of its intended task}{Eq.~\eqref{eq:softlabel}}
\symrow{$\widetilde y_i^I$ ($\widetilde y_r^I$, $\widetilde y^I$)}{Soft interpretation label: share of parsed probes that pick a wrong option}{Eq.~\eqref{eq:softlabel}}
\symrow{$y_r^A$}{Measured execution-failure label ($\yA$) for receiver $r$}{Sec.~\ref{sec:method}}
\symrow{$\mathcal E(z)$}{Set $\{T_z^{(1)},\ldots,T_z^{(R)}\}$ of validated descriptions of task $z$, where $R$ counts descriptions}{App.~\ref{app:probe}}
\symrow{$\mathcal E_{\rm send}$, $\mathcal E_{\rm probe}$, $\widetilde T_z$}{Disjoint banks of descriptions for messages and for probes, and a probe description of $z$}{App.~\ref{app:probe}}
\symrow{$s(T)$}{Task signature $(\mathcal I_T,\mathcal C_T,\operatorname{op}_T,\mathcal O_T,\mathcal F_T)$ of description $T$: visible input fields, extra context, semantic operation, output meaning, and output format or type}{App.~\ref{app:probe}}
\addlinespace[4pt]
\symgroup{Risk predictor}
\symrow{$\qI$, $\qC$, $\phi$}{Interpretation and capability heads, two separate multilayer perceptrons (MLPs, $q^I$ and $q^C$ for short), and their parameters}{Sec.~\ref{sec:method}}
\symrow{$q_r^I(m)$, $q_r^C(m)$}{Head outputs for receiver $r$, $\qI(m,z^\star,x_S,e_r)$ and $\qC(m,z^\star,x_S,e_r)$: predicted PIR and capability risk}{Sec.~\ref{sec:method}}
\symrow{$e_r$ ($e_R$)}{Learned 16-dimensional embedding of receiver type $r$}{Sec.~\ref{sec:method}, App.~\ref{app:labels}}
\symrow{$\mathcal L^I$, $\mathcal L^C$}{Losses of the heads, $\mathrm{BCE}(q^I,\widetilde y^I)$ and $(1-\widetilde y^I)\,\mathrm{BCE}(q^C,y^A)$, with $\mathcal L_r^I(m)$ and $\mathcal L_r^C(m)$ for one message and receiver}{Sec.~\ref{sec:method}, Alg.~\ref{alg:training}}
\symrow{$g$}{Scalar gap: root-mean-square difference of the standardised embeddings of $x_S$ and $m$ (not the intent map)}{App.~\ref{app:labels}}
\addlinespace[4pt]
\symgroup{Receiver inference}
\symrow{$\Rset=\{R_1,\ldots,R_K\}$}{Closed set of $K$ receiver types (\NumReceivers\ in the experiments), with $\Rset_{\rm train}=\Rset_{\rm test}$}{Sec.~\ref{sec:method}}
\symrow{$t$}{Decision step}{Sec.~\ref{sec:method}}
\symrow{$H_t$, $|H_t|$}{Behavioural history $\{(\xi_s,y_s)\}_{s<t}$ of calibration queries and responses, and its length}{Sec.~\ref{sec:method}}
\symrow{$H_0$}{Starting history for one-shot evaluation, built from stored responses of the hidden receiver}{Sec.~\ref{sec:method}}
\symrow{$\pi_t(r)$}{Calibrated posterior over receiver types, $P(R=r\mid H_t)$}{Sec.~\ref{sec:method}}
\symrow{$\bar q_t^I(m)$, $\bar q_t^C(m)$}{Posterior-averaged risks, $\sum_r\pi_t(r)q_r^I(m)$ and $\sum_r\pi_t(r)q_r^C(m)$}{Sec.~\ref{sec:method}}
\symrow{$\pi_\omega(r\mid H)$, $\omega$}{Receiver-type model, $\operatorname{softmax}_r(Wh_H+b)$ with output weights $W$ and bias $b$, and all its parameters}{App.~\ref{app:receiver}}
\symrow{$h_s$, $f_{\rm int}$}{Encoding $f_{\rm int}(\xi_s,y_s)$ of one interaction and the interaction encoder}{App.~\ref{app:receiver}}
\symrow{$h_H$, $n$}{History encoding $\operatorname{Pool}(h_1,\ldots,h_n)$ and the number of interactions in the history}{App.~\ref{app:receiver}}
\addlinespace[4pt]
\symgroup{Communication loss and message revision}
\symrow{$J_r(m)$}{Loss of sending $m$ to receiver $r$, $c_M(m)+L_Iq_r^I(m)+L_C(1-q_r^I(m))q_r^C(m)$}{Eq.~\eqref{eq:loss}}
\symrow{$c_M(m)$}{Cost of sending message $m$}{Eq.~\eqref{eq:loss}}
\symrow{$L_I$, $L_C$}{Weights of interpretation and execution failure, where $L_C=0$ isolates interpretation}{Eq.~\eqref{eq:loss}}
\symrow{$\bar J_t(m)$}{Posterior expected loss, $\sum_r\pi_t(r)J_r(m)$}{Eq.~\eqref{eq:posteriorloss}}
\symrow{$m_0$}{Initial (original) message}{Sec.~\ref{sec:method}}
\symrow{$k$, $d$}{Index of a predefined mutable message feature and the number of such features ($d=8$ in Experiment~4)}{Sec.~\ref{sec:method}, Alg.~\ref{alg:training}}
\symrow{$s_{rk}$}{Learned risk effect of mutable feature $k$ for receiver $r$}{Sec.~\ref{sec:method}, App.~\ref{app:rewrite}}
\symrow{$\bar s_{tk}$}{Posterior-weighted risk effect, $\sum_r\pi_t(r)s_{rk}$}{Sec.~\ref{sec:method}}
\symrow{$G_t$, $n_g$}{Guidance set: the at most $n_g$ mutable features with the largest positive $\bar s_{tk}$ ($n_g=3$ in Experiment~4), not the gain $G_t(B)$}{Alg.~\ref{alg:revision}}
\symrow{$K_m$, $m'_j$}{Rewrite candidates per request ($K_m=\RewriteCandidates$) and the $j$-th candidate}{Sec.~\ref{sec:method}}
\symrow{$V_{\rm sem}$, $v_j$}{Semantic-equivalence verifier, which returns \textsc{Pass} or \textsc{Fail}, and its decision on $m'_j$}{Sec.~\ref{sec:method}, Alg.~\ref{alg:revision}}
\symrow{$\Aset_t$}{Valid action set, $\{m_0\}\cup\{m'_j:V_{\rm sem}(m'_j)=\textsc{Pass}\}$}{Sec.~\ref{sec:method}}
\symrow{$\widehat m_t^\star$}{Revised message, $\arg\min_{m\in\Aset_t}\bar J_t(m)$}{Sec.~\ref{sec:method}}
\symrow{$\widehat m^\star$}{Message finally sent, after an optional query}{Alg.~\ref{alg:deployment}}
\symrow{$\Delta f_k(m)$}{Change in binary feature $k$ from $m_0$ to the rewrite $m$}{App.~\ref{app:rewrite}}
\symrow{$b_r$}{Intercept of receiver $r$ in the linear risk model of Experiment~4: the effect of rewriting as such}{App.~\ref{app:rewrite}}
\symrow{$\alpha$}{Ridge penalty ($\alpha=1$ in Experiment~4, $\alpha\in\{10,100,1000\}$ in Experiment~5), not the $\alpha$ of \cref{thm:conditioning}}{App.~\ref{app:rewrite}}
\addlinespace[4pt]
\symgroup{Receiver queries and VoII}
\symrow{$\xi$, $\Qset=\{\xi_1,\ldots,\xi_M\}$}{Calibration query and the fixed bank of $M$ queries, kept separate from the test tasks}{Sec.~\ref{sec:method}, App.~\ref{app:receiver}}
\symrow{$Y_\xi$, $\mathcal Y_\xi$, $y$}{Response signature of query $\xi$ (binary execution success in Experiment~5), its value set, and an observed response}{App.~\ref{app:receiver}}
\symrow{$\widehat P(y\mid R=r,\xi)$}{Smoothed likelihood $\widehat P(Y_\xi=y\mid R=r,\xi)$ of response $y$ for receiver $r$}{Sec.~\ref{sec:method}, App.~\ref{app:receiver}}
\symrow{$\widehat P(y\mid H_t,\xi)$}{Predictive response distribution, $\sum_r\pi_t(r)\widehat P(y\mid R=r,\xi)$}{Sec.~\ref{sec:method}}
\symrow{$\pi_{t+1}^{\xi,y}$, $\pi_{t+1}$}{Posterior after response $y$ to query $\xi$ (Bayes update) and the posterior after the query actually asked}{Eq.~\eqref{eq:bayesupdate}, Alg.~\ref{alg:deployment}}
\symrow{$\mathrm{IG}(\xi)$}{Information gain, the expected entropy reduction $H(\pi_t)-\E_yH(\pi_{t+1}^{\xi,y})$}{Sec.~\ref{sec:method}}
\symrow{$V_0$}{Value without a query, $\min_{m\in\Aset_t}\sum_r\pi_t(r)J_r(m)$}{Sec.~\ref{sec:method}}
\symrow{$V_\xi$}{Expected value after query $\xi$, with $\Aset_t$ held fixed}{Eq.~\eqref{eq:queryvalue}}
\symrow{$\operatorname{VoII}_t(\xi)$}{Value of interpretation information, $V_0-V_\xi$}{App.~\ref{app:proof_frontier}}
\symrow{$\operatorname{NetVoII}(\xi)$, $c_\xi$}{Net value of query $\xi$, $V_0-V_\xi-c_\xi$, and the cost of the query}{Eq.~\eqref{eq:netvoii}}
\symrow{$\xi^\star$}{Query with the largest NetVoII, asked only if this value is positive}{Sec.~\ref{sec:method}}
\addlinespace[4pt]
\symgroup{Evaluation}
\symrow{$\Delta\yI$}{Difference in $\yI$ between two conditions, in percentage points (ours minus the comparison)}{Tab.~\ref{tab:rewrite}}
\symrow{$p$}{Probability of the posterior mode}{App.~\ref{app:res_e3}}
\symrow{$c$}{Query cost in Experiment~5 ($c=\QueryCost$ unless stated otherwise)}{App.~\ref{app:protocol_e5}}
\symrow{$1-\yI-c\,r$, $r$}{Net utility of a query policy and its query rate, the share of episodes queried (not a receiver)}{Sec.~\ref{sec:results}, App.~\ref{app:metrics}}
\symrow{$B\%$}{Query quota: the share of episodes a policy may query (not the budget $B$ in nats)}{App.~\ref{app:alg_instances}}
\symrow{$\yI+(1-\yI)\yA$}{Combined loss that also scores execution}{App.~\ref{app:fullspec}}
\addlinespace[4pt]
\symgroup{Decision-theoretic results}
\symrow{$C$}{Context $(z^\star,x_S)$}{Sec.~\ref{sec:theory}}
\symrow{$E$, $e$, $\pi(e)$}{Receiver information available to the sender, a value of it, and its distribution}{Sec.~\ref{sec:theory}, App.~\ref{app:proof_conditioning}}
\symrow{$m_j$, $J$}{Candidate message $j$ and the number of candidates}{Sec.~\ref{sec:theory}}
\symrow{$F_j$}{Interpretation failure when $m_j$ is sent}{Sec.~\ref{sec:theory}}
\symrow{$p_j(e)$}{Failure probability of $m_j$ given $E=e$, $\Pp(F_j=1\mid E=e,C)$}{Sec.~\ref{sec:theory}}
\symrow{$c_j$, $J_j(e)$}{Cost of $m_j$ and its loss $c_j+L_Ip_j(e)$}{Sec.~\ref{sec:theory}}
\symrow{$\Delta_E$}{Value of receiver conditioning, $\min_j\E[J_j(E)]-\E[\min_jJ_j(E)]$}{Eq.~\eqref{eq:deltaE}}
\symrow{$j_0$, $j^\star(E)$}{Receiver-agnostic minimiser $\arg\min_j\E[J_j(E)]$ and receiver-aware minimiser $\arg\min_jJ_j(E)$}{Thm.~\ref{thm:conditioning}}
\symrow{$\gamma$, $\alpha$}{Loss margin and the probability of reaching it, in the bound $\Delta_E\ge\alpha\gamma$}{Thm.~\ref{thm:conditioning}}
\symrow{$I_t$}{Information state $(H_t,C)$}{Sec.~\ref{sec:theory}}
\symrow{$B$}{Budget, in nats, of receiver information that may influence the message}{Eq.~\eqref{eq:frontier}}
\symrow{$M$}{Message as a random variable (not the bank size)}{Eq.~\eqref{eq:frontier}}
\symrow{$\rho(m\mid r,I_t)$, $\bar\rho(m\mid I_t)$}{Message policy, a channel from receiver type to message, and its marginal $\sum_rP(r\mid I_t)\rho(m\mid r,I_t)$}{Eq.~\eqref{eq:frontier}, App.~\ref{app:proof_frontier}}
\symrow{$L(m,r)$, $P(r\mid I_t)$}{Loss of sending $m$ to receiver type $r$ and the belief over receiver types given $I_t$}{Eq.~\eqref{eq:frontier}, App.~\ref{app:proof_frontier}}
\symrow{$D_t(B)$}{Lowest expected loss when at most $B$ nats of receiver information may influence the message}{Eq.~\eqref{eq:frontier}}
\symrow{$G_t(B)$}{Gain from receiver information, $D_t(0)-D_t(B)$}{Sec.~\ref{sec:theory}}
\addlinespace[4pt]
\symgroup{Used only in the proofs}
\symrow{$p(e)$, $\bar p_j$}{Vector $(p_1(e),\ldots,p_J(e))$ of failure probabilities and the mean failure probability $\E[p_j(E)]$}{App.~\ref{app:proof_conditioning}}
\symrow{$g(u)$}{$\min_j\{c_j+L_Iu_j\}$ for $u\in[0,1]^J$, so that $\Delta_E=g(\E[p(E)])-\E[g(p(E))]$}{App.~\ref{app:proof_conditioning}}
\symrow{$r\in \Rset$, $m\in \Aset_t$}{Receiver types and messages}{App.~\ref{app:proof_frontier}}
\symrow{$B_1<B_2$, $\mathcal F_1$, $\mathcal F_2$}{Two information budgets and the sets of policies feasible under each}{App.~\ref{app:proof_frontier}}
\symrow{$\rho_1^\star$, $\rho_2^\star$, $P_1$, $P_2$}{Optimal policies under $B_1$ and $B_2$ and their joint distributions of $(m,r)$ (in \cref{app:proof_conditioning}, $P_1$ and $P_2$ are $\mathrm{Ber}(p_j(E))$ and $\mathrm{Ber}(\bar p_j)$)}{App.~\ref{app:proof_frontier}}
\symrow{$\mu$, $B_\mu$}{Mixing weight in $[0,1]$ and the mixed budget $\mu B_1+(1-\mu)B_2$}{App.~\ref{app:proof_frontier}}
\symrow{$P_\mu$, $D_{t,P_\mu}$}{Mixture $\mu P_1+(1-\mu)P_2$ of the two joint distributions and its expected loss}{App.~\ref{app:proof_frontier}}
\symrow{$\lambda$}{Cost per nat of receiver information in the rational-inattention relaxation}{App.~\ref{app:proof_frontier}}
\symrow{$\mathcal S[\rho]$, $\chi(r)$}{Objective of the relaxation with the normalisation constraint and the Lagrange multiplier of that constraint}{App.~\ref{app:proof_frontier}}
\symrow{$\rho_\lambda^\star$, $Z_\lambda^{I_t}(r)$}{Optimal policy of the relaxation, of Gibbs form, and its normalising constant}{App.~\ref{app:proof_frontier}}
\end{longtable}}

\section{Proofs and Additional Formal Results}
\label{app:proofs}
\newcommand{\norm}[1]{\mathopen{\Vert}#1\mathclose{\Vert}}

\subsection{Receiver-Conditioning Value}
\label{app:proof_conditioning}

This subsection proves \cref{thm:conditioning}:

\textbf{Decision-value identity $\Delta_E \ge 0$}: 

We rewrite $\Delta_E = \mathbb{E} [ J_{j^0}(E)] - \mathbb{E} [J_{j^{\star}} (E)]
\nonumber = \mathbb{E} [ J_{j^0}(E) - J_{j^{\star}}(E)]$ based on the definition of $\Delta_E$, receiver-agnostic minimiser $j_0$ and receiver-aware minimiser $j^{\star}(E)$\footnote{Throughout this subsection, expectations and variances
are taken over $E$ conditional on the fixed context $C$, and
this conditioning is suppressed in the notation.}. From the definition of expectation, we have:
\begin{align}
\label{eq:mr1-1-3}
\Delta_{E} = \sum_e \pi(e) \left(J_{j_0}(e) - J_{j^{\star}}(e)\right)
\nonumber = \sum_e \pi(e) \left(J_{j_0}(e) - \min_j J_{j}(e)\right)
\end{align}
where \(\pi(E)\) is the prior distribution over \(E\), and satisfies \(\pi(e) \geq 0\). By the definition of the minimising element \(j\), we have \(J_{j_0}(e) - \min_j J_{j}(e)\geq 0\), and hence $\Delta_E \geq 0 $.

\textbf{Characterisation of $\Delta_E = 0$}: 

Since \(\pi(E)\) is the distribution of receiver information available to the sender, we have  $J_{j_0} (E) = \min_j J_j(E)$ almost surely iff $\Delta_E = 0 $, meaning that the receiver-agnostic cost \(J_{j_0}(E)\) is equal to the receiver-aware cost for all receivers.

\textbf{Lower bound $\Delta_E\ge \alpha\gamma$}:

For \(\alpha, \gamma > 0\), we assume:
\begin{equation}
\label{eq:mr1-3-1}
\mathbb{P} \left( J_{j_0} (E) - \min_j J_j (E) \geq \gamma \right) \geq \alpha
\end{equation}
Since \(J_{j_0} (E) - J_{j^{\star}}(E) \geq 0\), according to  Markov's inequality, we have:
\begin{equation}
\label{eq:mr1-3-3}
\mathbb{P} \left( J_{j_0} (E) - \min_j J_j (E) \geq \gamma \right) \leq \frac{\mathbb{E} [ J_{j_0} (E) - \min_j J_j (E)]}{\gamma}
\end{equation}
The numerator on the right-hand side of \eqref{eq:mr1-3-3} is \(\Delta_E\). Bounding the left-hand side with the assumption \eqref{eq:mr1-3-1} and multiplying through by \(\gamma > 0\), we have $\Delta_E \geq \alpha \gamma $.

\textbf{Information-theoretic upper bound of $\Delta_E$:}

For each receiver state \(e\), we define $p_j(e):=\mathbb{P}(F_j=1\mid E=e,C)\in[0,1]$, and collect these scalar failure probabilities into $p(e):=(p_1(e),\ldots,p_J(e))\in[0,1]^J$. 

Defining $g(p_1,\ldots,p_J)=\min_j\{c_j+L_I p_j\}$, we can re-express the decision-value as $\Delta_E=g(\mathbb{E} [p(E)])-\mathbb{E}[g(p(E))]$, where the function \(g\) acts on the finite-dimensional probability vector defined as 
\[
g:[0,1]^J\rightarrow\mathbb{R},\ g(u):=\min_j\{c_j+L_Iu_j\}
\]

Since each \(u\mapsto c_j+L_Iu_j\) is affine, \(g\) is the pointwise minimum of finitely many affine functions and is therefore concave. 
Moreover, \(g\) is \(L_I\)-Lipschitz with respect to the \(\ell_\infty\)-norm, i.e. \(|g(p)-g(q)|\leq L_I\|p-q\|_\infty\). Therefore, we have:
\begin{align}
\label{eq:mr1-4-6}
\Delta_E &= g(\mathbb{E} [p(E)])-\mathbb{E}[g(p(E))]= \mathbb{E} [ g(\mathbb{E}[p(E)]) - g(p(E)) ] \nonumber\\
&\leq L_I \mathbb{E} [  \norm{\mathbb{E}[p(E)] - p(E)}_{\infty}]\leq L_I \mathbb{E} [  \norm{\mathbb{E}[p(E)] - p(E)}_2]
\end{align}

Noticing that for $\mathbb{E}[  \norm{\mathbb{E}[p(E)] - p(E)}_2]$, we have according to Jensen's inequality the bound
\begin{equation}
\label{eq:mr1-4-6a}
\left(\mathbb{E} \left[ \norm{p(E) - \mathbb{E} [p(E)]}_2\right] \right)^2 \leq \mathbb{E} \left[ \norm{p(E) - \mathbb{E} [p(E)]}_2^2 \right] = \sum_{j=1}^J \operatorname{Var}[p_j (E)]
\end{equation}
Taking the square root of \eqref{eq:mr1-4-6a} and inserting into \eqref{eq:mr1-4-6}, we find
\begin{equation}
\label{eq:mr1-4-res}
\Delta_E \leq L_I \sqrt{\sum_{j=1}^J\operatorname{Var}[p_j(E)]}
\end{equation}
Let \(\bar{p}_j=\mathbb{E}[p_j(E)]\). Message transmission is modelled as a Bernoulli trial with failure probability \(p_j(E)\) for the receiver-aware sender and \(\bar{p}_j\) for the receiver-agnostic sender. Taking \(P_1 := \mathrm{Ber}(p_j(E))\), \(P_2 := \mathrm{Ber}(\bar{p}_j)\), we have 
\[\norm{P_1-P_2}_1^2 \leq 2D_{\mathrm{KL}}(P_1\|P_2)\] 
via Pinsker's inequality. Since \(\norm{P_1-P_2}_2\leq \norm{P_1-P_2}_1\), we find:
\begin{equation}
\label{eq:mr1-5-4}
\norm{\mathrm{Ber}(p_j(E)) - \mathrm{Ber}(\bar{p}_j)}_2^2 
\leq  2 D_{\mathrm{KL}} (\mathrm{Ber}(p_j(E)) || \mathrm{Ber}(\bar{p}_j)) 
\end{equation}
The left-hand side of \eqref{eq:mr1-5-4} evaluates to
$2|p_j(E)-\bar p_j|^2$. Dividing both sides by two gives
$|p_j(E)-\bar p_j|^2
\leq D_{\mathrm{KL}}\left(
\mathrm{Ber}(p_j(E))\Vert\mathrm{Ber}(\bar p_j)
\right)$.

The variance of $p_j(E)$ is:
\begin{align}
\label{eq:mr1-5-7}
\mathrm{Var}[p_j(E)] &= \mathbb{E}[(p_j(E) - \bar{p}_j)^2] 
\leq \mathbb{E} \left( D_{\mathrm{KL}}(\mathrm{Ber}(p_j(E)) ~||~ \mathrm{Ber}(\bar{p}_j))\right)
\end{align}
The right-hand side of \eqref{eq:mr1-5-7} is the conditional mutual information $I(F_j; M| C)$, hence the variance is bounded by $\operatorname{Var}[p_j(E)] \leq I(F_j;M|C)$. Finally, from \eqref{eq:mr1-4-res}, the upper bound of the decision-value is given by
\begin{equation}
\label{eq:mr-1-6-3}
\Delta_E \leq L_I
\sqrt{\sum_{j=1}^J I(F_j;E\mid C)}\,.
\end{equation}

\subsection{Information-to-Communication Frontier}
\label{app:proof_frontier}

This subsection proves \cref{thm:frontier}:

In the sequel, we consider fixed $I_t=(H_t,C)$ and suppress the range of summations for clarity (they will be over the full space of the associated index unless otherwise indicated). 

To connect this result to the controller, take
$L(m,r)=J_r(m)$ and $P(r\mid I_t)=\pi_t(r)$.
All query-related expectations and mutual information are evaluated
under the same joint model
$\pi_t(r)\widehat P(y\mid R=r,\xi)$.
With estimated risks and likelihoods, the result bounds model-based
VoII and does not by itself guarantee realised gains under
model misspecification.

\textbf{Non-increasing}
\textit{The feasible set in \cref{eq:frontier} expands with $B$, so $D_t(B)$ is non-increasing:}

The mutual information from the receiver-type $r \in \Rset$ and messages $m \in \Aset_t$ conditioned on $I_t$ is given by
\begin{align} 
\label{eq:mr2a-1-2}
I(R; M  \mid  I_t) &= \sum_{m, r} P(r | I_t) \, \rho(m | r,I_t) \, \log \left(\frac{\rho(m | r,I_t)}{\bar{\rho}(m)}\right)
\end{align}
where
\begin{equation}
\label{eq:mr2a-1-3}
\bar{\rho}(m | I_{t}) = \sum_{r} P(r | I_t) \rho(m | r,I_t)
\end{equation}
Consider \(I(R; M  \mid  I_t)\) as a functional of the conditional probability \(\rho(m | r,I_t)\), for a fixed distribution \(P(r | I_t)\). 

Let \(B_1 < B_2\) and define the feasible sets
\begin{equation}
\label{eq:mr2a-1-4}
\mathcal{F}_1 := \{ \rho(m | r,I_t) ~:~ I(R ; M \mid I_t) \leq B_1 \}
\qquad
\mathcal{F}_2 := \{ \rho(m | r,I_t) ~:~ I(R ; M \mid I_t) \leq B_2 \}
\end{equation}
Because the mutual information \eqref{eq:mr2a-1-2} is positive semi-definite and continuous in \(\rho(m | r,I_t)\), then a given \(\rho(m | r,I_t) \in \mathcal{F}_1\) also lies in \(\mathcal{F}_2\), and
\begin{equation}
\label{eq:mr2a-1-5}
\mathcal{F}_1 \subseteq \mathcal{F}_2
\end{equation}
Consider \(D_t(B_1)\) and \(D_t(B_2)\), with 
\begin{equation}
\label{eq:mr2a-1-6}
D_t(B_i) = \min_{\rho_i(m | r,I_t) \in \mathcal{F}_i} \sum_{m, r} P(r| I_t)\, \rho_i(m | r,I_t) \, L(m, r)
\end{equation}
and let \(\rho_1^{\star}(m | r,I_t) \in \mathcal{F}_1\) be the conditional policy that provides the minimum \(D_t(B_1)\). In determining the value of \(D_t(B_2)\), there might be a \(\rho_2(m | r,I_t) \in \mathcal{F}_2\) but \(\notin \mathcal{F}_1\) might possibly provide a lower minimum of \(\mathbb{E}[L(M, R)]\) so that \(D_t(B_2) \leq D_t(B_1)\). If not, \(\rho_1^{\star}(m | r,I_t) \in \mathcal{F}_1\) also lies in \(\mathcal{F}_2\) by \eqref{eq:mr2a-1-5}, and so at worst \(D_t(B_2) = D_t(B_1)\).

This is true for all \(0 \leq B_1 < B_2\), and so \(D_t(B)\) must be non-increasing.

\textbf{Convexity} \textit{Convexity follows from time sharing between two receiver-state-to-message channels with an independent selector:}

For \(D_t(B)\) to be convex, we require that for  \(0 \leq B_1 < B_2\),  \(\mu \in [0,1]\) 
\begin{equation}
\label{eq:mr2a-1-7}
D_t(\mu B_1 + (1-\mu)B_2) \leq \mu D_t(B_1) + (1-\mu) D_t(B_2)
\end{equation}
Let \(\rho_1^{\star}(m|r,I_t)\) and \(\rho_2^{\star}(m|r,I_t)\) be the conditional distributions that produce \(D_t(B_1)\) and \(D_t(B_2)\), and let
\begin{equation}
\label{eq:mr2a-1-8}
P_{\mu}(m,r | I_t) := \mu \, P(r|I_t)\rho_1^{\star}(m|r,I_t) + (1-\mu)\, P(r|I_t)\rho_2^{\star}(m|r,I_t)
\end{equation}
be the joint distribution that interpolates between \(P_1(m, r | I_t)=P(r|I_t)\rho_1^{\star}(m|r,I_t)\) and \(P_2(m, r | I_t) = P(r|I_t)\rho_2^{\star}(m|r,I_t)\).

The expected loss is linear in \(\rho(m | r, I_t)\), so we define $D_{t,P_{\mu}} :=
\mathbb{E}_{P_{\mu}}[L(M,R)]$ and see that
\begin{align}
\label{eq:mr2a-1-9-1}
D_{t,P_{\mu}} 
&=  \mu \left(\sum_{m, r} P(r|I_t)\, \rho_1^{\star}(m|r,I_t) \, L(m, r) \right)
+
(1-\mu) \left(\sum_{m, r} P(r|I_t)\, \rho_2^{ * }(m|r,I_t) \, L(m, r) \right)
\nonumber \\
&=  \mu \left(\min_{\substack{\rho(m|r,I_t): \\ I(R; M \mid I_t) \leq B_1}} \sum_{m, r} P(r|I_t)\, \rho(m|r,I_t) \, L(m, r) \right)
\nonumber\\
&\qquad 
+
(1-\mu) \left(\min_{\substack{\rho(m|r,I_t): \\ I(R; M | I_t) \leq B_2}} \sum_{m, r} P(r|I_t)\, \rho(m|r,I_t) \, L(m, r) \right)
\nonumber \\
&= \mu D_t(B_1) + (1-\mu)D_t(B_2)
\end{align}
For fixed \(P(r|I_t)\), \(I(R;M \mid I_t)\) is known to be convex in \(\rho(m|r,I_t)\) (by suitably altering the analogous argument for the unconditioned mutual information in \citep{cover2006elements}), so
\begin{equation}
\label{eq:mr2a-1-10}
I_{P_{\mu}}(R; M \mid I_t) \leq \mu I_{P_1}(R; M \mid I_t) + (1-\mu) I_{P_2}(R; M \mid I_t) 
\end{equation}
where the right-hand side is at most
\(B_{\mu}:=\mu B_1+(1-\mu)B_2\),
since each policy satisfies its respective information constraint.

Defining
\begin{align}
\label{eq:mr2a-1-11}
D_t(B_{\mu}) := \min_{\rho(m|r,I_t) ~:~ I(R; M \mid I_t) \leq B_{\mu}} \mathbb{E}[L(M,R)]
\end{align}
we see that the inequality \eqref{eq:mr2a-1-10} implies that a \(\rho(m|r,I_t)\) can be found corresponding to a loss \(D_{t,P_{\mu}}\) needing \(B \leq B_{\mu}\). Since \(D_t(B)\) is non-increasing, this implies 
\begin{align}
\label{eq:mr2a-1-12}
D_t(B_{\mu}) &\leq D_{t,P_{\mu}} = \mu D_t(B_1) + (1-\mu) D_t(B_2)
\end{align}
and hence \(D_t(B)\) is convex.

\textit{Hence $G_t(B)=D_t(0)-D_t(B)$ is non-decreasing and concave.}

\textbf{Bound on VoII($\xi$)}:
Taking a fixed set of actions $\mathcal{A}_t$, the no-query decision value value $V_0 = \min_{m\in \mathcal{A}_t} \sum_r \pi_t(r)J_r(m)$, and the expected query value $V_{\xi}$ defined in  \eqref{eq:queryvalue}, we can construct the value of interpretation information for a given query $\xi$:
\[
\text{VoII}_{\xi} = V_0 - V_{\xi} 
\]
For a practical query $\xi$, any final message chosen only through its observed response satisfies the conditional Markov relation
\[
R\rightarrow Y_\xi\rightarrow M
\qquad\text{given }I_t.
\]
The data processing inequality \citep{cover2006elements} gives
\[
I(R;M\mid I_t)
\le
I(R;Y_\xi\mid I_t).
\]
The induced message policy is feasible at information budget
$I(R;Y_\xi\mid I_t)$, so
\[
V_\xi\ge D_t\!\left(I(R;Y_\xi\mid I_t)\right).
\]
Since $D_t(0)=V_0$, subtracting gives
\[
\operatorname{VoII}_t(\xi)
\le G_t\!\left(I(R;Y_\xi\mid I_t)\right).
\]

\textbf{Relation to rational inattention and information bottlenecks} \textit{We relate our formulation of VoII to some prior work in information and decision theory:}

Our sender $S$ acts as an agent facing the decision problem of picking a message $m \in \Aset_t$, given imperfect information about the payoff-relevant receive type $r \in \Rset$ with some prior belief $P(r | I_t)$. 

The action $S$ takes will come at some loss $L(m, r)$ depending on the mismatch between $m$ and $r$, while the information strategy $S$ takes comes at a cost which will be taken to be a constant $\lambda > 0$ times the mutual information $I(M;R|I_t)$. This setup is similar to the ``rational inattention" work described in \citep{sims2003rational, sims_rational_2010,matejka2015rational, RePEc:ecb:ecbwps:20212570} and elsewhere, and the ``information bottleneck" method from \citep{tishby2000ib}.

The optimisation problem facing a sender \(S\) is to find the action strategy \(\rho(m | r,I_t)\) satisfying 
\begin{equation}
\label{eq:mr2a-5-1}
\min_{\rho(m | r,I_t)} \mathbb{E}[ L(M, R)] + \lambda I(R; M \mid I_t)
\end{equation}
where the expected loss is defined as
\begin{equation}
\label{eq:mr2a-5-2}
\mathbb{E}[L(M, R)] = \sum_{m, r} P(r | I_t) \rho(m | r,I_t) L(m, r)
\end{equation}
and
\begin{align}
\label{eq:mr2a-5-3}
I(R;M \mid I_t) &= \sum_{m, r} P(m, r) \log \left( \frac{P(m, r)}{\bar{\rho}(m|I_t) P(r | I_t)}\right) 
= \sum_{m, r} P(r | I_t)\rho(m | r,I_t)  \log \left(\frac{\rho(m | r,I_t)}{ \bar{\rho}(m|I_t)}\right)
\end{align}
subject to the normalisation condition
\begin{equation}
\label{eq:mr2a-5-3-1}
\sum_m \rho(m|r,I_t) = 1\,, \qquad \forall r \in \Rset
\end{equation}
Here \(\bar{\rho}(m|I_t) = \sum_{r} P(r | I_t) \rho(m|r,I_t)\) is the conditional distribution of \(m\) given \(I_t\) having marginalised over all receiver types.

The optimisation problem can be expressed in terms of the following functional:
\begin{align}
\label{eq:mr2a-5-4}
\mathcal{S}[ \rho(m|r,I_t)] &= \sum_{m, r} P(r | I_t) \rho(m | r,I_t) \left( L(m, r) + \lambda \log \left(\frac{\rho(m | r,I_t)}{\bar{\rho}(m|I_t)}\right)\right) 
\nonumber \\
&\qquad + \sum_r P(r\mid I_t)\chi(r)
\left(\sum_m\rho(m\mid r,I_t)-1\right)
\end{align}
where \(\chi(r)\) acts as a Lagrange multiplier enforcing \eqref{eq:mr2a-5-3-1}.

Taking the variation with respect to \(\rho(m|r,I_t)\), we find
\begin{align}
\label{eq:mr2a-5-5}
\frac{\delta \mathcal{S}}{\delta \rho(m|r,I_t)} &=
P(r | I_t) \left( L(m,r) + \lambda \log \left(\frac{\rho(m|r,I_t)}{\bar{\rho}(m|I_t)}\right) + \lambda - \frac{\lambda}{\bar{\rho}(m|I_t)} \sum_{r^{\prime}} \left(P(r^{\prime} | I_t) \rho(m|r^{\prime},I_t)\right)\right) 
\nonumber \\
& \qquad + \chi(r) P(r | I_t)
\end{align}
where the third and fourth terms in brackets cancel.

Finding the extremum \(\rho^{\star}_{\lambda}(m, r | I_t)\) of \eqref{eq:mr2a-5-4} by setting \eqref{eq:mr2a-5-5} to zero, we have
\begin{equation}
\label{eq:mr2a-5-6}
\log \left(\frac{\rho_{\lambda}^{\star}(m|r,I_t) }{\bar{\rho}_{\lambda}^{\star}(m)}\right) = - \frac{L(m, r)}{\lambda} - \frac{\chi(r)}{\lambda}
\end{equation}
Defining \(\log Z^{I_t}_{\lambda}(r) := \chi(r)/\lambda\), the optimal action policy is
\begin{equation}
\label{eq:mr2a-5-7}
\rho_{\lambda}^{\star}(m|r,I_t) = \frac{\bar{\rho}_{\lambda}^{\star}(m | I_t)\exp \left( - L(m, r) /\lambda\right)}{Z^{I_t}_{\lambda}(r)}
\end{equation}
From \eqref{eq:mr2a-5-3-1}, we have
\[ 1 = \sum_{m^{\prime}} \frac{\bar{\rho}_{\lambda}^{\star}(m^{\prime} | I_t) \exp \left(- L(m^{\prime}, r)/\lambda\right)}{Z^{I_t}_{\lambda}(r)} \qquad \Rightarrow Z^{I_t}_{\lambda}(r) = \sum_{m^{\prime}} \bar{\rho}_{\lambda}^{\star}(m^{\prime} | I_t) \exp \left(- L(m^{\prime}, r)/\lambda\right)\]
so \(Z^{I_t}_{\lambda}(r)\) is a normalisation constant.

From equation \eqref{eq:mr2a-5-7}, we see that as one expects, when $\lambda\rightarrow\infty$ the policy becomes receiver-agnostic, and as $\lambda\rightarrow0$ it concentrates on receiver-specific loss minimisers, as expected from e.g.  \citep{tishby2000ib,matejka2015rational}.

\section{Interpretation-Probe Construction}
\label{app:probe}

\subsection{Task Identities and Surface Realisations}

The benchmark defines semantic task identities separately from their natural-language wordings. For task $z$, let
\[
\mathcal E(z)=\{T_z^{(1)},\ldots,T_z^{(R)}\}
\]
be validated descriptions of that task. The sender and probe banks are disjoint:
\[
m\sim\mathcal E_{\rm send}(z^\star),
\qquad
\widetilde T_z\sim\mathcal E_{\rm probe}(z),
\qquad
\mathcal E_{\rm send}\cap\mathcal E_{\rm probe}=\varnothing.
\]
The receiver chooses among task identities rather than matching the sender's wording.

When prompt-template resources give several transformations of the same record, we extract a task signature
\[
s(T)=
(\mathcal I_T,\mathcal C_T,\operatorname{op}_T,\mathcal O_T,\mathcal F_T),
\]
which lists the visible input fields, extra context, semantic operation, output meaning, and output format/type. Candidate pairs are first labelled as equivalent, extra-information variants, different tasks, or unclear. Only pairs already labelled as different tasks are then ranked by semantic closeness. This order keeps an opaque distance score from deciding task equivalence.

A strong distractor shares the same underlying example/interface and information, differs in one or two task-semantic components, and gives no trivial output-format cue. Easy/matched/hard difficulty is stored separately from the semantic relation.

\subsection{Task Bank and Splits}
\label{app:taskbank}

We apply this protocol to a bank of concrete task/message examples drawn from FreebaseQA, WebQSP, GrailQA, SQ-WD, PopQA, and Mintaka. Prompt-template resources such as PromptSource \citep{bach2022promptsource} help find neighbouring task transformations, but the benchmark consists of concrete instances with a known intended task, plausible nearby alternatives, and an executable or otherwise objective task score. Splits are made by normalised question group within each task family, so paraphrases and closely related transformations never cross the train, validation, and test sets.
Interpretation failures are rare (\NaturalPIRRate\ of test pairs). If extra failure cases are used to train or analyse the predictor, this enrichment stays in training/analysis data and is tagged by provenance. Calibration and headline evaluation always use untouched natural-prevalence validation and test sets.

The six datasets share one useful property: every question has a gold answer, so execution can be scored without a judge, while the questions range widely in style, from trivia to compositional queries. We split each dataset 70/15/15 by normalised question group, keeping related transformations within the same split.

\subsection{Probe Robustness Checks}
\label{app:position}

The headline measurement uses a correct task option, one contrast task, and \textsc{None of these}. Instead of a separate paraphrase study, every message--receiver pair gets six probe instances with different wording, and the label averages over them. Five temperature-0 repeats of the full test protocol reproduce every dataset's $\yI$ within \RepeatStabilityMaxDiff\ percentage points. Averaging over the six probes spreads wording and order effects over all messages but does not remove a systematic position effect, which we check below. A dedicated paraphrase-bank study and a construct-validation set of clear/easy, confusing/easy, and clear/hard messages are left for future work. \textsc{None of these} counts as a wrong task but is kept as a separate raw outcome. Unparsed probe responses are dropped from the denominator of \cref{eq:softlabel}.

\paragraph{Position of the intended task.}
The intended task is listed first, second, or last in 33.1, 33.6, and 33.2\% of the test probes, and \cref{tab:position} gives the probe error rate by this position. Nemotron misreads 24.5\% of the probes that list the intended task last, against 3.1\% and 4.5\% when it is first or second. It picks the last option in 25.8\% of probes, against 33.3\% for the other receivers, and nearly all of its errors at the last position are picks of the contrast task (24.43 of 24.49 points). The effect is one of position, not wording: under each of the six probe wordings, Nemotron's error rate is 18.5--31.0\% with the intended task last and 1.7--4.7\% with it first or second. The other receivers show small position effects in both directions. When only probes with the intended task first or second are kept, the error rate still ranges from 1.45\% (Qwen3.6) to 5.50\% (Mistral), a ratio of 3.8 (95\% CI 3.4--4.3, against 6.2 on all probes), and Mistral exceeds Qwen3.6 by 4.1 points [3.7, 4.4]. Within datasets the ratio is 1.7--2.6 on FreebaseQA, GrailQA, Mintaka, and WebQSP, 6.6 on PopQA, and 14.3 on SQ-WD. Qwen3.6 is the lowest receiver on all six datasets, and the highest is Mistral on four and Nemotron on two. Receiver heterogeneity therefore survives the position control, although it is smaller, and most of Nemotron's excess $\yI$ comes from the last position. Experiment~2 shows the same pattern. We rescore it against a label built only from probes with the intended task first or second, refitting on validation data either one global calibration map or one map per receiver. Per-receiver maps then add about 0.02 to across-receiver AUROC (A3: 0.750 to 0.770, full model: 0.754 to 0.772), against about 0.10 on the full label, and the full model still ranks slightly better within receivers (AUROC 0.756 vs.\ 0.751 for A3). Much of the across-receiver gain of Experiment~2 thus reflects Nemotron's position effect.

\begin{table}[!htbp]
\caption{Probe error rate (\%) by the position of the intended task (test split, parsed probes). Last column: probes with the intended task first or second, with question-cluster bootstrap 95\% CIs.}
\label{tab:position}
\centering
\scriptsize
\begin{tabular}{lccccc}
\toprule
Receiver & First & Second & Last & All & First or second [95\% CI]\\
\midrule
Qwen3.6 & 1.28 & 1.61 & 2.33 & 1.74 & 1.45 [1.26, 1.65]\\
Gemma & 1.59 & 2.49 & 2.57 & 2.22 & 2.04 [1.81, 2.28]\\
gpt-oss & 3.27 & 2.08 & 1.63 & 2.33 & 2.68 [2.42, 2.95]\\
Nemotron & 3.14 & 4.51 & 24.49 & 10.72 & 3.83 [3.57, 4.09]\\
Mistral & 5.26 & 5.74 & 4.45 & 5.16 & 5.50 [5.15, 5.85]\\
\midrule
Highest/lowest & 4.1 & 3.6 & 15.0 & 6.2 & 3.8 [3.4, 4.3]\\
\bottomrule
\end{tabular}
\end{table}

\section{Receiver-Type Inference and Query Bank}
\label{app:receiver}

\subsection{Calibration Bank}

The fixed query bank is
\[
\Qset=\{\xi_1,\ldots,\xi_M\},
\]
built from training/calibration instances only. Each item stores its task/message content, a fixed response parser, and repeated genuine responses from every receiver type. Responses are parsed into a discrete signature $Y_\xi\in\mathcal Y_\xi$ and smoothed likelihoods
\[
\widehat P(Y_\xi=y\mid R=r,\xi)
\]
are estimated by Laplace/Dirichlet smoothing. Raw responses are kept for rescoring.

\subsection{History Encoder}

A small receiver classifier encodes each interaction and pools over the history:
\[
h_s=f_{\rm int}(\xi_s,y_s),
\qquad
h_H=\operatorname{Pool}(h_1,\ldots,h_n),
\]
\[
\pi_\omega(r\mid H)
=\operatorname{softmax}_r(Wh_H+b).
\]
Mean or sum pooling is the default, and the main experiment needs no recurrent or transformer history encoder. Training uses receiver-type cross-entropy, and probabilities are calibrated on held-out histories. For $n=0$ we use an explicit receiver prior, which is uniform unless there is an independent reason to choose otherwise.

Histories are built by subsampling stored interactions from one receiver, and every response is a genuine output of that receiver. Query choice, history length, filenames, application programming interface (API) metadata, and other incidental fields are matched across receiver types, so the classifier cannot exploit leakage.

\section{Message-Revision Implementation}
\label{app:rewrite}

\paragraph{Estimators.}
PIR is the quantity being estimated, not one fixed predictor. Experiment~2 estimates it on natural messages with the heads of \cref{app:labels}. Rewrites differ in distribution from natural messages (for example, they state the output or turn the question into a request), so Experiments~4 and~5 use candidate-level estimators fitted to measured rewrites: a linear model of mutable features (Experiment~4) and paired risk heads (Experiment~5), both described below.

\paragraph{Feature scores.}
The feature set is fixed in advance: features are labelled as immutable/task-inherent or mutable/style-structural before rewriting. What is learned from training data is the risk effect $s_{rk}$ of mutable feature $k$ for receiver $r$, such as the difference in measured interpretation-failure rate when the feature/value is present versus absent. Under posterior $\pi_t$, use
\[
\bar s_{tk}=\sum_r\pi_t(r)s_{rk}
\]
and pass only positive, currently relevant mutable features to the rewriter.

\paragraph{Generation.}
Generate a fixed number $K_m=\RewriteCandidates$ of candidates, with the instruction to keep the exact task meaning while reducing the selected mutable features. All generate-and-select methods share this fixed budget.

\paragraph{Semantic verifier.}
The verifier is a separate stateless call to the same model version as the rewriter. It receives only $(z^\star,x_S,m_0,m'_1,\ldots,m'_{K_m})$ and returns \textsc{Pass}/\textsc{Fail} plus a short failure reason for each candidate. It never sees the receiver identity or posterior, PIR scores, risk-feature values, or candidate preference. The original $m_0$ is always kept, and failed candidates are logged, not regenerated.

\paragraph{Selection.}
For every valid candidate, evaluate every receiver-conditioned head and compute \cref{eq:posteriorloss}. Select by posterior expected loss, not by the MAP receiver. After a real query response, update the receiver posterior and re-score the same fixed action set.

\paragraph{Experiment~4 implementation.}
Experiment~4 uses a linear risk model with eight binary mutable features: the output is not stated, the question is replaced by a request, the instruction line is missing, surface errors (lower-case start or no question mark), a long question, a question word that is not first, a parenthetical, and a pronoun. For each receiver $r$, the change in soft $\yI$ from the original to a rewrite is regressed (ridge, $\alpha=1$) on the change $\Delta f$ in these features. The intercept $b_r$ is the effect of rewriting itself, and the slopes are the scores $s_{rk}$ (\cref{fig:e4_mechanism}a). The model is fitted on 4,255 rewrites of 600 training questions (100 per dataset), each measured by all five receivers. Guidance lists up to three features with the largest $\bar s_{tk}$, among those for which $\bar s_{tk}$ and its 2.5\% bootstrap quantile (500 question-cluster resamples) are both positive, each with a fixed edit instruction. Selection sends the message with the lowest predicted risk $\sum_r\pi_t(r)\,(b_r+\sum_k s_{rk}\Delta f_k)$ and keeps the original unless the predicted gain exceeds a margin of 0.001, chosen on 300 validation questions. The rewriter is \RewriteModel\ \citep{deepseekai2026deepseekv4} at temperature 0.4 with $K_m=4$. Each call holds eight requests from one condition only, so that conditions cannot copy one another. The instruction for an unstated output asks the rewriter to state the output in the question's own words (e.g., ``Give the name of the $\langle$thing asked for$\rangle$.'') and without task-type words such as kind, class, category, type, attribute, or label. Deterministic checks reject candidates that, for example, repeat the original, add a gold answer, change numbers, negations, or protected qualifiers, or add task-type words, and the verifier above then checks the rest. The posterior $\pi_t$ is the type model of Experiment~3 after five stored responses from the audit bank, and all conditions of a question share this history.

\paragraph{Paired risk heads for Experiment~5.}
Experiment~5 chooses among a fixed pool of generic rewrites. Candidates are generic clarity rewrites generated without feature guidance (four per message by \RewriteModel, English v1.1 prompts), verified by the stateless batched call above, and embedded with \EmbeddingModel\ \citep{zhang2025qwen3embedding}. The scoring head is trained on paired candidate measurements: for 600 training questions per dataset, every accepted candidate and the original were measured by all five receivers (one fresh repeat, six probes and one execution call each), giving one soft $\yI$ per question--message--receiver. Inputs are 64-component principal component analysis (PCA) projections (fitted on training texts only) of the embeddings of the message, the intended task, and the original message, plus the message--original difference, surface features, edit geometry, and a receiver one-hot. The MLP has hidden widths 128 and 64, dropout 0.15, learning rate $10^{-3}$, weight decay 0.01, and at most 100 epochs with early stopping (patience 12) on 80 held-out validation questions, and two seeds are ensembled. Three model families are compared: an absolute soft-risk head, a paired-loss head that adds eight times the squared error of the predicted within-question risk differences, and ridge regressions of the within-question risk change on text and edit features ($\alpha\in\{10,100,1000\}$). Scores are calibrated with a common positive slope and per-receiver intercepts on 60 calibration questions, and the family together with the minimum predicted improvement required to leave the original ($\{0,0.002,0.005\}$) was chosen on 60 further policy-selection questions by the measured $\yI$ of the posterior policy. The selections are the absolute head (FreebaseQA, WebQSP, SQ-WD), the paired head (GrailQA), and ridge (PopQA with $\alpha=1000$, Mintaka with $\alpha=10$). The posterior $\pi_t$ is the type model of Experiment~3 applied to five stored responses drawn from the audit bank. The same history and the same candidate pool serve all conditions of a question, and the true identity is used only to construct the environment and the true-identity reference. Weights, calibration, and margins were frozen before any outcome of the replication cohort of Experiment~5 was read. Its questions are the first 400 valid probe instruments per dataset in a frozen hash order over questions never used before (500 attempts per dataset, extended under a pre-registered amendment to at most 1,200 for Mintaka, where about half of the attempts gave a valid instrument, for 3,044 attempts in total). Rejected attempts were never selected on receiver outcomes.

\section{Experimental Protocols}
\label{app:protocols}

The experiments test the claims in the same order as the framework. Experiment~1 asks whether real LLM receivers reconstruct tasks differently and whether interpretation failure is distinct from execution failure. Experiment~2 asks whether this lower-dimensional risk is predictable before sending and whether privileged cross-receiver supervision adds information beyond the message alone. Experiment~3 asks how much ordinary behavioural history, rather than model identity, reveals about the latent receiver type. Experiment~4 then tests whether posterior-guided rewriting reduces \emph{measured} misinterpretation, and Experiment~5 tests whether VoII allocates queries better for the interpretation objective. Experiments~1--3 fix the message (one template, no sender model), so every receiver reads the same text and differences come from the receiver. Experiments~4 and~5 then let an LLM revise the message and control queries. Experiments~1--3 use the frozen test split, Experiment~4 uses 1,579 held-out questions from outside the test split, and Experiment~5 uses the test split as its primary cohort and a separate set of \ConfirmationQuestions\ new questions as a replication. This section gives the parts of the design in \cref{sec:experiments} that do not fit in the main text. Implementation details and hyperparameters are in \cref{app:fullspec}.

\subsection{Labels and Predictor}
\label{app:labels}

\textbf{Labels.}
Each probe offers three options: the intended task, the contrast task, and \textsc{None of these}. Probe text is generated per message by \RewriteModel, passed through deterministic checks and a verifier, and issued $K_p=\ProbeRepeats$ times with independently varied wording and option order. The label $\widetilde y^I$ is the fraction of parsed probes that choose a wrong option (\cref{eq:softlabel}). Execution is one separate answer call, scored by gold-alias containment. The test split alone uses \NumProbeCalls\ probe calls and \NumTrials\ answer calls.

\textbf{Predictor.}
The heads $q^I$ and $q^C$ are separate small MLPs on frozen \EmbeddingModel\ features of $m$ and of the original question $x_S$ (our proxy for sender context), with an optional difference vector $m-x_S$ and a 16-dimensional learned receiver embedding $e_R$. Each MLP output is added to a logistic offset fitted first on the training data: a per-receiver intercept and slope on the scalar gap $g$, the root-mean-square difference between the standardised embeddings of $x_S$ and $m$ (A3 uses one intercept and slope). $q^I$ is trained with BCE on $\widetilde y^I$, and $q^C$ with BCE on $\yA$ weighted by $1-\widetilde y^I$. Configurations (27 per dataset, 3 seeds) are selected on validation NLL only. A Platt map \citep{platt2000probabilities} with a per-receiver slope and intercept and shared terms in $g$ is then fitted on the calibration part of the validation split (A3 uses one global map). Candidates in the rewrite experiments are scored by the estimators of \cref{app:rewrite}, and only the combined objective of Experiment~5 reuses this $q^C$.

\subsection{Receiver Inference (Experiment~3)}
\label{app:protocol_e3}

Experiment~3 uses a calibration bank of 384 tasks separate from the test messages, with 9,600 archived responses from the five receivers, from which we build 20,800 histories of length 1 to 20. The type model embeds each response (PCA to 32 dimensions), scores it against its query with a small interaction MLP, mean-pools over the history, and outputs a five-way softmax with temperature calibration \citep{guo2017calibration}. Tasks are split by role (192 fit, 48 early stop, 48 calibration, 96 audit), and every reported number uses 960 audit histories per length. A shuffled-history control pairs each query with responses from a random receiver.

\subsection{Message Revision (Experiment~4)}
\label{app:protocol_e4}

Experiment~4 tests the full revision step of \cref{sec:rewrite}: learned risk guides the rewriter and then selects the sent message (\cref{app:rewrite}). All conditions share the rewriter, the intended task, four new candidates per request, the checks, and the verifier. They differ only in where learned risk is used. The conditions are the original message, the generic rewrite (the first valid candidate of a request for clarity only), selection only (the same generic pool, selected by the risk model), posterior only (the rewriter sees $\pi_t$ but no risk features, and the risk model selects), ours (guidance and selection), guidance only (our pool, first valid candidate), ours with a uniform belief or with the true identity in place of $\pi_t$ (a diagnostic, not a deployable method), one feature per step (four steps, each editing the top risk feature of the current message, then selection), and ours before the fix of the output instruction (\cref{app:res_e4}). The comparisons of ours with the original, the generic rewrite, selection only, and posterior only were pre-registered, with Holm correction and a non-inferiority margin of one point on $\yA$. Before the main run, a check on 300 validation questions (50 per dataset) had to show that no fixed condition adds task-type words, that candidate pools do not collapse to the original, and that ours lowers $\yI$ against the original. All three held. The main run uses 1,579 held-out questions, 300 per dataset in a frozen hash order (WebQSP has only 79 unused questions). Of these, 1,304 have the entity-versus-category contrast and 275 the value-versus-attribute contrast. They come from the training and validation splits of Experiments~1--3, not from the test split, and none was used to fit the risk model, to choose the margin, or to develop the prompts. The questions, all sent messages, the measurement protocol, and the analysis script were frozen before any outcome was read. Each distinct message is measured with each receiver by 36 probes (six wordings times six option orders, with a neutral \textsc{None} option) and one execution call. Here $\yI$ is the share of the 36 probes that choose a wrong option, with unparsed probes counted as wrong, so its level is not comparable with Experiments~1--3. A blind audit by Gemini 3.8 Flash judges whether each sent message keeps the task, and a pre-registered sensitivity analysis sends the original in place of every flagged message.

\subsection{Query Control (Experiment~5)}
\label{app:protocol_e5}

Experiment~5 tests the query rule of \cref{sec:voii_method}. The primary cohort is the frozen test split: \QueryQuestions\ questions, five receivers, and three fresh answer repeats (\QueryEpisodes\ episodes). A separate cohort of \ConfirmationQuestions\ new questions (\ConfirmationEpisodes\ episodes) is analysed as a replication. Every episode starts from five stored responses and the calibrated posterior $\pi_t$ of Experiment~3. The candidate pool is the original message and its four generic rewrites with frozen verifier decisions, scored by the frozen paired heads of \cref{app:rewrite}, and no head was refit. The query bank is the \QueryBankSize\ fit-role tasks of Experiment~3 with their archived responses. A query returns whether the receiver solved the task, and the posterior is updated by an exact Bayes step with Laplace-smoothed likelihoods. A policy may ask at most one query per episode and then sends the candidate with the lowest posterior expected interpretation risk. NetVoII (\cref{eq:netvoii}) is compared with IG (the query with the largest expected entropy reduction), random querying (three pre-declared seeds), and, as a secondary comparator, an adapted VoI rule \citep{dong2026voi} that keeps the same posterior, bank, candidates, and channel but replaces the learned heads by \RewriteModel\ utility estimates. Under every policy, ties in expected loss go to the earlier candidate, so the original wins a tie. Three references fix the scale: never query, always query, and the true receiver identity in place of $\pi_t$, a diagnostic reference rather than an upper bound (\cref{app:res_e5}). Budgets of 10, 20, 30, and 50\% of episodes are exact quotas filled in each policy's own ranking order. The strict rule instead queries only where $\operatorname{NetVoII}>0$ at $c=\QueryCost$ (five other costs in \cref{tab:e5_cost_sweep}), and IG and random querying are re-run at its query count. All schedules were frozen before any receiver call.

\subsection{Metrics and Statistics}
\label{app:metrics}

We report AUROC, AUPRC, Brier score \citep{brier1950verification}, NLL, and ECE \citep{naeini2015obtaining} for prediction, and accuracy, NLL, Brier, and ECE for receiver inference. Main-text prediction metrics are computed within each receiver and macro-averaged, so they measure how well messages are ranked for a known receiver. Pooled metrics, which also reward ranking across receivers, appear in \cref{tab:receiver_ablation}. Intervals are 95\% paired bootstraps \citep{efron1994introduction} over question groups (1,000 resamples) for Experiment~2, binomial intervals for Experiment~3, question-level normal intervals for Experiment~1, and paired bootstraps over questions (10,000 resamples, Holm-corrected over the four pre-registered comparisons) for Experiment~4. For Experiment~5 we report the share of episodes queried, measured $\yI$ and task success $1-\yA$, net utility $1-\yI-c\,r$ for query rate $r$, and the share of episodes whose posterior mode is the true receiver, with paired question-cluster bootstraps (\QueryBootstrapResamples\ resamples) Holm-corrected over the eight pre-registered net-utility contrasts (NetVoII against IG and against random querying at the four budgets). ECE uses 10 equal-mass bins in Experiment~2 and 10 equal-width bins in Experiment~3.

\subsection{Scope of the Reported Experiments}
\label{app:scope}

Three questions lie outside the reported experiments: a separate paraphrase/order robustness study (each pair already has six wording/order variants, five temperature-0 repeats reproduce $\yI$ within \RepeatStabilityMaxDiff\ points, and \cref{app:position} controls for option position), a step-by-step input ladder (because $x_S$ is a proxy that fixes $z^\star$, the $m$ and $(m,z^\star)$ steps are not separately identifiable), and open-set receivers. Experiment~4 uses one rewriter and does not compare the posterior with a MAP receiver. Experiment~5 compares NetVoII with never, always, and random querying, IG, an adapted VoI rule (secondary), and the true-identity reference, each with at most one query per episode. A self-confidence control and rankings by $q^I$, $q^C$, generic action risk, or receiver-agnostic PIR are natural extensions.

\section{Baselines and Ablations}
\label{app:ablations}

\subsection{Essential Prediction Baselines}

\begin{table*}[t]
\caption{Baseline definitions.}
\label{tab:baseline_defs}
\centering
\small
\begin{tabular}{p{0.10\textwidth}p{0.22\textwidth}p{0.60\textwidth}}
\toprule
ID & Method & Definition \\
\midrule
A0 & Base-rate priors & Global $P(Y^I=1)$ and receiver-specific $P(Y^I=1\mid R)$ estimated on training data.\\
A1 & Message-only ambiguity & Fixed strong judge sees $m$ only and scores whether materially different task interpretations are possible.\\
A2 & Semantic equivalence & Judge sees $(m,z^\star,x_S)$ and predicts whether $m$ faithfully specifies the intended task, with no receiver information.\\
A3 & Receiver-agnostic PIR & Same learned PIR architecture but remove $e_R$ / receiver posterior information.\\
A4 & Action-failure predictor & Same architecture/input budget trained against $Y^A$ rather than $Y^I$, evaluated against both targets.\\
A5 & Listener/ToM & Predict $P(\widehat Z_R=z\mid m,x,e_R)$ over probe task identities and use $1-P(\widehat Z_R=z^\star)$ as risk.\\
-- & Shuffled receiver & Full model trained with receiver identities randomly permuted across pairs, to test whether the gain requires the correct receiver.\\
-- & Receiver-calibrated A3 & A3 passed through the full model's calibration map (per-receiver slope and intercept, shared gap terms), refitted on the same validation questions, to test whether receiver information adds more than receiver-specific risk levels.\\
\bottomrule
\end{tabular}
\end{table*}

The reported runs use A0--A5 plus the shuffled-receiver control and the receiver-calibrated A3, and all learned methods use the same question-group split and validation-only hyperparameter selection. A1 and A2 use \RewriteModel\ as judge. Of 26,528 judge calls, 54 were unparsable and fall back to the training mean. Other candidate baselines (semantic entropy \citep{farquhar2024semantic}, a zero-shot receiver-aware judge, sampled receiver entropy, intent entropy, structured expected value of perfect information \citep{suri2026structured}, and AgentAsk-style clarification \citep{agentask2026}) are outside the reported comparison.

\subsection{Essential Ablations}

The input-information ladder $q^I(m)\rightarrow q^I(m,z^\star)\rightarrow q^I(m,z^\star,x_S)\rightarrow q^I(m,z^\star,x_S,e_R)$ cannot be split step by step here: the sender-context input $x_S$ is the original benchmark question, which already fixes $z^\star$, so the first two steps are not separately identifiable. We therefore report the last step, which adds the receiver information, in four forms: full receiver information, receiver information removed (A3), receiver information shuffled (\cref{tab:receiver_ablation}), and receiver information used only to recalibrate A3 (\cref{tab:e2_recal}), together with $Y^I$ versus $Y^A$ supervision (A4). For the receiver belief, the rewrite experiment compares the history posterior with a uniform prior and with the true identity.

For query control, Experiment~5 compares NetVoII with IG, random querying, an adapted VoI rule with LLM utility estimates (a secondary comparator), never and always querying, and the true-identity reference, at exact query quotas and under the strict rule (\cref{sec:res_e5}). The contrast between IG and NetVoII directly tests the claim that interpretation-specific uncertainty and decision value are distinct.

For rewriting, Experiment~4 turns each part of the method off in turn: selection only and posterior only drop the guidance, guidance only drops the selection, and a uniform belief or the true identity replaces the posterior (\cref{app:protocol_e4}). One feature per step tests a sequential form of the guidance, at about twice the tokens and four dependent rounds (\cref{app:res_e4}).

\FloatBarrier
\section{Additional Results}
\label{app:results}

This section gives the full results behind \cref{sec:results}, one subsection per experiment. Experiments~1--3 test the premises of the framework: interpretation risk differs across receivers (Experiment~1), it can be predicted before sending (Experiment~2), and behaviour carries information about the receiver type (Experiment~3). Experiments~4 and~5 test the two decisions built on these premises, message revision and receiver queries, and we relate their outcomes to \cref{thm:conditioning,thm:frontier}.

\subsection{Experiment~1: Interpretation and Execution Failures}
\label{app:res_e1}

\begin{figure}[!htb]
\centering
\includegraphics[width=\textwidth]{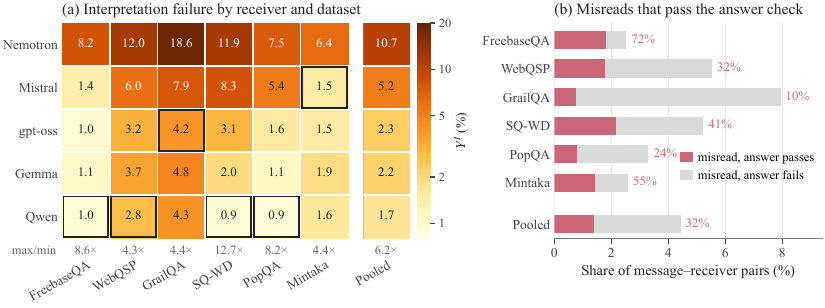}
\caption{Interpretation failure depends on the receiver and the dataset (test split). (a)~$\yI$ (\%) per receiver and dataset on a log colour scale. Boxes mark the least risky receiver in each dataset, and the bottom row gives the ratio of the highest to the lowest receiver. (b)~Misread pairs split by the answer check. Labels give the share of misreads that pass.}
\label{fig:e1_landscape}
\end{figure}

Interpretation risk depends on the pair of receiver and message, not on the receiver alone. \Cref{fig:e1_landscape}a gives $\yI$ for every receiver and dataset. Within a dataset, the riskiest receiver fails 4.3 to 12.7 times as often as the least risky one (1.7 to 14.3 times when the intended task is listed first or second, \cref{app:position}). Nemotron is the riskiest receiver on every dataset, but the least risky one changes: Qwen on four datasets, gpt-oss on GrailQA, and Mistral on Mintaka, although Mistral is the second riskiest receiver overall. A single score per receiver would miss these reversals, which is why the predictor conditions on both the message and the receiver. Heterogeneity alone does not make receiver information useful: by \cref{thm:conditioning}, it pays off only when it changes the preferred message, which Experiments~4 and~5 test.

Answer checks miss many interpretation failures. \Cref{fig:e1_landscape}b splits the misread pairs of each dataset by the answer check, and the last two columns of \cref{tab:decomp} give both off-diagonal cells. Pooled over datasets, 1.4\% of all pairs misread the task yet pass the answer check, which is \MisreadPassShare\ of all misreads, and on FreebaseQA the share is 72\%. The reverse case is far more common: 56.4\% of pairs read the task correctly and still fail. An answer-based score therefore mixes the two failures, and $\yI$ has to be measured on its own, as the probes do.

\begin{table}[!htb]
\caption{Interpretation ($\yI$) and execution ($\yA$) failures per dataset (test split, pooled over five receivers, 95\% question-level normal intervals).}
\label{tab:decomp}
\centering
\scriptsize
\setlength{\tabcolsep}{2.5pt}
\begin{tabular}{lrcccc}
\toprule
& & \multicolumn{2}{c}{Marginal rates (\%)} & \multicolumn{2}{c}{Off-diagonal cells (\%)}\\
\cmidrule(lr){3-4}\cmidrule(lr){5-6}
Dataset & Pairs & $\yI$ [95\% CI] & $\yA$ [95\% CI] & misread, pass & read, fail\\
\midrule
FreebaseQA & 6,550 & 2.5 [1.7, 3.4] & 24.0 [21.7, 26.4] & 1.8 & 23.3\\
WebQSP & 1,095 & 5.5 [2.5, 8.6] & 54.1 [47.5, 60.7] & 1.8 & 50.3\\
GrailQA & 6,430 & 7.9 [6.5, 9.4] & 89.6 [87.9, 91.3] & 0.8 & 82.4\\
SQ-WD & 7,090 & 5.2 [4.1, 6.4] & 61.9 [59.3, 64.4] & 2.2 & 58.8\\
PopQA & 7,255 & 3.3 [2.4, 4.2] & 75.7 [73.5, 77.9] & 0.8 & 73.2\\
Mintaka & 4,725 & 2.6 [1.6, 3.6] & 40.3 [37.2, 43.4] & 1.4 & 39.1\\
\midrule
Pooled & 33,145 & 4.4 & 59.5 & 1.4 & 56.4\\
\bottomrule
\end{tabular}
\end{table}

\FloatBarrier
\subsection{Experiment~2: Prospective PIR Prediction}
\label{app:res_e2}

\begin{figure}[!htb]
\centering
\includegraphics[width=\textwidth]{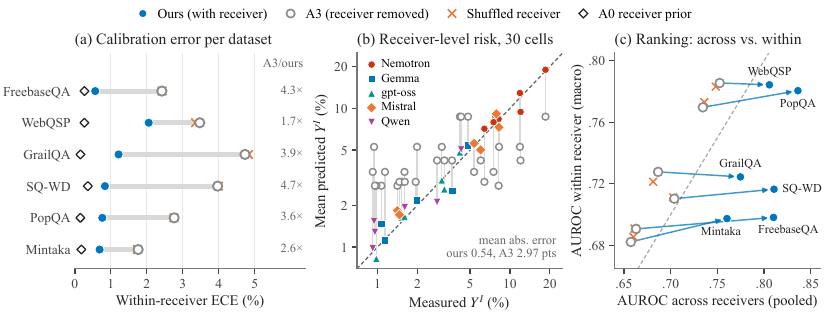}
\caption{Receiver information calibrates the predicted risk (test split, seed 0 in (a, b), three-seed means in (c)). (a)~Within-receiver ECE per dataset. (b)~Mean predicted versus measured $\yI$ in each of the 30 receiver--dataset cells. (c)~AUROC across versus within receivers, with arrows from A3 to ours.}
\label{fig:calibration}
\end{figure}

Receiver information makes the predicted risk calibrated. Without it (A3), within-receiver ECE is 1.7 to 4.7 times higher on every dataset (\cref{fig:calibration}a). The shuffled-receiver control stays at the A3 level, so the gain needs the correct receiver, not extra inputs. The per-receiver base rate (A0) is calibrated by construction but cannot rank messages (AUROC 0.500), whereas the full model is both calibrated and discriminative.

\Cref{fig:calibration}b shows where the gap comes from. A3 predicts almost the same mean risk for every receiver of a dataset, so it overestimates the risk of the low-risk receivers and underestimates that of Nemotron. The full model tracks the measured level in each of the 30 receiver--dataset cells, with a mean absolute error of 0.54 points against 2.97 for A3. This matters for the decision rule because posterior expected loss compares each message across receivers, so systematically compressing receiver-specific risk levels can distort message selection. The full model learns these levels from privileged supervision: during training, the same messages are measured on every receiver.

\Cref{fig:calibration}c separates the two ways in which the prediction is used. Within a receiver, removing the receiver changes AUROC by at most 0.015 on any dataset, so the message features alone rank messages for a fixed receiver. Across receivers, AUROC rises from 0.66--0.75 to 0.76--0.84 on every dataset (\cref{tab:receiver_ablation}). This across-receiver ranking is what the decision rule needs, since the expected loss $\bar J_t(m)=\sum_r\pi_t(r)J_r(m)$ weighs each message across receivers.

\paragraph{Receiver levels or interaction?}
A3 predicts almost the same risk for every receiver, so part of the gain of the full model could come from receiver-specific base rates alone. We test this by passing A3's predictions through the calibration map of the full model, with a per-receiver slope and intercept and shared terms in the gap $g$, refitted on the same validation questions (\cref{tab:e2_recal}). This receiver-calibrated A3 recovers 93\% of the ECE reduction, 90--93\% of the Brier and NLL gains, and 96\% of the across-receiver AUROC gain. A per-receiver intercept alone recovers 79\% of the ECE reduction, and a richer map fitted on all validation data, which uses more calibration data than the full model, recovers 99\%. The full model keeps an advantage within receivers, where recalibration barely changes the ranking: AUROC $+0.004$ [$+0.001$, $+0.008$] and AUPRC $+0.007$ [$+0.002$, $+0.016$]. These two differences also hold against the richer map, whereas its Brier, NLL, and ECE differences from the full model include zero. Over three seeds, the receiver-calibrated A3 again recovers 93\% of the ECE reduction and 96\% of the across-receiver AUROC gain. Receiver information therefore acts mainly through receiver-specific risk levels, and the message--receiver interaction adds a small gain in ranking messages within a receiver. Both enter the expected loss $\bar J_t$.

\begin{table}[!htbp]
\caption{Receiver levels or interaction (Experiment~2, test split, six-dataset means, seed 0). Maps are refitted on the validation questions used by the full model ($^\ast$\,on all validation data). Difference: ours minus A3 + recv.\ calib., with paired question-group bootstrap 95\% CIs.}
\label{tab:e2_recal}
\centering
\scriptsize
\setlength{\tabcolsep}{3pt}
\begin{tabular}{@{}lccccccc@{}}
\toprule
& \multicolumn{2}{c}{Per receiver$\uparrow$} & \multicolumn{2}{c}{Across receivers$\uparrow$} & \multicolumn{3}{c@{}}{Calibration$\downarrow$}\\
\cmidrule(lr){2-3}\cmidrule(lr){4-5}\cmidrule(l){6-8}
Method & AUROC & AUPRC & AUROC & AUPRC & Brier & NLL & ECE\\
\midrule
A3 (no receiver) & .729 & .128 & .701 & .115 & .0414 & .168 & .032\\
A3 + recv.\ intercept & .729 & .128 & .792 & .158 & .0401 & .156 & .015\\
A3 + recv.\ calib. & .728 & .128 & .796 & .158 & .0399 & .155 & .012\\
A3 + recv.\ map$^\ast$ & .728 & .128 & .797 & .160 & .0398 & .154 & .010\\
Ours & .732 & .135 & .800 & .164 & .0398 & .154 & .010\\
\midrule
Difference & $+.004$ & $+.007$ & $+.004$ & $+.006$ & $-.0002$ & $-.0010$ & $-.0015$\\
\quad 95\% CI & [.001, .008] & [.002, .016] & -- & -- & [$-$.0003, $-$.0001] & [$-$.0015, $-$.0006] & [$-$.0022, $-$.0001]\\
\bottomrule
\end{tabular}
\end{table}

\Cref{tab:e2_per_dataset} gives all metrics per dataset, and its AUROC intervals use 1,000 paired bootstrap resamples over frozen question groups, paired across methods and receivers. \Cref{tab:receiver_ablation} gives the receiver-information ablation. There, pooled AUROC ranks all receivers together, whereas macro AUROC and ECE are computed within each receiver and averaged. Brier and NLL are the same under both views.

\begin{table*}[!t]
\caption{Experiment~2 results per dataset (test split, seed 0, with metrics computed within each receiver and macro-averaged). AUROC has a paired 95\% bootstrap CI. Bold: best value per dataset among learned methods.}
\label{tab:e2_per_dataset}
\centering
\scriptsize
\setlength{\tabcolsep}{4pt}
\renewcommand{\arraystretch}{0.94}
\begin{tabular}{llccccc}
\toprule
Dataset & Method & AUROC [95\% CI] & AUPRC & Brier$\downarrow$ & NLL$\downarrow$ & ECE$\downarrow$\\
\midrule
\multirow{9}{*}{FreebaseQA}
 & A0 global base rate & 0.500 & 0.025 & 0.0246 & 0.118 & 0.024\\
 & A0 per-receiver base rate & 0.500 & 0.025 & 0.0238 & 0.105 & 0.003\\
 & A1 message-only judge & 0.529 [0.481, 0.578] & 0.029 & 0.0246 & 0.118 & 0.024\\
 & A2 equivalence judge & 0.500 & 0.025 & 0.0246 & 0.118 & 0.024\\
 & A3 receiver-agnostic PIR & 0.690 [0.645, 0.730] & 0.085 & 0.0243 & 0.113 & 0.024\\
 & A4 action-failure predictor & 0.558 [0.518, 0.599] & 0.032 & 0.115 & 0.377 & 0.234\\
 & A5 listener / ToM & 0.658 [0.614, 0.699] & 0.043 & 0.0241 & 0.105 & 0.007\\
 & Shuffled receiver & 0.693 [0.649, 0.733] & 0.084 & 0.0243 & 0.113 & 0.024\\
 & Receiver-conditioned PIR & \textbf{0.696} [0.652, 0.735] & \textbf{0.095} & \textbf{0.0234} & \textbf{0.100} & \textbf{0.006}\\
\midrule
\multirow{9}{*}{WebQSP}
 & A0 global base rate & 0.500 & 0.055 & 0.0523 & 0.214 & 0.028\\
 & A0 per-receiver base rate & 0.500 & 0.055 & 0.0512 & 0.204 & 0.003\\
 & A1 message-only judge & 0.599 [0.521, 0.673] & 0.080 & 0.0524 & 0.215 & 0.034\\
 & A2 equivalence judge & 0.500 & 0.055 & 0.0525 & 0.217 & 0.025\\
 & A3 receiver-agnostic PIR & \textbf{0.791} [0.736, 0.829] & 0.187 & 0.0500 & 0.194 & 0.035\\
 & A4 action-failure predictor & 0.617 [0.541, 0.682] & 0.076 & 0.378 & 1.128 & 0.516\\
 & A5 listener / ToM & 0.762 [0.704, 0.800] & 0.144 & 0.0513 & 0.195 & 0.022\\
 & Shuffled receiver & 0.788 [0.736, 0.828] & \textbf{0.197} & 0.0497 & 0.193 & 0.034\\
 & Receiver-conditioned PIR & 0.787 [0.731, 0.826] & 0.191 & \textbf{0.0484} & \textbf{0.182} & \textbf{0.021}\\
\midrule
\multirow{9}{*}{GrailQA}
 & A0 global base rate & 0.500 & 0.079 & 0.0731 & 0.277 & 0.042\\
 & A0 per-receiver base rate & 0.500 & 0.079 & 0.0701 & 0.260 & 0.002\\
 & A1 message-only judge & 0.570 [0.540, 0.601] & 0.096 & 0.0730 & 0.277 & 0.046\\
 & A2 equivalence judge & 0.502 [0.501, 0.504] & 0.080 & 0.0731 & 0.278 & 0.045\\
 & A3 receiver-agnostic PIR & 0.726 [0.700, 0.750] & \textbf{0.198} & 0.0701 & 0.261 & 0.047\\
 & A4 action-failure predictor & 0.473 [0.444, 0.503] & 0.072 & 0.739 & 2.548 & 0.810\\
 & A5 listener / ToM & 0.705 [0.680, 0.729] & 0.145 & 0.0693 & 0.250 & 0.019\\
 & Shuffled receiver & 0.725 [0.701, 0.748] & 0.188 & 0.0706 & 0.262 & 0.048\\
 & Receiver-conditioned PIR & \textbf{0.729} [0.705, 0.753] & 0.193 & \textbf{0.0672} & \textbf{0.241} & \textbf{0.012}\\
\midrule
\multirow{9}{*}{SQ-WD}
 & A0 global base rate & 0.500 & 0.052 & 0.0496 & 0.205 & 0.039\\
 & A0 per-receiver base rate & 0.500 & 0.052 & 0.0478 & 0.188 & 0.004\\
 & A1 message-only judge & 0.552 [0.521, 0.581] & 0.063 & 0.0496 & 0.206 & 0.038\\
 & A2 equivalence judge & 0.500 & 0.052 & 0.0496 & 0.206 & 0.038\\
 & A3 receiver-agnostic PIR & 0.709 [0.686, 0.734] & 0.113 & 0.0483 & 0.192 & 0.040\\
 & A4 action-failure predictor & 0.467 [0.437, 0.494] & 0.048 & 0.442 & 1.292 & 0.560\\
 & A5 listener / ToM & 0.695 [0.672, 0.717] & 0.098 & 0.0476 & 0.180 & 0.013\\
 & Shuffled receiver & 0.716 [0.693, 0.739] & 0.114 & 0.0484 & 0.193 & 0.040\\
 & Receiver-conditioned PIR & \textbf{0.718} [0.693, 0.742] & \textbf{0.116} & \textbf{0.0456} & \textbf{0.172} & \textbf{0.008}\\
\midrule
\multirow{9}{*}{PopQA}
 & A0 global base rate & 0.500 & 0.033 & 0.0320 & 0.145 & 0.025\\
 & A0 per-receiver base rate & 0.500 & 0.033 & 0.0313 & 0.134 & 0.002\\
 & A1 message-only judge & 0.509 [0.465, 0.552] & 0.036 & 0.0319 & 0.145 & 0.025\\
 & A2 equivalence judge & 0.500 & 0.033 & 0.0320 & 0.145 & 0.025\\
 & A3 receiver-agnostic PIR & 0.775 [0.738, 0.802] & 0.116 & 0.0308 & 0.133 & 0.028\\
 & A4 action-failure predictor & 0.408 [0.382, 0.435] & 0.028 & 0.617 & 1.941 & 0.718\\
 & A5 listener / ToM & \textbf{0.787} [0.753, 0.815] & \textbf{0.138} & 0.0300 & \textbf{0.118} & \textbf{0.006}\\
 & Shuffled receiver & 0.780 [0.747, 0.805] & 0.121 & 0.0308 & 0.132 & 0.028\\
 & Receiver-conditioned PIR & 0.777 [0.743, 0.804] & 0.133 & \textbf{0.0296} & 0.119 & 0.008\\
\midrule
\multirow{9}{*}{Mintaka}
 & A0 global base rate & 0.500 & 0.026 & 0.0252 & 0.120 & 0.016\\
 & A0 per-receiver base rate & 0.500 & 0.026 & 0.0248 & 0.114 & 0.002\\
 & A1 message-only judge & 0.554 [0.487, 0.617] & 0.036 & 0.0252 & 0.121 & 0.016\\
 & A2 equivalence judge & 0.500 & 0.026 & 0.0252 & 0.120 & 0.016\\
 & A3 receiver-agnostic PIR & 0.681 [0.629, 0.729] & 0.067 & 0.0249 & 0.116 & 0.018\\
 & A4 action-failure predictor & 0.541 [0.482, 0.601] & 0.034 & 0.246 & 0.687 & 0.400\\
 & A5 listener / ToM & 0.687 [0.641, 0.733] & 0.056 & 0.0249 & 0.112 & 0.010\\
 & Shuffled receiver & 0.676 [0.625, 0.723] & 0.077 & 0.0249 & 0.116 & 0.017\\
 & Receiver-conditioned PIR & \textbf{0.688} [0.636, 0.737] & \textbf{0.083} & \textbf{0.0244} & \textbf{0.109} & \textbf{0.007}\\
\bottomrule
\end{tabular}
\end{table*}

\begin{table*}[!htbp]
\caption{Receiver-information ablation (test split, three-seed means). $\Delta$: full minus receiver-agnostic, with paired 95\% CIs in the row below. Last row per dataset: A4 scored on $\yI$, with its AUROC on $\yA$ in parentheses.}
\label{tab:receiver_ablation}
\centering
\scriptsize
\setlength{\tabcolsep}{3.5pt}
\renewcommand{\arraystretch}{0.94}
\begin{tabular}{lccccc}
\toprule
Variant & AUROC (pooled)$\uparrow$ & AUROC (macro)$\uparrow$ & Brier$\downarrow$ & NLL$\downarrow$ & ECE (macro)$\downarrow$\\
\midrule
\multicolumn{6}{l}{\textit{FreebaseQA}}\\
Full model (with $e_R$) & 0.810 & 0.698 & 0.0234 & 0.100 & 0.005\\
Receiver removed (A3) & 0.663 & 0.691 & 0.0243 & 0.113 & 0.024\\
Receiver shuffled & 0.661 & 0.691 & 0.0243 & 0.113 & 0.024\\
$\Delta$ full $-$ A3 & $+$0.147 & $+$0.008 & $-$0.0009 & $-$0.013 & \\
\quad 95\% CI & [0.124, 0.170] & & [$-$0.0012, $-$0.0007] & [$-$0.015, $-$0.011] & \\
A4 on $\yI$ (on $\yA$) & 0.582 (0.802) & 0.561 & & & \\
\midrule
\multicolumn{6}{l}{\textit{WebQSP}}\\
Full model (with $e_R$) & 0.806 & 0.784 & 0.0486 & 0.183 & 0.021\\
Receiver removed (A3) & 0.752 & 0.786 & 0.0500 & 0.194 & 0.034\\
Receiver shuffled & 0.748 & 0.783 & 0.0501 & 0.195 & 0.034\\
$\Delta$ full $-$ A3 & $+$0.053 & $-$0.001 & $-$0.0014 & $-$0.012 & \\
\quad 95\% CI & [0.028, 0.088] & & [$-$0.0020, $-$0.0008] & [$-$0.016, $-$0.008] & \\
A4 on $\yI$ (on $\yA$) & 0.599 (0.774) & 0.602 & & & \\
\midrule
\multicolumn{6}{l}{\textit{GrailQA}}\\
Full model (with $e_R$) & 0.775 & 0.725 & 0.0673 & 0.241 & 0.014\\
Receiver removed (A3) & 0.687 & 0.728 & 0.0700 & 0.260 & 0.048\\
Receiver shuffled & 0.681 & 0.721 & 0.0706 & 0.262 & 0.048\\
$\Delta$ full $-$ A3 & $+$0.088 & $-$0.003 & $-$0.0027 & $-$0.019 & \\
\quad 95\% CI & [0.073, 0.104] & & [$-$0.0033, $-$0.0021] & [$-$0.021, $-$0.016] & \\
A4 on $\yI$ (on $\yA$) & 0.507 (0.779) & 0.469 & & & \\
\midrule
\multicolumn{6}{l}{\textit{SQ-WD}}\\
Full model (with $e_R$) & 0.811 & 0.717 & 0.0456 & 0.172 & 0.008\\
Receiver removed (A3) & 0.704 & 0.710 & 0.0483 & 0.192 & 0.040\\
Receiver shuffled & 0.703 & 0.711 & 0.0484 & 0.193 & 0.039\\
$\Delta$ full $-$ A3 & $+$0.107 & $+$0.006 & $-$0.0027 & $-$0.020 & \\
\quad 95\% CI & [0.096, 0.118] & & [$-$0.0031, $-$0.0023] & [$-$0.022, $-$0.018] & \\
A4 on $\yI$ (on $\yA$) & 0.469 (0.844) & 0.463 & & & \\
\midrule
\multicolumn{6}{l}{\textit{PopQA}}\\
Full model (with $e_R$) & 0.836 & 0.781 & 0.0296 & 0.119 & 0.007\\
Receiver removed (A3) & 0.734 & 0.770 & 0.0309 & 0.133 & 0.027\\
Receiver shuffled & 0.736 & 0.773 & 0.0309 & 0.133 & 0.027\\
$\Delta$ full $-$ A3 & $+$0.102 & $+$0.011 & $-$0.0012 & $-$0.014 & \\
\quad 95\% CI & [0.086, 0.117] & & [$-$0.0016, $-$0.0009] & [$-$0.016, $-$0.012] & \\
A4 on $\yI$ (on $\yA$) & 0.445 (0.860) & 0.388 & & & \\
\midrule
\multicolumn{6}{l}{\textit{Mintaka}}\\
Full model (with $e_R$) & 0.760 & 0.697 & 0.0244 & 0.108 & 0.006\\
Receiver removed (A3) & 0.657 & 0.682 & 0.0249 & 0.115 & 0.017\\
Receiver shuffled & 0.660 & 0.686 & 0.0248 & 0.115 & 0.017\\
$\Delta$ full $-$ A3 & $+$0.103 & $+$0.015 & $-$0.0005 & $-$0.007 & \\
\quad 95\% CI & [0.078, 0.128] & & [$-$0.0007, $-$0.0002] & [$-$0.009, $-$0.005] & \\
A4 on $\yI$ (on $\yA$) & 0.549 (0.801) & 0.538 & & & \\
\bottomrule
\end{tabular}
\end{table*}

\FloatBarrier
\subsection{Experiment~3: Receiver-Type Inference}
\label{app:res_e3}

\begin{figure}[!htb]
\centering
\includegraphics[width=\textwidth]{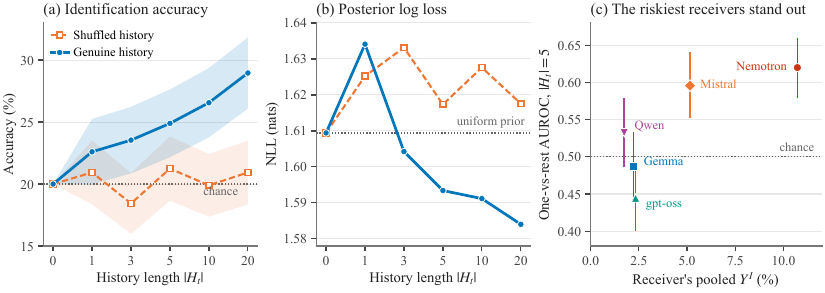}
\caption{Behavioural history sharpens the receiver belief (960 audit histories per length). (a)~Accuracy and (b)~calibrated NLL by history length for genuine and shuffled history, with binomial 95\% CIs in (a). (c)~One-vs-rest AUROC at $|H_t|=5$ against each receiver's pooled $\yI$, with 95\% bootstrap CIs.}
\label{fig:history}
\end{figure}

Behavioural evidence about the receiver accumulates with history (\cref{fig:history}a,b). Accuracy rises from 20.0\% under the uniform prior to 29.0\% after twenty responses, and from three responses onward, the calibrated NLL stays below the prior's 1.609, reaching 1.584. The posterior sharpens, but identification stays weak in absolute terms. The shuffled-history control, which pairs each query with the responses of a random receiver, stays near chance (18.4--21.3\%) and above the prior's NLL at every length, so the evidence comes from the receiver's own behaviour. After one response the calibrated NLL (1.634) is still above the prior's. \Cref{tab:e3_history} adds Brier scores (summed over classes), ECE (10 equal-width bins), and the NLL before calibration.

The posterior separates the receivers whose risk differs most (\cref{fig:history}c). At $|H_t|=5$, the one-vs-rest AUROC is 0.62 for Nemotron and 0.60 for Mistral, with 95\% CIs above chance, and these are the two riskiest receivers. The three low-risk receivers are not told apart (AUROC 0.45--0.53), but their pooled $\yI$ lie within 0.6 points of each other, so mistaking one for another barely changes the predicted level of risk. This is the pattern the decision rule needs: by \cref{thm:conditioning}, receiver information has value only where it changes the preferred message, which requires receivers whose risk differs. The confusion matrix in \cref{tab:confusion} shows the same pattern: predictions concentrate on Nemotron and Mistral, and Gemma and gpt-oss are almost never predicted. The same table gives the results per dataset.

\begin{table*}[!htb]
\caption{Receiver-type inference by history length $|H_t|$ (960 audit histories per length). Length 0 is the uniform prior over five receivers, and ``calibrated'' uses temperature scaling fitted on the calibration split.}
\label{tab:e3_history}
\centering
\small
\begin{tabular}{lcccccccc}
\toprule
& \multicolumn{4}{c}{Calibrated, genuine history} & Raw & \multicolumn{2}{c}{Shuffled history}\\
\cmidrule(lr){2-5}\cmidrule(lr){6-6}\cmidrule(lr){7-8}
$|H_t|$ & Accuracy$\uparrow$ & NLL$\downarrow$ & Brier$\downarrow$ & ECE$\downarrow$ & NLL$\downarrow$ & Accuracy & NLL\\
\midrule
0 & 20.0\% & 1.609 & 0.800 & 0.000 & -- & -- & --\\
1 & 22.6\% & 1.634 & 0.811 & 0.056 & 1.626 & 20.9\% & 1.625\\
3 & 23.5\% & 1.604 & 0.798 & 0.013 & 1.603 & 18.4\% & 1.633\\
5 & 24.9\% & 1.593 & 0.794 & 0.014 & 1.594 & 21.3\% & 1.617\\
10 & 26.6\% & 1.591 & 0.793 & 0.031 & 1.593 & 19.9\% & 1.628\\
20 & 29.0\% & 1.584 & 0.790 & 0.063 & 1.587 & 20.9\% & 1.618\\
\bottomrule
\end{tabular}
\end{table*}

\begin{table*}[!htb]
\caption{Receiver-type inference at $|H_t|=5$: by dataset of the calibration tasks (top, 960 histories each) and row-normalised confusion matrix (bottom, \%, true receiver in rows).}
\label{tab:confusion}
\centering
\footnotesize
\begin{tabular}{lcccc}
\toprule
Dataset & Accuracy$\uparrow$ & NLL$\downarrow$ & Brier$\downarrow$ & ECE$\downarrow$\\
\midrule
FreebaseQA & 21.5\% & 1.604 & 0.798 & 0.025\\
WebQSP & 24.6\% & 1.598 & 0.796 & 0.028\\
GrailQA & 20.5\% & 1.608 & 0.799 & 0.040\\
SQ-WD & 18.3\% & 1.610 & 0.801 & 0.051\\
PopQA & 25.9\% & 1.583 & 0.790 & 0.008\\
Mintaka & 25.9\% & 1.588 & 0.792 & 0.017\\
\bottomrule
\end{tabular}
\par\vspace{6pt}
\begin{tabular}{lccccc}
\toprule
True $\backslash$ predicted & Nemotron & Gemma & gpt-oss & Mistral & Qwen\\
\midrule
Nemotron & \textbf{71.4} & 0.0 & 0.5 & 18.8 & 9.4\\
Gemma & 61.5 & \textbf{0.0} & 0.0 & 31.8 & 6.8\\
gpt-oss & 50.5 & 0.0 & \textbf{0.5} & 42.7 & 6.3\\
Mistral & 47.4 & 0.5 & 0.0 & \textbf{44.3} & 7.8\\
Qwen & 53.6 & 0.0 & 0.0 & 38.0 & \textbf{8.3}\\
\bottomrule
\end{tabular}
\label{tab:e3_by_dataset}
\end{table*}

\paragraph{Posterior versus MAP receiver.}
Even after twenty responses, the posterior spreads its mass over several receivers (\cref{tab:e3_history}). Collapsing it to its MAP receiver would discard this remaining uncertainty. The controller therefore averages over $\pi_t$ throughout, and the rewrite experiment compares this posterior with a uniform prior and with the true identity.

\FloatBarrier
\subsection{Experiment~4: Message Revision}
\label{app:res_e4}

\begin{figure}[!htb]
\centering
\includegraphics[width=\textwidth]{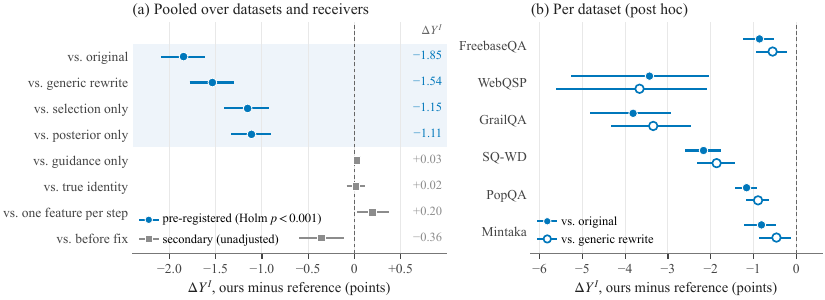}
\caption{Risk-guided revision lowers measured interpretation failure on the 1,579 held-out questions. Paired differences in $\yI$, ours minus each reference (points, 95\% CIs, negative favours ours). (b)~Per dataset against the original (filled) and the generic rewrite (hollow), post hoc.}
\label{fig:rewrite}
\vspace{6pt}
\end{figure}

\Cref{fig:rewrite} shows the paired differences behind \cref{tab:rewrite}, pooled over datasets and receivers (a) and per dataset (b). \Cref{tab:e4_arms} lists every condition and \cref{tab:e4_receivers} gives the results per receiver. Intervals use 10,000 bootstrap resamples clustered by question. The four pre-registered contrasts are Holm-adjusted, and all other contrasts are unadjusted.

Risk-guided revision beats every reference on every dataset. All four pre-registered contrasts lie well below zero (\cref{fig:rewrite}a), and in each of the six datasets ours is below both the original, by 0.81 to 3.81 points, and the generic rewrite, by 0.46 to 3.66 points, with every interval below zero (\cref{fig:rewrite}b, post hoc). The secondary contrasts place ours with the other guided conditions: guidance only and the true identity are within 0.03 points of it, one feature per step is 0.20 points lower, and the method before the fix is 0.36 points higher.

\begin{figure}[!htb]
\centering
\includegraphics[width=\textwidth]{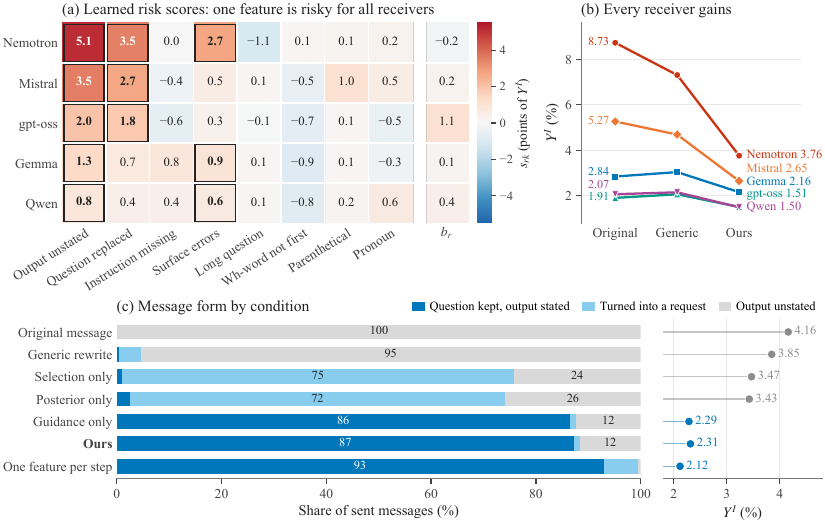}
\caption{One risk feature is shared by all receivers, and guidance removes it. (a)~Learned scores $s_{rk}$ and intercepts $b_r$ (points of $\yI$). Boxed: the score and its 2.5\% bootstrap quantile are both positive, the guidance rule for a known receiver. (b)~$\yI$ per receiver. (c)~Form of all sent messages (unchanged originals included) and $\yI$ per condition, with guided conditions in blue.}
\label{fig:e4_mechanism}
\vspace{6pt}
\end{figure}

\paragraph{Why guidance works.}
\Cref{fig:e4_mechanism} traces the gain to one risk feature that all receivers share. The learned scores differ in size across receivers, from 5.1 points for Nemotron to 0.8 for Qwen, but an unstated output is the top feature for every receiver and passes the guidance rule for all five (panel a). The same instruction, to state the output, therefore lowers risk for every receiver, and receiver information barely changes the preferred message: a uniform belief gives the same $\yI$ (2.31\%, \cref{tab:e4_arms}). This is the case $\Delta_E\approx0$ of \cref{thm:conditioning}, in which receivers differ in how much risk they carry but hardly in which message is best. The framework thus shows that receiver-specific adaptation is not needed here, rather than assuming that it always is. It would be needed where receivers disagree about the best message. The intercepts $b_r$, the effect of rewriting itself, are small or positive (up to 1.1 points for gpt-oss), so a rewrite helps only when it removes a risky feature. The gain follows each receiver's original risk (panel b): 4.96 points for Nemotron (from 8.73\%), 2.62 for Mistral (from 5.27\%), and 0.40 to 0.69 for the three low-risk receivers. The generic rewrite, in contrast, raises $\yI$ for Gemma, gpt-oss, and Qwen by 0.09 to 0.20 points (\cref{tab:e4_receivers}).

\paragraph{Message form.}
\Cref{fig:e4_mechanism}c shows what each condition does to the message. Among changed messages, 95\% of generic rewrites leave the output unstated and 84\% drop the instruction line. Selection only and posterior only mostly turn the question into a request such as ``Name the \ldots'' (83--86\% of changed messages), which states the output but replaces the question. The guided conditions keep the question and add a line that states the output (86--93\% of all sent messages and 97\% of our changed messages), and they have the lowest $\yI$ in the panel (2.12--2.31\%). Guidance thus improves the candidate set itself: selection among generic candidates gains only 0.38 points over the generic rewrite, whereas guided candidates sent without selection gain 1.56.

\begin{table}[!htbp]
\caption{Experiment~4 results by condition. $\yI$, $\yA$, \textsc{None}, and Distr.\ (probes that choose the contrast task) are percentages on the 1,579 held-out questions, and Changed is the share of messages rewritten. Val.\ $\yI$ is on the 300 validation questions of the check before the main run.}
\label{tab:e4_arms}
\centering
\scriptsize
\setlength{\tabcolsep}{4pt}
\begin{tabular}{lcccccc}
\toprule
Condition & $\yI$ & $\yA$ & \textsc{None} & Distr. & Changed & Val.\ $\yI$\\
\midrule
Original message & 4.16 & 59.33 & 0.93 & 3.20 & 0.0 & 5.01\\
Generic rewrite & 3.85 & 58.15 & 0.88 & 2.95 & 97.2 & 4.97\\
Selection only & 3.47 & 58.20 & 0.86 & 2.59 & 86.7 & 4.60\\
Posterior only & 3.43 & 58.59 & 0.81 & 2.60 & 86.1 & 4.23\\
\midrule
\textbf{Ours} & \textbf{2.31} & 58.33 & 0.72 & 1.56 & 90.0 & 2.83\\
Guidance only & 2.29 & 58.33 & 0.70 & 1.54 & 91.5 & 2.66\\
Uniform belief & 2.31 & 58.33 & 0.72 & 1.56 & 89.9 & 2.83\\
True identity & 2.30 & 58.58 & 0.72 & 1.54 & 88.0 & 2.81\\
One feature per step & 2.12 & 58.75 & 0.70 & 1.37 & 99.4 & 2.44\\
Before fix & 2.67 & 59.34 & 0.74 & 1.88 & 87.0 & 2.69\\
\bottomrule
\end{tabular}
\end{table}

\begin{table}[!htbp]
\caption{Experiment~4 results per receiver on the held-out questions. Ours minus each reference and the generic rewrite minus the original are paired differences in $\yI$ (points), and Value is ours minus the original on the 275 value-versus-attribute questions.}
\label{tab:e4_receivers}
\centering
\scriptsize
\setlength{\tabcolsep}{3.5pt}
\begin{tabular}{lcccccccc}
\toprule
& \multicolumn{2}{c}{$\yI$ (\%)} & \multicolumn{4}{c}{Ours minus} & Generic minus & \\
\cmidrule(lr){2-3}\cmidrule(lr){4-7}
Receiver & Original & Ours & original [95\% CI] & generic & sel.\ only & post.\ only & original & Value\\
\midrule
Nemotron & 8.73 & 3.76 & $-4.96$ [$-5.34$, $-4.59$] & $-3.54$ & $-2.70$ & $-2.66$ & $-1.42$ & $-5.60$\\
Gemma & 2.84 & 2.16 & $-0.69$ [$-1.04$, $-0.35$] & $-0.88$ & $-0.55$ & $-0.42$ & $+0.20$ & $+0.07$\\
gpt-oss & 1.91 & 1.51 & $-0.40$ [$-0.62$, $-0.17$] & $-0.56$ & $-0.63$ & $-0.52$ & $+0.16$ & $+0.01$\\
Mistral & 5.27 & 2.65 & $-2.62$ [$-3.00$, $-2.25$] & $-2.04$ & $-1.50$ & $-1.60$ & $-0.58$ & $-4.09$\\
Qwen & 2.07 & 1.50 & $-0.57$ [$-0.85$, $-0.31$] & $-0.66$ & $-0.40$ & $-0.36$ & $+0.09$ & $-0.38$\\
\bottomrule
\end{tabular}
\end{table}

\paragraph{Further analysis.}
Four results add detail. First, the gain is smallest on value-versus-attribute questions for Gemma and gpt-oss. There, ours matches the original ($+0.07$ [$-0.60$, $+0.77$] and $+0.01$ [$-0.62$, $+0.72$], \cref{tab:e4_receivers}), while the generic rewrite, selection only, and posterior only raise $\yI$ by 1.0 to 1.4 points (post hoc). Only one feature per step falls below the original there ($-0.77$ for Gemma and $-0.57$ for gpt-oss, the latter not significant). Second, the fix matters. Before the fix, the instruction for an unstated output pointed to the intended task, whose text for value questions uses words such as ``kind'' and ``attribute''. The rewriter copied such words into 7.0\% of sent messages (40\% on value questions), and gpt-oss read them literally and chose the contrast task. The method before the fix has 0.36 points higher $\yI$ [$+0.11$, $+0.59$] and 1.01 points higher $\yA$ [$+0.43$, $+1.61$] than ours, and on gpt-oss value questions it raises $\yI$ by 2.65 points over the original [$+1.18$, $+4.19$]. The fix was designed on earlier development questions, not on the held-out questions. Third, the receiver belief does not change the result. Under every posterior in the cohort, guidance lists the same three features as under a uniform belief (output unstated, question replaced, surface errors), and selection differs in only 2 of 7,895 decisions. With the true identity, four receivers get a different, shorter list (\cref{fig:e4_mechanism}a), yet $\yI$ does not change ($-0.02$ [$-0.12$, $+0.08$]). Fourth, the audit supports the measured gain. It judged 12,402 distinct sent messages: 98.1\% keep the task and 1.1\% add a hint. Flagged messages make up 2.7--3.6\% of the unguided conditions and 0.5--0.7\% of the guided ones, and sending the original in their place leaves the four pre-registered contrasts at $-1.81$, $-1.50$, $-1.14$, and $-1.07$.

\paragraph{Validation check and cost.}
On the 300 validation questions, ours lowered $\yI$ from 5.01\% to 2.83\% ($-2.18$ [$-2.95$, $-1.51$]), but its $\yA$ was not shown to be non-inferior to the original there ($-0.47$ [$-2.07$, $+1.07$]) because the interval is wide. The held-out run measured 51,953 distinct message--receiver pairs with 1,870,308 probe calls and 51,953 execution calls. Both stages together used 10.99 A100 hours for the receivers and \$11.41 of API calls for generation, verification, and the audit. One feature per step is the most expensive condition: it needs four dependent generate-and-verify rounds instead of one, 2,435 API tokens per question against about 1,220 for ours (\$0.53 against about \$0.32 for the 1,579 questions), and about four times the latency by its call structure, which we did not time. Its lower $\yI$ (2.12\% vs.\ 2.31\%, unadjusted $+0.20$ [$+0.03$, $+0.37$]) comes from a secondary contrast, so we keep the single-call method as the default.

\FloatBarrier
\subsection{Experiment~5: Query Control}
\label{app:res_e5}

\begin{figure}[!htb]
\centering
\includegraphics[width=\textwidth]{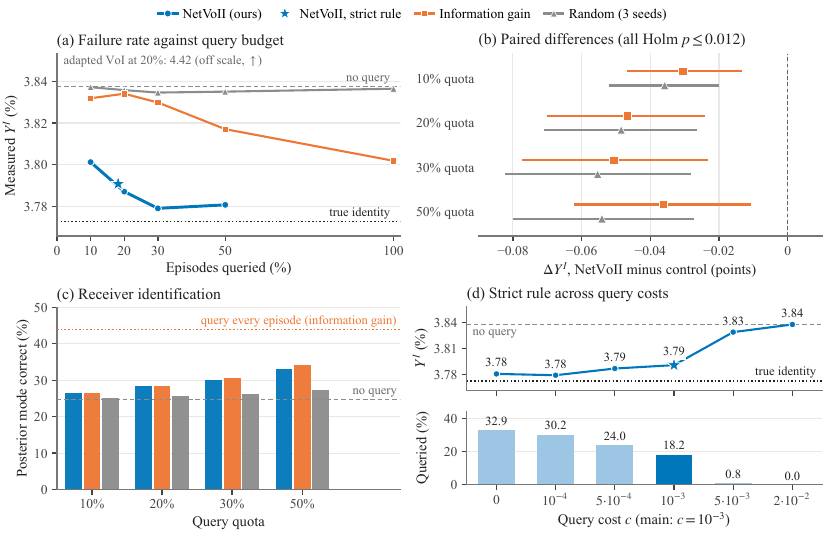}
\caption{NetVoII allocates queries by decision value (primary cohort, interpretation objective, $c=\QueryCost$). (a)~$\yI$ against the share of episodes queried. (b)~Paired differences at matched quotas with 95\% CIs, equal to the net-utility contrasts with the sign reversed. (c)~Share of episodes whose posterior mode is the true receiver. (d)~Strict rule at six query costs $c$: measured $\yI$ (top) and share of episodes queried (bottom).}
\label{fig:e5_query}
\end{figure}

\begin{figure}[!htbp]
\centering
\includegraphics[width=\textwidth]{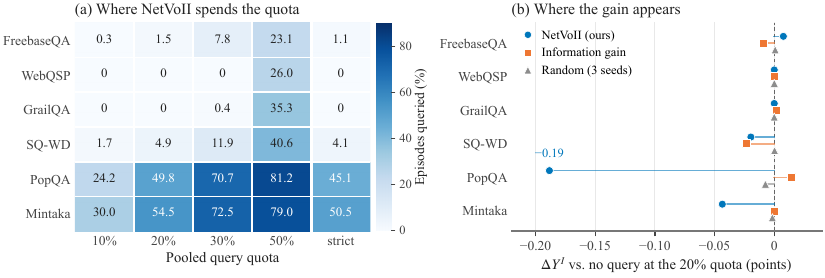}
\caption{NetVoII spends a pooled quota where predicted value is high (primary cohort, interpretation objective). (a)~Share of each dataset's episodes queried at pooled quotas of 10--50\% and under the strict rule. (b)~Change in $\yI$ against no query at the 20\% quota.}
\label{fig:e5_datasets}
\end{figure}

\Cref{fig:e5_query} shows the query-control results of \cref{tab:e5_policies} in detail. NetVoII reaches 3.78--3.80\% at every quota, below both controls. IG reaches 3.80\% only when it queries every episode, and random querying stays within 0.01 points of no query (panel a). All eight paired differences exclude zero (panel b). At a matched quota the query costs cancel, so each net-utility contrast equals the $\yI$ contrast with its sign reversed, and all eight survive Holm correction ($p\le0.012$). NetVoII does not identify the receiver better than IG (panel c), so its gain does not come from identification. Intervals use \QueryBootstrapResamples\ bootstrap resamples clustered by question, and random querying is always the mean of three seeds. \Cref{tab:e5_budget_grid} gives the budget grids for both cohorts, at exact quotas of 10, 20, 30, and 50\% of episodes and for querying every episode with the IG or random choice of query (Always). \Cref{tab:e5_cost_sweep} gives the cost sweep, and \cref{tab:e5_per_dataset,fig:e5_datasets} give the per-dataset and per-receiver levels.

\begin{figure}[!htb]
\centering
\includegraphics[width=\textwidth]{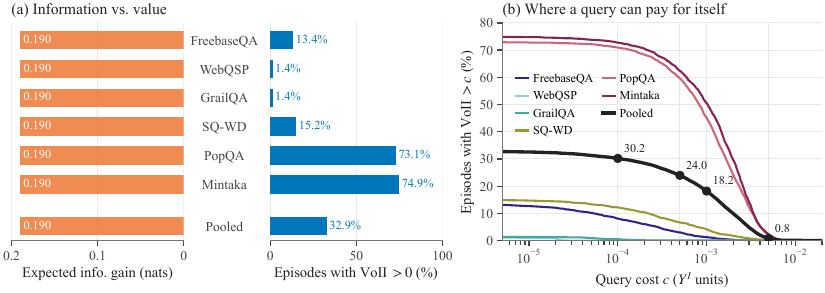}
\caption{Information is not value (primary cohort). (a)~Expected IG of the most informative query (nats) and share of episodes with positive predicted VoII, per dataset. (b)~Share of episodes whose best query is worth more than a cost $c$, with pooled values marked at four of the tested costs.}
\label{fig:e5_value}
\end{figure}

\paragraph{Information is not value.}
\Cref{fig:e5_value} shows why the two criteria diverge. The most informative query carries almost the same expected information in every episode, 0.19 nats (range 0.186--0.193), yet the share of episodes with positive predicted VoII ranges from 1.4\% on WebQSP and GrailQA to 73--75\% on PopQA and Mintaka (panel a). \Cref{thm:frontier} explains the gap: it bounds VoII by $G_t$, the largest loss reduction that the query's information can buy, and $G_t$ is zero when one message is best for every receiver type, however informative the query. IG cannot see this difference, since every episode offers about the same information. NetVoII spends its queries where the answer can change the message.

\paragraph{Sparsity of predicted value.}
Predicted value is sparse and small. Only \VoIIPositiveShare\ of episodes have positive predicted VoII, the median positive value is 0.0012, close to the main cost $c=\QueryCost$, and 0.8\% of episodes exceed 0.005 (\cref{fig:e5_value}b). Because NetVoII ranks episodes by predicted value, returns fall as the quota grows: of the episodes added from 30\% to 50\%, 86\% have zero predicted value, and the added queries change $\yI$ by $+0.002$ points ($[-0.005,+0.008]$). The strict rule follows the same curve. At each of six frozen costs $c$ it queries exactly the episodes whose predicted VoII exceeds $c$ (\cref{fig:e5_query}d, \cref{tab:e5_cost_sweep}): as $c$ rises from 0 to 0.02, the query rate falls from 32.9\% to zero and $\yI$ moves from 3.78\% back to \NeverQueryPIRRate, so the controller adjusts smoothly to the cost rather than switching off. A higher cost drops the queries with the lowest predicted value first, and these barely matter: raising $c$ from 0 to $10^{-4}$ lowers the query rate by 2.7 points and leaves $\yI$ at 3.78\%.

\paragraph{Scope of the gain.}
Three results set the scope of the gain. First, its size is set by how much decision-relevant receiver information is left after five stored responses. The posterior is still broad (its mode is correct in 24.6\% of episodes), but little of this uncertainty can change the message under the current candidate set: in half of the questions (50.4\%) one candidate is best for every receiver, so no answer can change the choice. On a base of \NeverQueryPIRRate, the true identity would lower $\yI$ by only 0.06 points, and NetVoII lowers it by about 0.05. Second, the gain is in interpretation, the target VoII optimises. On the primary cohort, task success is 35.7\% for every policy but the adapted VoI rule (35.3\%), and all paired differences between NetVoII and the controls lie within 0.1 points of zero. On the replication cohort, NetVoII's task success is 0.2 to 0.3 points below both controls at matched budgets (unadjusted intervals exclude zero), and the strict rule lowers it by 0.23 points relative to no query ($[-0.40,-0.05]$). Under a combined objective that also scores execution, $\yI+(1-\yI)\yA$, none of the eight contrasts survives correction (Holm $p=1$). Third, the replication is directional but not significant: NetVoII is at or below both controls at every budget, but every interval includes zero, and the strict rule's net utility there is slightly below no query.

\paragraph{Adapted VoI comparator.}
The adapted VoI rule shares the posterior, query bank, candidates, verifier decisions, and argmin with NetVoII. Only the risk estimates differ, coming from \RewriteModel\ instead of the learned heads (122 of 9,029 estimate items fell back to training base rates). These estimates are coarse: five values (0, 0.01, 0.02, 0.05, and 0.10) make up 92\% of them. As a result, one candidate is best, or tied for best, for every receiver in every question, against 50\% (primary) and 55\% (replication) of questions for the heads, so its VoII is zero everywhere and the strict rule never queries. The minimum expected loss is tied in 94.5\% of episodes (92.5\% in the replication), and since ties go to the earlier candidate under every policy, the rule sends the original in 81\% of episodes, against 18\% (primary) and 21\% (replication) for the heads without a query. Tied episodes account for 0.56 of its 0.58-point gap to the heads (0.30 of 0.34 in the replication). In the 5.5\% of episodes without a tie, its messages give 4.91\% against 4.44\% for the heads. A random tie-break would narrow the gap: on the tied episodes whose tied candidates were all measured (5.9\% of tied episodes), a random tied candidate gives 3.86\%, against 3.99\% for the original and 3.62\% for the heads' choice (replication, 7.4\%: 3.39\%, 3.65\%, and 3.55\%), but these subsets are small and selected by what was sent. The comparison therefore tests the LLM utility estimates inside the same VoI machinery, not the VoI rule itself, and we treat it as a secondary comparator.

\paragraph{Per-dataset allocation.}
Because the quota is pooled, NetVoII spends it where predicted value is high (\cref{fig:e5_datasets}a). At 20\% it queries about half of the PopQA and Mintaka episodes and none of WebQSP or GrailQA (\cref{tab:e5_per_dataset}). At 50\% it queries about 80\% of PopQA and Mintaka, and only at this quota does it query more than 0.4\% of WebQSP or GrailQA. The gain appears where the queries go (\cref{fig:e5_datasets}b): at 20\%, NetVoII lowers $\yI$ by 0.19 points on PopQA and 0.04 on Mintaka and stays within 0.02 points of no query elsewhere, while IG and random querying move $\yI$ by at most 0.03 points on any dataset. The true identity in \cref{tab:e5_per_dataset} is a diagnostic, not a lower bound: it can sit above no query where the learned risk heads are imperfectly calibrated for a receiver (e.g., WebQSP, 5.14\% vs.\ 5.07\%) and above NetVoII on PopQA (2.52\% vs.\ 2.42\%) and for Gemma and Mistral.

\begin{table*}[!htbp]
\caption{Experiment~5 budget grid (interpretation objective): measured $\yI$ (\%) by exact query quota, with net utility at $c=\QueryCost$ in parentheses. Bold: net-utility contrasts against both IG and random querying survive Holm correction. $^\dagger$\,Secondary comparator: its estimates never let a query change the message.}
\label{tab:e5_budget_grid}
\centering
\scriptsize
\setlength{\tabcolsep}{4pt}
\begin{tabular}{lccccc}
\toprule
Policy & 10\% & 20\% & 30\% & 50\% & Always\\
\midrule
\multicolumn{6}{l}{\textit{Primary cohort (test split). No query: 3.84 (0.9616), true identity: 3.77 (0.9623), adapted VoI$^\dagger$ without queries: 4.42 (0.9558).}}\\
Random (3-seed mean) & 3.84 (0.9615) & 3.84 (0.9614) & 3.83 (0.9614) & 3.83 (0.9612) & 3.84 (0.9606)\\
IG & 3.83 (0.9616) & 3.83 (0.9615) & 3.83 (0.9614) & 3.82 (0.9613) & 3.80 (0.9610)\\
Adapted VoI$^\dagger$ & 4.42 (0.9557) & 4.42 (0.9556) & 4.42 (0.9555) & 4.42 (0.9553) & --\\
\textbf{NetVoII (ours)} & \textbf{3.80 (0.9619)} & \textbf{3.79 (0.9619)} & \textbf{3.78 (0.9619)} & \textbf{3.78 (0.9617)} & --\\
\midrule
\multicolumn{6}{l}{\textit{Replication cohort. No query: 3.93 (0.9607), true identity: 3.91 (0.9609), adapted VoI$^\dagger$ without queries: 4.27 (0.9573).}}\\
Random (3-seed mean) & 3.93 (0.9606) & 3.93 (0.9605) & 3.93 (0.9604) & 3.93 (0.9602) & 3.92 (0.9598)\\
IG & 3.94 (0.9605) & 3.94 (0.9604) & 3.94 (0.9603) & 3.96 (0.9599) & 3.96 (0.9594)\\
Adapted VoI$^\dagger$ & 4.27 (0.9572) & 4.27 (0.9571) & 4.27 (0.9570) & 4.27 (0.9568) & --\\
NetVoII (ours) & 3.92 (0.9607) & 3.91 (0.9607) & 3.92 (0.9605) & 3.92 (0.9603) & --\\
\bottomrule
\end{tabular}
\end{table*}

\begin{table}[!htbp]
\caption{Strict rule of Experiment~5 across query costs $c$ (interpretation objective, net utility at each row's cost). The last two rows of each block re-run IG and random querying at the query count of the $c=\QueryCost$ rule.}
\label{tab:e5_cost_sweep}
\centering
\scriptsize
\setlength{\tabcolsep}{5pt}
\begin{tabular}{lcccc}
\toprule
Cost $c$ & Queried (\%) & $\yI$ (\%) & $1-\yA$ (\%) & Net utility\\
\midrule
\multicolumn{5}{l}{\textit{Primary cohort (test split)}}\\
0 & 32.9 & 3.78 & 35.66 & 0.9622\\
0.0001 & 30.2 & 3.78 & 35.68 & 0.9622\\
0.0005 & 24.0 & 3.79 & 35.62 & 0.9620\\
0.001 (main) & 18.2 & 3.79 & 35.68 & 0.9619\\
0.005 & 0.8 & 3.83 & 35.72 & 0.9617\\
0.02 & 0.0 & 3.84 & 35.73 & 0.9616\\
IG at the $c=0.001$ count & 18.2 & 3.83 & 35.72 & 0.9615\\
Random at the $c=0.001$ count & 18.2 & 3.84 & 35.72 & 0.9615\\
\midrule
\multicolumn{5}{l}{\textit{Replication cohort}}\\
0 & 31.2 & 3.92 & 36.67 & 0.9608\\
0.0001 & 28.8 & 3.92 & 36.65 & 0.9608\\
0.0005 & 23.4 & 3.92 & 36.66 & 0.9607\\
0.001 (main) & 17.8 & 3.92 & 36.68 & 0.9606\\
0.005 & 0.8 & 3.94 & 36.88 & 0.9606\\
0.02 & 0.0 & 3.93 & 36.90 & 0.9607\\
IG at the $c=0.001$ count & 17.8 & 3.94 & 36.90 & 0.9604\\
Random at the $c=0.001$ count & 17.8 & 3.93 & 36.91 & 0.9605\\
\bottomrule
\end{tabular}
\end{table}

\begin{table}[!htbp]
\caption{Experiment~5 results per dataset and per receiver (primary cohort, interpretation objective): $\yI$ (\%) at the pooled 20\% quota and under the strict rule, with the share of episodes that NetVoII queries.}
\label{tab:e5_per_dataset}
\centering
\scriptsize
\setlength{\tabcolsep}{4pt}
\begin{tabular}{lcccccccc}
\toprule
& & \multicolumn{3}{c}{$\yI$ at the 20\% quota} & \multicolumn{2}{c}{Queried by NetVoII (\%)} & & \\
\cmidrule(lr){3-5}\cmidrule(lr){6-7}
Unit & No query & NetVoII & Inf.\ gain & Random & 20\% quota & Strict & $\yI$ strict & True identity\\
\midrule
FreebaseQA & 2.16 & 2.16 & 2.15 & 2.16 & 1.5 & 1.1 & 2.16 & 2.10\\
WebQSP & 5.07 & 5.07 & 5.07 & 5.07 & 0.0 & 0.0 & 5.07 & 5.14\\
GrailQA & 6.96 & 6.96 & 6.96 & 6.96 & 0.0 & 0.0 & 6.96 & 6.96\\
SQ-WD & 4.53 & 4.51 & 4.50 & 4.53 & 4.9 & 4.1 & 4.51 & 4.45\\
PopQA & 2.61 & 2.42 & 2.62 & 2.60 & 49.8 & 45.1 & 2.42 & 2.52\\
Mintaka & 2.49 & 2.45 & 2.49 & 2.49 & 54.5 & 50.5 & 2.46 & 2.34\\
\midrule
Nemotron & 8.53 & 8.43 & 8.54 & 8.53 & 19.9 & 17.9 & 8.43 & 8.31\\
Gemma & 2.39 & 2.34 & 2.37 & 2.39 & 19.9 & 18.2 & 2.34 & 2.37\\
gpt-oss & 2.40 & 2.40 & 2.40 & 2.40 & 19.9 & 18.3 & 2.40 & 2.39\\
Mistral & 4.16 & 4.06 & 4.16 & 4.15 & 20.1 & 18.4 & 4.07 & 4.12\\
Qwen & 1.72 & 1.72 & 1.70 & 1.71 & 20.1 & 18.2 & 1.71 & 1.68\\
\bottomrule
\end{tabular}
\end{table}

\FloatBarrier

\section{Pseudocode and Full Experimental Specification}
\label{app:pseudocode}

\Cref{alg:revision,alg:deployment} are the two decisions made at deployment (\cref{sec:method}), and \cref{alg:training} is the offline stage that fits their learned parts. Where a line implements an equation of the main text, its comment cites it. \Cref{app:alg_instances} states how Experiments~4 and~5 instantiate these steps, and \cref{app:fullspec} lists every setting.

\subsection{End-to-End Deployment Pseudocode}
\label{app:alg_deploy}

At deployment the receiver identity is hidden, and the sender knows the receiver only through the behavioural history $H_t$. \Cref{alg:deployment} calls \cref{alg:revision} once. The action set $\Aset_t$ it returns stays fixed through the query decision, so a query can change which candidate is sent but never adds a new one. \textsc{Rewrite} is one request to the rewrite model for $K_m$ candidates that keep the exact task while reducing the features in $G_t$. \textsc{Verify} is a separate stateless call to the same model version. It receives only its arguments and never sees $\pi_t$, risk scores, feature values, or which candidate is preferred. In the experiments, the probes and the execution call run only after the message is sent, and their labels are used only for evaluation.

\begin{algorithm}[!htb]
\caption{Posterior-guided message revision}
\label{alg:revision}
\begin{algorithmic}[1]
\Require intended task $z^\star$, context $x_S$, initial message $m_0$, receiver posterior $\pi_t$, heads $\qI,\qC$ and risk effects $s_{rk}$ from \cref{alg:training}, loss terms $c_M,L_I,L_C$, candidate budget $K_m$, and guidance size $n_g$
\Ensure fixed valid action set $\Aset_t$ and revised message $\widehat m_t^\star$
\Function{Revise}{$z^\star,x_S,m_0,\pi_t$}
  \State $\bar s_{tk}\gets\sum_r\pi_t(r)\,s_{rk}$ for every mutable feature $k$ \Comment{posterior-weighted risk effect}
  \State $G_t\gets$ the at most $n_g$ mutable features with the largest positive $\bar s_{tk}$ \label{alg:rev:guide}
  \State $(m'_1,\ldots,m'_{K_m})\gets\Call{Rewrite}{m_0,z^\star,x_S,G_t}$ \Comment{fixed budget, no regeneration}
  \State $(v_1,\ldots,v_{K_m})\gets\Call{Verify}{z^\star,x_S,m_0,m'_1,\ldots,m'_{K_m}}$ \Comment{one stateless batched call}
  \State $\Aset_t\gets\{m_0\}\cup\{m'_j:v_j=\textsc{Pass}\}$ \Comment{the original is always kept}
  \For{each $m\in\Aset_t$ and each $r\in\Rset$}
    \State $J_r(m)\gets c_M(m)+L_I\,q_r^I(m)+L_C\,\bigl(1-q_r^I(m)\bigr)\,q_r^C(m)$ \Comment{Eq.~\eqref{eq:loss}}
  \EndFor
  \State $\bar J_t(m)\gets\sum_r\pi_t(r)\,J_r(m)$ for every $m\in\Aset_t$ \Comment{Eq.~\eqref{eq:posteriorloss}}
  \State \Return $\Aset_t$ and $\widehat m_t^\star=\arg\min_{m\in\Aset_t}\bar J_t(m)$ \label{alg:rev:select} \Comment{posterior, not MAP receiver}
\EndFunction
\end{algorithmic}
\end{algorithm}

\begin{algorithm}[!htb]
\caption{Deployment: revision, VoII-gated receiver query, and send}
\label{alg:deployment}
\begin{algorithmic}[1]
\Require intended task $z^\star$, context $x_S$, initial message $m_0$, behavioural history $H_t$, query bank $\Qset$ with likelihoods $\widehat P(y\mid R=r,\xi)$ and costs $c_\xi$, and frozen outputs of \cref{alg:training}
\Ensure the sent message $\widehat m^\star$
\State $\pi_t\gets\pi_\omega(\cdot\mid H_t)$ \Comment{calibrated posterior, the prior if $H_t=\emptyset$}
\State $(\Aset_t,\widehat m_t^\star)\gets\Call{Revise}{z^\star,x_S,m_0,\pi_t}$ \Comment{\cref{alg:revision}}
\State $V_0\gets\min_{m\in\Aset_t}\sum_r\pi_t(r)\,J_r(m)$ \Comment{value without a query}
\For{each query $\xi\in\Qset$} \Comment{the bank may be shortlisted by $\mathrm{IG}(\xi)$}
  \For{each response $y\in\mathcal Y_\xi$}
    \State $\widehat P(y\mid H_t,\xi)\gets\sum_r\pi_t(r)\,\widehat P(y\mid R=r,\xi)$ \Comment{predictive distribution}
    \State $\pi_{t+1}^{\xi,y}(r)\gets\widehat P(y\mid R=r,\xi)\,\pi_t(r)\,/\,\widehat P(y\mid H_t,\xi)$ for every $r$ \Comment{Eq.~\eqref{eq:bayesupdate}}
  \EndFor
  \State $V_\xi\gets\sum_y\widehat P(y\mid H_t,\xi)\,\min_{m\in\Aset_t}\sum_r\pi_{t+1}^{\xi,y}(r)\,J_r(m)$ \Comment{same fixed $\Aset_t$, Eq.~\eqref{eq:queryvalue}}
  \State $\operatorname{NetVoII}(\xi)\gets V_0-V_\xi-c_\xi$ \Comment{Eq.~\eqref{eq:netvoii}}
\EndFor
\State $\xi^\star\gets\arg\max_{\xi\in\Qset}\operatorname{NetVoII}(\xi)$
\If{$\operatorname{NetVoII}(\xi^\star)>0$} \label{alg:dep:gate} \Comment{ask only if the query pays for itself}
  \State ask $\xi^\star$, observe the response $y$, and set $\pi_{t+1}\gets\pi_{t+1}^{\xi^\star,y}$
  \State $\widehat m^\star\gets\arg\min_{m\in\Aset_t}\sum_r\pi_{t+1}(r)\,J_r(m)$ \Comment{re-score the same $\Aset_t$}
\Else
  \State $\widehat m^\star\gets\widehat m_t^\star$
\EndIf
\State send $\widehat m^\star$ to the receiver
\State \Return $\widehat m^\star$
\end{algorithmic}
\end{algorithm}

\subsection{Training Pseudocode}
\label{app:alg_training}

Offline, every message of the training and validation splits is sent to every frozen receiver, which gives the privileged labels of \cref{sec:probe} (lines~\ref{alg:tr:labels}--\ref{alg:tr:exec} of \cref{alg:training}). All learned parts are fitted on training data, selected and calibrated on validation data, and frozen before any headline evaluation. Line~\ref{alg:tr:loss} gives the two losses of \cref{sec:method}: the heads are separate networks, and the capability loss weights each pair by $1-\widetilde y^I$, the share of probes read correctly, a soft form of the condition $\yI=0$ in \cref{eq:caprisk}.

\begin{algorithm}[!htb]
\caption{Offline training with privileged receiver supervision}
\label{alg:training}
\begin{algorithmic}[1]
\Require training and validation messages $m$ (each with intended task $z^\star$, context $x_S$, and contrast tasks), frozen receivers $\Rset=\{R_1,\ldots,R_K\}$, calibration bank $\Qset$ disjoint from the test tasks, predefined mutable features $k=1,\ldots,d$, and probe repeats $K_p$
\Ensure calibrated heads $q_r^I(m)=\qI(m,z^\star,x_S,e_r)$ and $q_r^C(m)=\qC(m,z^\star,x_S,e_r)$, type model $\pi_\omega$, likelihoods $\widehat P(y\mid R=r,\xi)$, and learned risk effects $s_{rk}$
\AlgStage{Stage 1: privileged labels (offline only)}
\For{each message $m$ and each receiver $r\in\Rset$} \label{alg:tr:labels}
  \State $\widetilde y^I_r(m)\gets$
the mean error over parsed probe responses
\Comment{Eq.~\eqref{eq:softlabel}}
  \State $y^A_r(m)\gets\1\{\operatorname{score}(A_r,\theta)<\delta\}$ \label{alg:tr:exec} \Comment{separate execution rollout, Eq.~\eqref{eq:labels}}
\EndFor
\AlgStage{Stage 2: receiver-conditioned risk heads}
\State $\mathcal L^I_r(m)\gets\mathrm{BCE}\bigl(q_r^I(m),\widetilde y^I_r(m)\bigr)$,\quad $\mathcal L^C_r(m)\gets\bigl(1-\widetilde y^I_r(m)\bigr)\,\mathrm{BCE}\bigl(q_r^C(m),y^A_r(m)\bigr)$ \label{alg:tr:loss}
\State fit $\qI$ and $\qC$ as separate networks minimising $\sum_{m,r}\mathcal L^I_r(m)$ and $\sum_{m,r}\mathcal L^C_r(m)$ on the training split \Comment{configuration selected by validation NLL}
\State calibrate $\qI$ and $\qC$ on natural-prevalence validation data \Comment{per-receiver Platt map}
\AlgStage{Stage 3: query likelihoods and receiver-type model}
\For{each query $\xi\in\Qset$ and each receiver $r\in\Rset$}
  \State $\widehat P(y\mid R=r,\xi)\gets$ Laplace-smoothed frequency of $y$ in the stored responses of $r$ to $\xi$
\EndFor
\State build histories $H=\{(\xi_s,y_s)\}$, each from the stored responses of one receiver
\State $\omega\gets\arg\min_\omega\sum_{(H,r)}-\log\pi_\omega(r\mid H)$, then calibrate $\pi_\omega$ on held-out histories \Comment{\cref{app:receiver}}
\AlgStage{Stage 4: learned risk effects of predefined mutable features}
\For{each receiver $r\in\Rset$ and each mutable feature $k$}
  \State $s_{rk}\gets$ training-set risk effect of feature $k$ on the interpretation failure of $r$ \Comment{\cref{app:rewrite}}
\EndFor
\State freeze $\phi$, $\omega$, $\widehat P$, $\{s_{rk}\}$, all prompts, and all thresholds before headline evaluation
\State \Return $\qI,\ \qC,\ \pi_\omega,\ \widehat P,\ \{s_{rk}\}$
\end{algorithmic}
\end{algorithm}

\subsection{Instantiation in Experiments~4 and~5}
\label{app:alg_instances}

Experiment~4 runs \cref{alg:revision} without a query, with the posterior of Experiment~3 after five stored responses. Its risk scorer is the linear model of \cref{app:rewrite}, which predicts the change in soft $\yI$ from the original to a rewrite $m$ as $\sum_r\pi_t(r)\,(b_r+\sum_k s_{rk}\Delta f_k(m))$. In line~\ref{alg:rev:select} the original is kept unless this predicted drop exceeds 0.001, and in line~\ref{alg:rev:guide} $G_t$ holds at most $n_g=3$ features, each with a positive 2.5\% bootstrap quantile of $\bar s_{tk}$. Deterministic checks run before \textsc{Verify}. The other conditions change single steps: the generic rewrite and selection only set $G_t=\emptyset$, the generic rewrite and guidance only send the first valid candidate instead of the $\arg\min$, posterior only passes $\pi_t$ to \textsc{Rewrite} in place of $G_t$, the uniform-belief and true-identity diagnostics replace $\pi_t$, and one feature per step runs \textsc{Rewrite} four times with $n_g=1$, each time on the current message, before selection.

Experiment~5 runs \cref{alg:deployment} with $G_t=\emptyset$ (generic rewrites), frozen verifier decisions, and the frozen paired heads of \cref{app:rewrite} in place of $\qI$. Its bank holds the \QueryBankSize\ fit-role tasks of Experiment~3, a response $y$ is binary execution success, and $L_C=0$ except under the combined objective, which adds the frozen $\qC$. The strict rule is line~\ref{alg:dep:gate} of \cref{alg:deployment} with $c_\xi=\QueryCost$. A budget of $B\%$ instead queries the $B\%$ of episodes with the largest $\operatorname{NetVoII}(\xi^\star)$. The IG control ranks queries and episodes by $\mathrm{IG}(\xi)$ instead, random querying ranks them at random, and under the strict rule both controls are matched to its query count.

\FloatBarrier

\subsection{Full Experimental Details for Reproducibility}
\label{app:fullspec}

\paragraph{Datasets.}
Six KBQA collections in their public releases: FreebaseQA, WebQSP, GrailQA, SQ-WD, PopQA, and Mintaka. Questions are normalised and grouped, and the groups are split 70/15/15 into train/validation/test. Test questions per dataset: 1,310 / 219 / 1,286 / 1,418 / 1,451 / 945 (6,629 in total), giving 33,145 message--receiver pairs. Each question yields one message from the PromptSource template \texttt{qa\_template\_basic}, one intended task, and one contrast task from the fixed contrast set (entity-versus-category or value-versus-attribute). No synthetic enrichment is used at any stage.

\paragraph{Models.}
Receivers: Nemotron-3 Nano 30B, Gemma 4 26B, gpt-oss-20b, Mistral Small 3.2, and Qwen3.6 35B, all served frozen through vLLM at temperature 0 with default system prompts and no tools. Probe generation, probe verification, and the A1/A2 judges use \RewriteModel. Predictor features come from \EmbeddingModel\ (4,096 dimensions).

\paragraph{Interpretation probe.}
Three options per probe (intended task, contrast task, \textsc{None of these}) and six differently worded instances per message--receiver pair, each a separate temperature-0 call. Probe text is generated per message, checked deterministically (option distinctness, no answer leakage), and verified by a second call. Parsed responses that choose the contrast task or \textsc{None of these} count as wrong, and unparsed responses are dropped from the denominator. Execution is one separate answer call scored by gold-alias containment.

\paragraph{Receiver histories and query bank.}
Calibration bank of 384 tasks disjoint from the test messages, answered by all five receivers (9,600 archived responses). Histories are sampled from the archive at lengths 1, 3, 5, 10, and 20 (20,800 in total), and length 0 is the uniform prior. Tasks are split by role into 192 fit / 48 early-stop / 48 calibration / 96 audit, and every reported number uses 960 audit histories per length. The shuffled-history control pairs each query with responses from a randomly chosen receiver.

\paragraph{Training details.}
PIR heads: two separate MLPs ($q^I$ and $q^C$) with hidden widths in $\{[256,64],[64,16],[512,128]\}$, dropout 0.3--0.7, and a 16-dimensional receiver embedding. Inputs are the embeddings of $m$ and $x_S$ and optionally their difference, chosen per dataset. The MLP output is added to a fixed logistic offset fitted first on the training data (per-receiver intercept and slope on the gap $g$, one intercept and slope for A3). 27 configurations times 3 seeds per dataset, selected on validation NLL only, followed by a Platt map with per-receiver slope and intercept and shared terms in $g$ (one global map for A3), fitted on the calibration part of the validation split. $q^C$ is trained on $\yA$ with example weights $1-\widetilde y^I$. Receiver-type model: PCA to 32 dimensions, an interaction MLP between query and response embeddings, mean pooling over the history, and a five-way softmax, with 3 seeds times 2 regularisation strengths, early-stopped on the early-stop role split and temperature-calibrated on the calibration role split.

\paragraph{Rewrite module.}
Number of generated candidates $K_m=\RewriteCandidates$ per request, with no regeneration. In Experiment~4 the rewriter (\RewriteModel, temperature 0.4, eight requests of one condition per call) receives the guidance of its condition, the verifier is one stateless batched call, and selection uses the linear risk model with no further LLM call. In Experiment~5 the pool is four generic rewrites per message, verified with the frozen v1.1 rules (acceptance \SemanticVerifierPassRate\ of candidates) and scored by the paired heads, which add one embedding call per candidate. Both models are described in \cref{app:rewrite}.

\paragraph{Query module.}
Query bank: the \QueryBankSize\ fit-role tasks of Experiment~3 with their archived responses from all five receivers. Signature: binary execution success against the known bank answer. Likelihoods: Laplace-smoothed per receiver and task. Update: one exact Bayes step from the frozen posterior of Experiment~3 after five stored responses, with at most one query per episode. Candidate pool: the original message and its four generic rewrites with the frozen verifier decisions (rejected slots map to the original). Scores: the six frozen paired heads of \cref{app:rewrite}, and the frozen $q^C$ for the combined objective, which was not refit on rewrites. Budgets are exact quotas over pre-response scores within each cohort, random schedules use three pre-declared seeds, and the strict rule uses six frozen costs. The adapted VoI rule \citep{dong2026voi} enumerates the same finite posterior with \RewriteModel\ estimates of candidate and receiver utilities (2,260 calls, with 122 of 9,029 estimate items falling back to training base rates under a pre-declared rule). Ties in expected loss go to the earlier candidate (the original first), as for every policy (\cref{app:res_e5}). All schedules were frozen before any receiver call. Experiment~5 then issued \QueryReceiverCalls\ receiver calls (642,210 queries, 359,627 executions, and 2,157,762 probes over the 135,435 episodes of both cohorts).

\paragraph{Baselines and ablations.}
A0--A5 as defined in \cref{tab:baseline_defs}, plus the shuffled-receiver control and the receiver-calibrated A3 (A3 predictions passed through the full model's calibration map, refitted on the same validation questions). A1 and A2 are \RewriteModel\ judge calls with a fixed rubric. Of 26,528 calls, 54 returned unparsable output and fall back to the training-mean score. A4 shares the architecture, inputs, and selection protocol of the full model but is trained on $\yA$. A5 is a listener classifier over the three probe options with the same inputs as the full model. Rewrite ablations follow \cref{app:ablations} and are reported in \cref{sec:res_e4}, and query controls are reported in \cref{sec:res_e5} and \cref{tab:e5_budget_grid,tab:e5_cost_sweep,tab:e5_per_dataset}.

\paragraph{Evaluation metrics.}
Prediction: AUROC, AUPRC, Brier, NLL, and ECE (10 equal-mass bins), computed within each receiver and macro-averaged in the main text, and pooled over receivers in \cref{tab:receiver_ablation}. Receiver inference: accuracy, NLL, Brier (summed over classes), and ECE (10 equal-width bins). Rewrite: measured $\yI$ and $\yA$ per condition, change rate, and verifier acceptance, with paired question-cluster bootstrap intervals. Query control: share of episodes queried, measured $\yI$ and $1-\yA$, net utility $1-\yI-c\,r$ at the stated cost, the combined loss $\yI+(1-\yI)\yA$, and the share of posteriors whose mode is the true receiver, with paired question-cluster bootstrap intervals.

\paragraph{Statistical reporting.}
Main-text values of Experiment~2 use seed 0 with 1,000 paired bootstrap resamples over frozen question groups, paired across methods and receivers. Three-seed means appear in \cref{tab:receiver_ablation}. Intervals are binomial with $n=960$ in Experiment~3 and question-level normal approximations in Experiment~1. In Experiment~4 they are 10,000 paired bootstrap resamples over the 1,579 held-out questions, with Holm correction over the four pre-registered comparisons. Per-receiver, per-dataset, and secondary contrasts are unadjusted. In Experiment~5 they are \QueryBootstrapResamples\ paired bootstrap resamples over question clusters that keep all five receivers and three repeats together, with Holm correction over the eight pre-registered net-utility contrasts on the primary cohort (NetVoII against IG and against random querying at four budgets). The replication cohort, the combined objective, the strict-rule comparison with no query, and per-dataset contrasts are unadjusted. Model selection, calibration, and thresholds use validation data only.

%

\section{Implementation and Reproducibility Checklist}
\label{app:implementation}

For each receiver--message rollout, log: run ID, code commit, timestamp, seed, dataset/split/task-family/item identifiers, intended-task and distractor IDs, sender/receiver model versions, receiver-history IDs, current posterior $\pi_t$, raw message and template ID, complete probe options and order, raw/parsed probe response, $Y^I$, raw execution response and $Y^A$, raw and calibrated head outputs, query-bank scores and response likelihoods, entropy reduction, gross/net VoII, rewrite candidates, semantic-verifier decisions/reasons, final message, token counts, latency, and monetary/API cost where available.

Raw model outputs are stored separately from parsed labels so that parsers and scoring rules can be rerun without repeating inference. No private chain-of-thought is requested or treated as ground truth.

\paragraph{Statistical reporting.}
For each headline table, report the point estimate and a paired 95\% bootstrap CI over question groups, paired across methods and receivers. Report macro receiver averages together with pooled results. Thresholds, calibration, cost parameters, and model-selection decisions are fixed using validation data only.

\paragraph{Natural prevalence.}
All splits are frozen at natural prevalence, and the reported experiments use no oversampling and no synthetic examples. If enrichment is used in later training, the predictor must be recalibrated on the untouched natural validation set before Brier/NLL/ECE or probability-based decision rules are evaluated.

\section{Discussion and Limitations}
\label{app:discussion}

This section interprets the results of \cref{sec:results,app:results} in the light of \cref{sec:theory}, states the limits of the evidence, and outlines extensions.

\subsection{What the Results Show}
\label{app:disc_results}

\paragraph{The premises of PIR hold.}
The five experiments support the framework step by step. Receivers read the same messages differently: pooled $\yI$ ranges from \ReceiverYIMin\ to \ReceiverYIMax, the best and worst receiver differ by 4--13$\times$ within each dataset, and the lowest-risk receiver changes across datasets (Experiment~1). Interpretation failure is its own target: \MisreadPassShare\ of misread pairs still pass the answer check (Experiment~1), and a predictor trained on execution failure is near chance on $\yI$ (Experiment~2). The risk can be predicted before sending (AUROC \PIRAUROC\ within receivers), and receiver information cuts ECE from \AgnosticECE\ to \PIRECE, mostly by learning each receiver's risk level (Experiment~2). A short behavioural history sharpens the posterior over receiver types without naming the receiver (Experiment~3). On these premises, risk-guided revision lowers measured $\yI$ from 4.16\% to 2.31\% while execution failure stays within a 1-point non-inferiority margin of every reference (Experiment~4). At a 20\% query budget, NetVoII closes 78\% of the gap between no query and the true identity, against 5\% for IG and 3\% for random querying (Experiment~5).

\paragraph{When receiver information pays off.}
Heterogeneity alone does not make receiver information useful, as \cref{thm:conditioning} states. Receivers differ several-fold in how often they misread, yet in Experiment~4 the receiver belief barely matters: a uniform belief changes 2 of 7,895 selection decisions, and the true identity gives almost the same $\yI$ (2.30\% vs.\ 2.31\%). The learned risk scores show why. An unstated output is the top risk feature for all five receivers, so the best message hardly depends on the receiver and $\Delta_E\approx0$. Receiver information pays off where it changes a decision. In Experiment~2 it corrects the risk level of each receiver and, when one message is ranked across receivers, lifts AUROC from 0.701 to 0.800. A receiver-specific recalibration of A3 recovers most of both gains, and much of the ranking gain reflects Nemotron's position effect (\cref{app:res_e2,app:position}). In Experiment~5 it decides which episodes deserve a query.

\paragraph{Why IG is not enough.}
Experiment~5 shows the gap between information and value that \cref{thm:frontier} describes. The most informative query carries about the same information in every episode (0.19 nats), yet predicted value is positive in only \VoIIPositiveShare\ of episodes, from 1.4\% on WebQSP and GrailQA to 73--75\% on PopQA and Mintaka. IG identifies the receiver at least as well as NetVoII: at a 20\% budget, its posterior mode is the true receiver in 28.5\% of episodes, against 28.3\% for NetVoII. Yet its answers rarely change the preferred message, so $\yI$ barely moves. A query has value only when its answer can move the posterior into the decision region of another message, and NetVoII spends its budget on the episodes where it predicts such a move.

\paragraph{Implications for LLM systems.}
Three lessons follow for systems in which models pass messages to one another. First, interpretation should be scored separately from execution: an answer-based check counts \MisreadPassShare\ of misreads as successes. Second, most of the revision gain comes from one fix shared by all receivers, stating the expected output, which the learned risk scores rank first for every receiver. A sender can apply such a fix before it knows who will read the message, and receiver-specific adaptation is needed only where receivers disagree about the best message. Third, the sender should keep its uncertainty about the receiver and query only when the expected gain exceeds the cost. Predicted value is concentrated in a minority of episodes, and querying every episode gives lower net utility than never querying (\cref{tab:e5_policies}). In large agent networks, querying by default can also flood the system with queries.

\subsection{Limitations}
\label{app:limitations}

\paragraph{Measurement.}
The probe measures interpretation through behaviour, not through internal representations: it records which of three options the receiver picks, the intended task, one contrast task, or \textsc{None of these}. A misreading other than the offered contrast can register only as \textsc{None of these}, so each message tests one direction of misreading. Option order can bias the pick. Nemotron tends to avoid the last listed option: it misreads 24.5\% of probes that list the intended task last, against 3.8\% of the other probes. The gap appears under each of the six probe wordings, so it is an effect of position. Receiver heterogeneity survives when only probes with the intended task first or second are kept (3.8$\times$ between the highest and lowest receivers, \cref{app:position}), but most of Nemotron's excess $\yI$, and much of the across-receiver ranking gain of Experiment~2, reflect this effect. Execution is scored by gold-alias containment, which can miss paraphrased answers and accept answers that only mention the gold entity. Finally, misreadings are rare, about \NaturalPIRRate\ of test pairs, so estimates for a single receiver, dataset, or question type have wide intervals.

\paragraph{Setting.}
The evidence comes from a narrow setting. All messages are knowledge-base questions rendered by one fixed template, so no sender model writes them, and each carries one of two contrast types (entity versus category, value versus attribute). Real handoffs between agents are longer, carry intermediate results, and can be misread in more ways. The receivers are five open-weight models run at temperature 0, and the type model is closed-set: a receiver from a new family would be assigned to one of the five known types. Identification is also weak in absolute terms. After twenty stored responses, top-1 accuracy is \HistoryLongAcc\ against \HistoryZeroAcc\ for the uniform prior, so the posterior stays broad (\cref{app:res_e3}).

\paragraph{Control.}
Revision uses one rewriter, \RewriteModel, and the semantic verifier is the same model version, called separately and without any optimisation signal. An audit of 12,402 sent messages found that 98.1\% keep the task and 1.1\% add a hint (\cref{app:res_e4}). The candidate set is finite, so a safer equivalent message can be missed. Our evaluation therefore concerns selection among the generated, verified candidates rather than global prompt optimality.  Queries come from a fixed bank of \QueryBankSize\ tasks, with at most one query per episode and no open-ended dialogue. The query gain is small in absolute terms: the true identity would lower $\yI$ by 0.06 points, and NetVoII lowers it by about 0.05. The gain is in interpretation rather than task success, and on the replication cohort its intervals include zero (\cref{app:res_e5}). The adapted VoI rule is a weak comparator: its LLM estimates are too coarse to let a query change the message, so it tests these estimates rather than the VoI rule. \Cref{app:scope} lists the studies that lie outside the reported experiments.

\subsection{Future Directions}
\label{app:future}

Three extensions follow from these limits. Open-set receiver modelling would detect a receiver that matches none of the known types instead of forcing it into one. Multi-round interaction would let the sender query, revise, and query again, and longer communication chains would show how interpretation errors spread across hops. Richer tasks, such as tool calls, code, and multi-step plans, would test whether PIR and VoII carry over when the set of possible readings is open-ended.

\end{document}